\documentclass[onecolumn,10pt]{article}
\usepackage{amssymb}
\usepackage{times}
\usepackage{named}
\usepackage{makeidx}
\makeindex
\newcommand{\braket}[2]{\langle#1|#2\rangle}
\newcommand{\abs}[1]{|#1|}
\newcommand{\norm}[1]{\| #1\|}
\newcommand{\alg}[1]{\mbox{$\mathfrak{#1}$}}
\newcommand{\trace}[1]{\mbox{$\mathrm{Tr}$}(#1)}
\newcommand{\ptrace}[2]{\mbox{$\mathrm{Tr}_{#1}$}(#2)}

\newcommand{\hil}[1]{\mbox{$\mathcal{#1}$}}
\newcommand{\dimn}{\mbox{$\mathrm{dim}$}}
\newcommand{\ket}[1]{| #1 \rangle}
\newcommand{\bra}[1]{\langle #1 |}

\begin{document}

\title{\textbf{Quantum Information\\ and Computation}}
  \vspace{1in}

\author{\textbf{Jeffrey Bub}\\
\footnotesize Department of Philosophy, University of Maryland,
College Park, MD 20742\footnote{\textit{E-mail address:}
jbub@umd.edu}}

\bibliographystyle{named}

\maketitle

\begin{abstract}
This Chapter deals with  theoretical developments in the subject of quantum information and quantum computation, and includes an overview of classical information and some relevant quantum mechanics. The discussion covers topics in quantum communication, quantum cryptography, and quantum computation, and  concludes by considering whether a perspective in terms of quantum information sheds new light on the conceptual problems of quantum mechanics.
\end{abstract}


\tableofcontents

\section{Introduction}

The subject of quantum information has its roots in the debate about conceptual issues in the foundations of quantum mechanics.

The story really begins with the dispute between Einstein \index{Einstein}  and Bohr \index{Bohr} about the interpretation of quantum states, in particular the interpretation of so-called `entangled states,' which exhibit peculiar nonlocal statistical correlations for widely separated quantum systems. See, for example, \cite[p. 283]{Bohr} and Einstein's reply in the same volume \cite{Schilpp}. Einstein took the position that quantum mechanics is simply an incomplete theory. On the basis of a certain restricted set of correlations for a pair of systems in a particular entangled state, Einstein, Podolsky, and Rosen (EPR) \index{Einstein, Podolsky, and Rosen (EPR)} argued in a seminal paper \cite{EPR}  that the phenomenon of entanglement conflicts with certain basic realist principles of separability and locality that all physical theories should respect, unless we regard quantum states as incomplete descriptions. 

Bohr's view, which he termed `complementarity,' 
\index{complementarity} eventually became entrenched as the orthodox Copenhagen 
interpretation\index{Copenhagen interpretation}, a patchwork of reformulations by Heisenberg, Pauli, von Neumann, Dirac, Wheeler, and others. (For a discussion, see \cite{Howard} and Landsman, this vol., ch. 5.) As Pauli\index{Pauli}  put it in correspondence with Max Born \cite[p. 218]{Born}, a `detached observer' description\index{detached observer} of the sort provided by classical physics is precluded by the nature of quantum phenomena, and a quantum description of events is as complete as it can be (in principle). Any application of quantum theory requires a `cut' between the observer and the observed, or the macroscopic measuring instrument and the measured system, so that the description is in a certain sense contextual, where the relevant context is defined by the whole macroscopic experimental arrangement. So, for example, a `position measurement context' provides information about position but excludes, in principle, the possibility of simultaneously obtaining momentum information, because there is no fact of the matter about momentum in this context: the momentum value is indeterminate. The Copenhagen interpretation\index{Copenhagen interpretation} conflicts with Einstein's realism, his `philosophical prejudice,' as Pauli characterized it in a letter to  Born \cite[p. 221]{Born}, that lies at the heart of the dispute between Einstein and Bohr about the significance of the transition from classical to quantum mechanics.

The 1990's saw the development of a quantum theory of information, based on the realization that entanglement\index{entanglement}, rather than being a minor source of embarrassment for physics that need only concern philosophers, can actually be exploited as a nonclassical communication channel to perform information-processing tasks that would be impossible in a classical world. In a two-part commentary on the EPR\index{EPR} paper, Schr\"{o}dinger\index{Schr\"{o}dinger} \shortcite[p. 555]{Schr1} identified entanglement as `\textit{the} characteristic trait of quantum theory, the one that enforces its entire departure from classical lines of thought.' This has led to an explosive surge of research among physicists and computer scientists on the application of information-theoretic ideas to quantum computation\index{quantum!computation} (which exploits entanglement in the design of a quantum computer\index{quantum!computer}, so as to enable the efficient performance of certain computational tasks), to quantum communication\index{quantum!communication} (new forms of `entanglement-assisted' communication, such as quantum teleportation\index{quantum!teleportation}), and to quantum cryptography\index{quantum!cryptography} (the identification of cryptographic protocols that are guaranteed to be unconditionally secure against eavesdropping or cheating, by the laws of quantum mechanics, even if all parties have access to quantum computers).

Some milestones: Bell's\index{Bell} analysis \shortcite{BellEPR} turned the EPR argument on its head by showing that Einstein's assumptions of separability and locality, applicable in classical physics and underlying the EPR incompleteness argument, are incompatible with certain quantum statistical correlations (not explicitly considered by EPR) of separated systems in EPR-type entangled states. Later experiments  \cite{Aspect1,Aspect2} confirmed these nonclassical correlations in set-ups that excluded the possibility of any sort of physically plausible, non-superluminal, classical communication between the separated systems.

In the 1980s, various authors, e.g.,  Wiesner\index{Wiesner}, Bennett\index{Bennett}, and Brassard\index{Brassard} \cite{Wiesner,BB84,BBBW82} pointed out that one could exploit features of the measurement process in quantum mechanics to thwart the possibility of undetected eavesdropping in certain cryptographic procedures, specifically in key distribution\index{key distribution}---a procedure where two parties, Alice and Bob, who initially share no information end up each holding a secret random key which can be used to send encrypted messages between them.  No third party, Eve, can obtain any information about the communications between Alice and Bob that led to the establishment of the key, without Alice and Bob becoming aware of Eve's interference, because Eve's measurements necessarily disturb the quantum states of the systems in the communication channel. 

Bennett\index{Bennett}  \shortcite{Bennett73} showed how to make a universal Turing machine\index{Turing machine!universal} reversible for any computation, a required step in the design of a quantum computer
that evolves via unitary (and hence reversible) state transformations, and Benioff\index{Benioff} \shortcite{Benioff} developed Hamiltonian models for computer computers. Feynman\index{Feynman} \shortcite{Feynman} considered the problem of efficiently simulating the evolution of physical systems using quantum resources (noting that the classical simulation of a quantum process would be exponentially costly), which involves the idea of a quantum computation, but it was Deutsch\index{Deutsch} \shortcite{Deutsch1985,Deutsch1989} who characterized the essential features of a universal quantum computer and formulated the first genuinely quantum algorithm.

Following Duetsch's work on quantum logic gates and quantum networks, several quantum algorithms were proposed for performing computational tasks more efficiently than any known classical algorithm, or in some cases more efficiently than any classical algorithm. The most spectacular of these is Shor's\index{Shor} algorithm\index{Shor} \shortcite{Shor94,Shor97} for finding the two prime factors of a positive integer $N = pq$, which is exponentially faster than the best-known classical algorithm. Since prime factorization\index{prime factorization} is the basis of the most widely used public key encryption scheme (currently universally applied in communications between banks and commercial transactions over the internet), Shor's result has enormous practical significance.  

In the following, I present an account of some of the theoretical developments in quantum information, quantum communication, quantum cryptography, and  quantum computation. I conclude by considering whether a perspective in terms of quantum information suggests a new way of resolving the foundational problems of quantum mechanics that were the focus of the debate between Einstein and Bohr. 

My discussion is heavily indebted to Michael Nielsen and Isaac Chuang's illuminating and comprehensive  \textit{Quantum Computation and Quantum Information}  \shortcite{NielsenChuang}, and to several insightful review articles:  `The Joy of Entanglement'  by Sandu Popescu and Daniel Rohrlich \shortcite{PopRohrlich},  `Quantum Information and its Properties'  by Richard Jozsa \shortcite{Jozsa98}, and `Quantum Computing'  by Andrew Steane \shortcite{Steane98}. 

\section{Classical Information\index{information!classical}}

\label{sec:Shannon}

\subsection{Classical Information Compression\index{information!compression!classical} and Shannon Entropy\index{Shannon entropy}\index{entropy!Shannon}}

\label{sec:classinfocomp}

In this section, I review the basic elements of classical information theory\index{information!classical}. In \S 2.1, I introduce the notion of the Shannon entropy\index{Shannon entropy}\index{entropy!Shannon} of an information source and the fundamental idea of information compression in Shannon's source coding theorem\index{Shannon's source coding theorem (noiseless channel coding theorem)} (or noiseless channel coding theorem). In \S 2.2, I define some information-theoretic concepts relevant to Shannon's noisy channel coding theorem.

The classical theory of information was initially developed to deal with certain questions in the communication of electrical signals. Shannon's \index{Shannon}ground-breaking paper `A Mathematical Theory of Communication'  \cite{Shannon} followed earlier work by people like Nyquist \shortcite{Nyquist} and Hartley \shortcite{Hartley} in the 1920s. The basic problem was the representation of messages, selected from an ensemble generated by a stochastic process at the message source, in such a way as to ensure their efficient transmission over an electrical circuit such as a noisy telegraph wire.

A communication set-up involves a transmitter or source of information\index{information!source}, a (possibly noisy) channel, and a receiver. The source produces messages in the form of sequences of symbols from some alphabet, which Shannon represented mathematically as sequences of values of independent, identically distributed random variables. In later idealizations, the source is represented as \textit{stationary}\index{information!source!stationary}, in the sense (roughly) that the probability of any symbol (or $n$-tuple of symbols) appearing at any given position in a (very long) sequence, when that position is considered with respect to an ensemble of possible sequences, is the same for all positions in the sequence, and \textit{ergodic}\index{information!source!ergodic}, in the sense that this `ensemble average' probability is equal to the `time average'  probability, where the time average refers to the probability of a symbol (or $n$-tuple of symbols) in a given (very long) sequence.

The fundamental question considered by Shannon\index{Shannon} was how to quantify the minimal physical resources required to store messages produced by a source, so that they could be communicated via a channel without loss and reconstructed by a receiver. Shannon's source coding theorem (or noiseless channel coding theorem) answers this question. 

To see the idea behind the theorem, consider a source that produces long sequences (messages) composed of symbols from a finite alphabet $a_{1}, a_{2}, \ldots, a_{k}$, where the individual symbols are produced  with probabilities $p_{1}, p_{2},\ldots, p_{k}$. A given sequence of symbols is represented as a sequence of values of independent, identically distributed, discrete random variables $X_{1}, X_{2},\ldots$. A \textit{typical} sequence\index{typical sequence} of length $n$, for large $n$, will contain close to $p_{i}n$ symbols $a_{i}$, for $ i = 1, \ldots, n$. So the probability of a sufficiently long typical sequence (assuming independence) will be:
\begin{equation}
p(x_{1},x_{2}, \ldots, x_{n}) = p(x_{1})p(x_{2})\ldots p(x_{n}) \approx 
p_{1}^{p_{1}n}p_{2}^{p_{2}n}\ldots p_{k}^{p_{k}n}.
\end{equation}
Taking the logarithm of both sides (conventionally, in information theory, to the base 2) yields:
\begin{equation}
\log p(x_{1},\ldots, x_{n}) \approx n\sum_{i}p_{i}\log p_{i} := -nH(X)
\end{equation}
where $H(X) :=  -\sum_{i}p_{i}\log p_{i}$ is the \textit{Shannon entropy}\index{Shannon entropy}\index{entropy!Shannon} of the source.

We can think about information in Shannon's sense in various ways. We can take $-\log p_{i}$, a decreasing function of $p_{i}$ with a minimum value of 0 when $p_{i}=1$ for some $i$, as a measure of the information associated with identifying the symbol $a_{i}$ produced by an information source. Then $H(X) = -\sum_{i}p_{i} \log p_{i}$ is the average information gain, or the expectation value of the information gain associated with ascertaining the value of the random variable $X$. Alternatively, we can think of the entropy as a measure of the amount of uncertainty about $X$ before we ascertain its value. A source that produces one of two distinguishable symbols with equal probability, such as the toss of a fair coin, is said to have a Shannon entropy of 1 bit\index{bit}: ascertaining which symbol is produced, or reducing one's uncertainty about which symbol is produced, is associated with an amount of information equal to 1 bit.\footnote{Note that the term `bit'\index{bit} (for `binary digit') is used to refer to the basic unit of classical information in terms of Shannon entropy, and to an elementary two-state classical system considered as representing the possible outputs of an elementary classical information source.} If we already know which symbol will be produced (so the probabilities are 1 and 0), the entropy is 0: there is no uncertainty, and no information gain.

Since
\begin{equation}
p(x_{1},\ldots, x_{n}) = 2^{-nH(X)}
\end{equation}
for sufficiently long typical sequences\index{typical sequence}, and the probability of all the typical $n$-length sequences is less than 1, it follows that there are at most $2^{nH(X)}$ typical sequences. In fact, if the $p_{i}$ are not all equal, the typical sequences comprise an exponentially small set $T$ (of equiprobable typical sequences\index{typical sequence}) in the set of all sequences as $n \rightarrow \infty$, but since the probability that the source produces an atypical sequence tends to zero as $n \rightarrow \infty$, the set of typical sequences has probability close to 1. So each typical $n$-sequence could be encoded as a distinct binary number of $nH(X)$ binary digits or bits before being sent through the channel to the receiver, where the original sequence could then be reconstructed by inverting the 1--1 encoding map. (The reconstruction would fail, with low probability, only for the rare atypical sequences, each of which could be encoded as, say, a string of 0's.) 
 
Notice that if the probabilities $p_{i}$ are all equal ($p_{i} = 1/k$ for all $i$), then $H(X) = \log k$, and if some $p_{j} = 1$ (and so $p_{i} = 0$ for $i \neq j$), then $H(X) = 0$ (taking $0\log 0 = \lim_{x \rightarrow 0}x\log x = 0)$. It can easily be shown that:
\begin{equation}
0 \leq H(X) \leq \log k.
\end{equation}
If we encoded each of the $k$ distinct symbols as a distinct binary number, i.e., as a distinct string of 0's and 1's, we would need binary numbers composed of $\log k$ bits to represent each symbol ($2^{\log k} = k$).  So Shannon's analysis shows that messages produced by a stochastic source can be compressed, in the sense that (as $n \rightarrow \infty$ and the probability of an atypical $n$-length sequence tends to zero) $n$-length sequences can be encoded without loss of information using $nH(X)$ bits rather than the $n\log k$ bits required if we encoded each of the $k$ symbols $a_{i}$ as a distinct string of 0's and 1's: this \emph{is} a compression, since $nH(X) < n\log k$ except for equiprobable distributions. 

More precisely, let $\overline{X} = \frac{1}{n}(X_{1} + X_{2} + \ldots + X_{n})$, where $X_{1}, X_{2}, \ldots, X_{n}$ are $n$ independent and identically distributed random variables with mean $<X>$ and finite variance. The weak law of large numbers tells us that, for any $\epsilon, \delta > 0$,
\begin{equation}
\mbox{Pr}(|\overline{X} -<X>| \geq \delta) < \epsilon
\end{equation}
for sufficiently large $n$. 

Now consider a random variable $X$ that takes values $x$ in an alphabet $\hil{X}$ with probabilities $p(x) = \mbox{Pr}(X=x), x \in \hil{X}$. \footnote{Note that $p(x)$ is an abbreviation for $p_{X}(x)$, so $p(x)$ and $p(y)$ refer to two different random variables. The expression $\mbox{Pr}(X \in S)$ = $\sum_{x \in S}p(x)$ denotes the probability that the random variable $X$ takes a value in the set $S$, and $\mbox{Pr}(X = x)$ denotes the probability that $X$ takes the value $x$. The expression $p(x_{1}, x_{2}, \ldots, x_{n})$ denotes the probability that the sequence of random variables $X_{1}, X_{2}, \ldots, X_{n}$ takes the sequence of values $(x_{1}, x_{2}, \ldots, x_{n})$. The discussion here follows Cover and Thomas \shortcite{CoverThomas} and I use their notation.} Let 
\begin{equation}
Z = -\log p(X)
\end{equation}
be a function of $X$ that takes the value $-\log p(x)$ when $X$ takes the value $x$. Then
\begin{equation}
<Z> = - \sum_{x \in \mathcal{X}}p(x)\log p(x) = H(X)
\end{equation}
and for a sequence of $n$ independent and identically distributed random variables \newline $X_{1}, X_{2}, \ldots, X_{n}$:
\begin{eqnarray}
-\frac{1}{n}\log p(X_{1}, \ldots, X_{n}) & = & -\frac{1}{n}\sum_{i}\log p(X_{i}) \nonumber \\
& = & \frac{1}{n}(Z_{1}+ \ldots +Z_{n}) \nonumber \\
& = & \overline{Z}.
\end{eqnarray}
So, by the weak law of large numbers, for $\epsilon, \delta > 0$ and sufficiently large $n$:
\begin{equation}
\mbox{Pr}(|\overline{Z} - <Z>| \geq\delta) < \epsilon
\end{equation}
i.e.,
\begin{equation}
\mbox{Pr}(|-\frac{1}{n}\log p(X_{1}, \ldots, X_{n}) - H(X)| \geq\delta) < \epsilon
\end{equation}
or equivalently,
\begin{equation}
\mbox{Pr}(|-\frac{1}{n}\log p(X_{1}, \ldots,X_{n}) - H(X)| < \delta) \geq 1-\epsilon 
\end{equation}
and hence, with probability greater than or equal to $1-\epsilon$:
\begin{equation}
-n(H(X)+\delta) < \log p(X_{1}, \ldots, X_{n}) < -n(H(X)-\delta).
\end{equation}

A `$\delta$-typical $n$-length sequence' $(x_{1}, \ldots, x_{n}) \in \mathcal{X}^{n}$ of values of the random variables $X_{1}, \ldots, X_{n}$ is defined as a sequence of symbols of $\mathcal{X}$ satisfying:
\begin{equation}
2^{-n(H(X)+\delta)} \leq p(x_{1}, \ldots, x_{n})  \leq 2^{-n(H(X)-\delta)}. \label{eq:typical}
\end{equation}
Denote the set of $\delta$-typical $n$-length sequences by $T^{(n)}_{\delta}$ and the number of sequences in $T^{(n)}_{\delta}$ by $|T^{(n)}_{\delta}|$. Then, for sufficiently large n,
\begin{equation}
  \mbox{Pr}(\{X_{1}, \ldots, X_{n}\} \in T^{(n)}_{\delta}) \geq 1-\epsilon;
\end{equation}
and it can be shown that
\begin{equation}
(1-\epsilon)2^{n(H(X)-\delta)} \leq |T^{(n)}_{\delta}| \leq 2^{n(H(X)+\delta)}.
\end{equation}
So, roughly, $T^{(n)}$ contains $2^{nH}$ equiprobable sequences, each having a probability of $2^{-nH}$.

Shannon's source coding theorem\index{Shannon's source coding theorem (noiseless channel coding theorem)} applies the above result about typical sequences to show  that the compression rate\index{information!compression rate} of $H(X)$ bits per symbol produced by a source of independent and identically distributed random variables is optimal. The source produces $n$-length sequences of symbols $x_{1}, x_{2}, \ldots, x_{n}$ with probability $p(x_{1}, x_{2}, \ldots, x_{n}) = p(x_{1})p(x_{2})\ldots p(x_{n})$, where each symbol is chosen from an alphabet $\hil{X}$. If there are $k$ symbols in \hil{X}, these $n$-sequences can be represented as sequences of $n\log k$ bits. Suppose there is a `block coding' compression scheme\index{information!compression!`block coding'} that encodes each `block' or $n$-length sequence (for sufficiently large $n$) as a shorter sequence of $nR$ bits, where $0 \leq R \leq \log k$. Suppose also that the receiver has a decompression scheme for decoding sequences of $nR$ bits into sequences of $n$ symbols. Then one speaks of a compression/decompression scheme\index{compression/decompression scheme} of \emph{rate $R$}. 

The source coding theorem states that
\begin{quote}
if the Shannon entropy of a source is $H(X)$, then there exists a reliable compression/decompression scheme\index{compression/decompression scheme} of rate $R$ if and only if $R \geq H(X)$, where a scheme is said to be reliable if it reproduces the original sequence with a probability that tends to 1 as $n \rightarrow \infty$.
\end{quote}

For reliable communication, we want the compression and decompression of a sequence of symbols to yield the original sequence, but in general there will be a certain probability, $q(x_{1}, \ldots, x_{n})$, of decoding a given sequence of $nR$ encoded bits received by the receiver as the original $n$-sequence produced by the source. The average \index{fidelity}\textit{fidelity}\footnote{\label{foot:NielsenChuang}Note that this definition of fidelity is different from the definition proposed by Nielsen and Chuang \shortcite[p.400]{NielsenChuang} for the fidelity between two probability distributions $\{p_{x}\}$ and $\{q_{x}\}$ as a  `distance measure` between the distributions.  They define $F_{NC}(p_{x},q_{x}) := \sum_{x}\sqrt{p_{x}q_{x}}$, so $F_{NC}(p_{x},q_{x}) = 1$ if $p_{x} = q_{x}$.} of a compression/decompression scheme for $n$-length blocks is defined as:
\begin{equation}
F_{n} = \sum_{\mbox{all }n\mbox{-sequences}}p(x_{1},\ldots, x_{n})q(x_{1},\ldots, x_{n})
\end{equation}
If all the probabilities $q(x_{1},\ldots, x_{n})$ are 1, $F_{n} = 1$; otherwise $F_{n} < 1$. In terms of the fidelity as a measure of reliability of correct decoding, the source coding theorem states that \begin{quote}
for any $\epsilon, \delta > 0$: (i) there exists a compression/decompression scheme using $H(X) + \delta$ bits per symbol for $n$-length sequences produced by the source that can be decompressed by the receiver with a fidelity $F_{n} > 1-\epsilon$, for sufficiently large $n$, and (ii) any compression/decompression scheme using $H(X) - \delta$ bits per symbol for $n$-length sequences will have a fidelity $F_{n} < \epsilon$, for sufficiently large $n$.
\end{quote}

As a simple example of compression, consider an information source\index{information!source} that produces sequences of symbols from a 4-symbol alphabet $a_{1}, a_{2}, a_{3}, a_{4}$ with probabilities 1/2, 1/4, 1/8, 1/8. Each symbol can be represented by a distinct 2-digit binary number:
\begin{eqnarray*}
a_{1} & :  & 00 \\
a_{2} & :  & 01 \\
a_{3} & : & 10 \\
a_{4} & :  & 11
\end{eqnarray*}
so without compression \index{information!compression} we need two bits per symbol of storage space to store the output of the source. The Shannon entropy of the source is $H(X) = -\frac{1}{2}\log \frac{1}{2} -\frac{1}{4}\log \frac{1}{4} -\frac{1}{8}\log \frac{1}{8} -\frac{1}{8}\log \frac{1}{8}  = \frac{7}{4}$. Shannon's source coding theorem tells us that there is a compression scheme  that uses an average of 7/4 bits per symbol rather than two bits per symbol, and that such a compression scheme is optimal. The optimal scheme is provided by the following encoding:
\begin{eqnarray*}
a_{1} & :  & 0 \\
a_{2} & :  & 10 \\
a_{3} & :  & 110  \\
a_{4} & :  & 111
\end{eqnarray*}
for which the \textit{average} length of a compressed sequence is: $\frac{1}{2}\cdot 1 + \frac{1}{4}\cdot 2 + \frac{1}{8}\cdot 3 + \frac{1}{8}\cdot 3 = \frac{7}{4}$ bits per symbol.

The significance of Shannon's source coding theorem\index{Shannon's source coding theorem (noiseless channel coding theorem)} lies is showing that there is an optimal or most efficient way of compressing messages produced by a source (assuming a certain idealization) in such a way that they can be reliably reconstructed by a receiver. Since a message is abstracted as a sequence of distinguishable symbols produced by a stochastic source, the only relevant feature of a message with respect to reliable compression and decompression is the sequence of probabilities associated with the individual symbols: the nature of the physical systems embodying the representation of the message through their states is irrelevant to this notion of compression (provided only that the states are reliably distinguishable), as is the content or meaning of the message. The Shannon entropy\index{Shannon entropy}\index{entropy!Shannon} $H(X)$ is a measure of the minimal physical resources, in terms of the average number of bits per symbol, that are necessary and sufficient to reliably store the output of a source of messages. In this sense, it is a measure of the amount of information per symbol produced by an information source.

The essential notion underlying Shannon's measure of information is compressibility: information as a physical resource is something that can be compressed, and the amount of information produced by an information source is measured by its optimal compressibility. 

\subsection{Conditional Entropy\index{entropy!conditional}, Mutual Information\index{mutual information}\index{information!mutual}, Channel Capacity\index{channel capacity}} 

\label{sec:entropy}

The analysis so far assumes a noiseless channel between the source and the receiver. I turn now to a 
brief sketch of some concepts relevant to a noisy channel\index{noisy channel}, and a statement of  Shannon's noisy channel coding theorem\index{Shannon's noisy channel coding theorem}.

An information channel\index{information!channel}  maps inputs consisting of values of a random variable $X$ onto outputs consisting of values of a random variable $Y$, and the map will generally not be 1-1 if the channel is noisy\index{noisy channel}. Consider the conditional probabilities $p(y|x)$ of obtaining an output value $y$ for a given input value $x$, for all $x,y$. From the probabilities $p(x)$ we can calculate $p(y)$ as:
\[
p(y) = \sum_{x}p(y|x)p(x)
\]
and we can also calculate $p(x|y)$ by Bayes' rule from the probabilities $p(y|x)$ and $p(x)$, for all $x,y$, and hence the Shannon entropy of the conditional distribution $p(x|y)$, for all $x$ and a fixed $y$, denoted by $H(X|Y=y)$. 

The quantity
\begin{equation}
H(X|Y) = \sum_{y}{p(y)H(X|Y=y)} 
\end{equation}
is known as the \textit{conditional entropy}\index{entropy!conditional}. It is the expected value of $H(X|Y=y)$ for all $y$. If we think of $H(X)$, the entropy of the distribution 
$\{p(x): x \in \hil{X}\}$, as a measure of the uncertainty of the $X$-value, then $H(X|Y=y)$ is a measure of the uncertainty of the $X$-value, given the $Y$-value $y$, and $H(X|Y)$ is a measure of the average uncertainty of the $X$-value, given a $Y$-value. 

Putting it differently, the  number of input sequences of length $n$  that are consistent with a given output sequence (as 
$n \rightarrow \infty$) is $2^{nH(X|Y)}$, i.e., $H(X|Y)$ is the number of bits per symbol of additional information 
needed, on average, to identify an input $X$-sequence from a given $Y$-sequence. This follows 
because there are $2^{nH(X,Y)}$ typical sequences of pairs $(x,y)$, where the \emph{joint entropy}\index{joint entropy} $H(X,Y)$
is calculated from the joint probability $p(x,y)$. So there are
\begin{equation}
\frac{2^{nH(X,Y)}}{2^{nH(Y)}} = 2^{n(H(X,Y)-H(Y))} = 2^{nH(X|Y)}
\end{equation}
typical $X$-sequences associated with a given $Y$-sequence. 

The `chain rule' equality 
\begin{equation}
H(X,Y) =  H(X) + H(Y|X) = H(Y) + H(X|Y) = H(Y,X) \label{eq:chainrule}
\end{equation}
follows immediately from the logarithmic definitions of the quantities:
\begin{eqnarray}
H(X\!:\!Y) & := & -\sum_{x,y}p(x,y)\log p(x,y) \nonumber \\
& = & -\sum_{x,y}p(x) p(y|x) \log \left(p(x)p(y|x)\right) \nonumber \\
& = & -\sum_{x,y}p(x)p(y|x) \log p(x) - \sum_{x,y}p(x)p(y|x)\log p(y|x) \nonumber \\
& = & -\sum_{x}p(x) \log p(x) + \sum_{x}p(x) \left(-\sum_{y}p(y|x)\log p(y|x)\right) \nonumber \\
& = & H(X) + H(Y|X)
\end{eqnarray}
Note that $H(X|Y) \neq H(Y|X)$.

The \textit{mutual information}\index{mutual information}\index{information!mutual} measures the average amount of information gained about $X$ by ascertaining a $Y$-value, i.e., the amount of information one random variable contains about another, or the reduction in uncertainty of one random variable obtained by measuring another. 

Mutual information\index{mutual information}\index{information!mutual} can be defined in terms of the concept of \textit{relative entropy}\index{entropy!relative}, which is a measure of something like the distance between two probability distributions (although it is not a true metric, since it is not symmetric and does not satisfy the triangle inequality). The relative entropy\index{entropy!relative} between distributions $p(x)$ and $q(x)$ is defined as:
\begin{equation}
D(p\parallel q) = \sum_{x \in \mathcal{A}}p(x)\log \frac{p(x)}{q(x)}.
\end{equation}
The mutual information can now be defined as:
\begin{eqnarray}
H(X\!:\!Y) & = & D(p(x,y)\parallel p(x)p(y)) \nonumber  \\
& = & \sum_{x}\sum_{y}p(x,y)\log \frac{p(x,y)}{p(x)p(y)}. \label{eq:mutinfo1}
\end{eqnarray}
It follows that
\begin{equation}
H(X\!:\!Y) = H(X) - H(X|Y) = H(Y) - H(Y|X), \label{eq:mutinfo2}
\end{equation}
 i.e., the mutual information\index{mutual information}\index{information!mutual} of two random variables represents the average information gain about one random variable obtained by measuring the other:  the difference between the initial uncertainty of one of the random variables, and the average residual uncertainty of that random variable after ascertaining the value of the other random variable. Also, since $H(X,Y) = H(X) + H(Y|X)$, it follows  that
\begin{equation}
H(X\!:\!Y) = H(X) + H(Y) - H(X,Y); \label{eq:mutinfo3}
\end{equation}
i.e., the mutual information\index{mutual information}\index{information!mutual} of two random variables is a measure of how much information they have in common: the sum of the information content of the two random variables, as measured by the Shannon entropy\index{Shannon entropy}\index{entropy!Shannon} (in which joint information is counted twice), minus their joint information. Note  that $H(X\!:\!X) = H(X)$, as we would expect.

For a noisy channel\index{noisy channel}, if $X$ represents the input to the channel and $Y$ represents the output of the channel, $H(X\!:\!Y)$ represents the average amount of information gained about the input $X$ by ascertaining the value of the output $Y$. The \textit{capacity} of a channel\index{channel capacity}, $C$, is defined as the supremum of $H(X\!:\!Y)$ over all input distributions. 

Shannon's noisy channel coding theorem\index{Shannon's noisy channel coding theorem} shows, perhaps surprisingly, that up to $C$ bits of information can be sent through a noisy channel with arbitrary low error rate. That is, 
\begin{quote}
there exists an optimal coding for an information source with entropy $H \leq C$ such that $n$-length sequences produced by the source can be transmitted faithfully over the channel: the error rate tends to zero as $n \rightarrow \infty$. The probability of error tends to 1 if we attempt to transmit more than $C$ bits through the channel. 
\end{quote}
This means that there are two ways of improving the transmission rate over a noisy channel such as a telephone cable. We can improve the channel capacity by replacing the cable with a faster one, or we can improve the information processing (the data compression).

\section{Quantum Information\index{quantum!information}\index{information!quantum}}

The physical notion of information, discussed in \S \ref{sec:Shannon}, is profoundly transformed by the transition from classical mechanics to quantum mechanics. The aim of this section is to bring out the nature of this transformation. In \S 3.1, I develop some core concepts of quantum mechanics relevant to quantum information: entangled states, the Schmidt decomposition, the density operator formalism for the representation of pure and mixed states, the `purification' of mixed states, generalized quantum measurements in terms of positive operator valued measures (POVMs), and the evolution of open systems represented by quantum operations. I assume throughout Hilbert spaces of finite dimension (and so avoid all the technicalities of functional analysis required for the treatment of infinite-dimensional Hilbert spaces). In fact, there is no loss of generality here, since both classical and quantum information sources are considered to produce messages consisting of sequences of symbols from some finite alphabet, which we represent in terms of a finite set of classical or quantum states. Moreover, all the conceptual issues relevant to the difference between classical and quantum information show up in finite-dimensional Hilbert spaces. In \S 3.2, I introduce von Neumann's generalization of the Shannon entropy and related notions for quantum information. In \S 3.3 and \S 3.4, I discuss some salient features that distinguish quantum information from classical information: \S 3.3 deals with the limitations on copying quantum information imposed by the `no cloning' theorem, and \S 3.4 deals with  the limited accessibility of quantum information defined by the Holevo bound. Finally, in \S 3.5 I show how the notion of compressibility applies to quantum information, and I outline Schumacher's generalization of Shannon's source coding theorem for quantum information, noting a distinction between `visible' and `blind' compression applicable to quantum information.

\subsection{Some Relevant Quantum Mechanics}

\label{sec:quantum}

\subsubsection{Entangled States\index{entangled state}}

\label{sec:entanglement}

Consider a quantum system $Q$ which is part of a compound system $QE$; $E$ for `environment,' although $E$ could be any quantum system of which $Q$ is a subsystem. Pure states of $QE$ are represented as rays or unit vectors in a tensor product Hilbert space $\hil{H}^{Q} \otimes \hil{H}^{E}$. A general pure state of $QE$ is a state of the form:
\begin{equation}
\ket{\Psi} = \sum c_{ij}\ket{q_{i}}\ket{e_{j}}
\end{equation}
where $\ket{q_{i}} \in \hil{H}^{Q}$ is a complete set of orthonormal states (a basis) in  $\hil{H}^{Q}$ and $\ket{e_{j}} \in \hil{H}^{E}$ is a basis in $\hil{H}^{E}$. If the coefficients $c_{ij}$ are such that $\ket{\Psi}$ cannot be expressed as a product state $\ket{Q}\ket{E}$, then $\ket{\Psi}$ is called an \emph{entangled state}\index{entangled state}. 

For any state $\ket{\Psi}$ of $QE$, there exist orthonormal bases $\ket{i} \in \hil{H}^{Q}$, $\ket{j} \in \hil{H}^{E}$ such that $\ket{\Psi}$ can be expressed in a biorthogonal correlated form as:
\begin{equation}
\ket{\Psi} = \sum_{i}\sqrt{p_{i}}\ket{i}\ket{i}
\end{equation}
where the coefficients $\sqrt{p_{i}}$ are real and non-negative, and $\sum p_{i}
= 1$. This representation is referred to as the \emph{Schmidt decomposition}\index{Schmidt decomposition}. The Schmidt decomposition is unique if and only if the $p_{i}$ are all distinct.

An example is the biorthogonal EPR\index{EPR} state\footnote{Einstein, Podolsky and Rosen considered a more complicated state entangled over position and momentum values. The spin example is due to Bohm \shortcite[pp, 611--623]{Bohm}.}
\begin{equation}
\ket{\Psi} = (\ket{0}\ket{1} - \ket{1}\ket{0})/\sqrt{2};
\end{equation}
say, the singlet state of two spin-1/2 particles (the Schmidt form with positive coefficients is obtained by asborbing the relative phases in the definition of the basis vectors). In the singlet state, $\ket{0}$ and $\ket{1}$ can be taken as representing the two eigenstates of spin in the $z$-direction, but since the state is symmetric, $\ket{\Psi}$ retains the same form for spin in any direction. The EPR argument\index{EPR argument} exploits the fact that spin measurements in the same direction on the two particles, which could be arbitrarily far apart, will yield outcomes that are perfectly anti-correlated for any spin direction. Bell's counterargument\index{Bell's argument} exploits the fact that when the spin is measured on one particle in a direction $\theta_{1}$ to the $z$-axis, but on the other particle in a direction $\theta_{2}$ to the $z$-axis, the probability of finding the same outcome for both particles (both 1 or both 0) is $\frac{1}{2}\sin^{2}(\theta_{1} - \theta_{2})$. It follows that the outcomes are perfectly correlated when $\theta_{1} - \theta_{2} = \pi$ and that 3/4 of the outcomes are the same when $\theta_{1} - \theta_{2} = 2\pi/3$. On the other hand, from Bell's inequality\index{Bell's inequality}, derived under Einstein's realist assumptions of separability and locality, we see that the correlation for $\theta_{1} - \theta_{2} = 2\pi/3$ cannot exceed 2/3. See Dickson (this vol., ch. 4) for further discussion.

This means that the dynamical evolution of a quantum system can result in a state representing correlational information that no classical computer can simulate. That is, no classical computer can be programmed to perform the following task: for any pair of input angles, $\theta_{1}, \theta_{2}$, at different locations, output a pair of values (0 or 1) for these locations such that the values are perfectly correlated when $\theta_{1} - \theta_{2} = \pi$, perfectly anti-correlated when $\theta_{1} = \theta_{2}$, and 75\% correlated when $\theta_{1} - \theta_{2} = 2\pi/3$, where the response time between being given the input and producing the output in each case is less than the time taken by light to travel between the two locations. 

Notice that the four states:
\begin{eqnarray}
\ket{00} & = & \frac{1}{\sqrt{2}}(\ket{0}\ket{0} + \ket{1}\ket{1})  \\
\ket{01} & = & \frac{1}{\sqrt{2}}(\ket{0}\ket{1} + \ket{1}\ket{0}) \\
\ket{10} & = & \frac{1}{\sqrt{2}}(\ket{0}\ket{0} - \ket{1}\ket{1}) \\
\ket{11} & = & \frac{1}{\sqrt{2}}(\ket{0}\ket{1} - \ket{1}\ket{0}) 
 \end{eqnarray}
form an orthonormal basis, called the Bell basis\index{Bell basis}\index{Bell states}, in the 2 x 2-dimensional Hilbert space. Any Bell state can be transformed into any other Bell state by a local unitary transformation, $X, Y$, or $Z$, where $X, Y, Z$ are the Pauli spin matrices:
\begin{equation}
X =  \sigma_{x}  = \ket{0}\bra{1} + \ket{1}\bra{0} =  \left( \begin{array}{lr}
	0 & 1 \\
	1 & 0
	\end{array} \right) 
\end{equation}
\begin{equation}
Y =  \sigma_{y} =  i\ket{0}\bra{1} - i\ket{1}\bra{0} =  \left( \begin{array}{lr}
	0 & -i \\
	i & 0
	\end{array} \right)
\end{equation}
\begin{equation}
Z =  \sigma_{z} =  \ket{0}\bra{0} + \ket{1}\bra{1} =  \left( \begin{array}{lr}
	1 & 0 \\
	0 & -1
	\end{array} \right).
\end{equation}
For example:
\begin{equation}
X\otimes I \cdot \frac{1}{\sqrt{2}}(\ket{0}\bra{1} - \ket{1}\ket{0} = \frac{1}{\sqrt{2}}(\ket{0}\bra{0} - \ket{1}\ket{1}.
\end{equation}

If $QE$ is a closed system in an entangled pure state represented by 
\begin{equation}
\ket{\Psi} = \sum_{i}\sqrt{p_{i}}\ket{i}\ket{i} \label{eq:Schmidt}
\end{equation}
in the Schmidt decomposition\index{Schmidt decomposition}, the expected value of any $Q$-observable $A$ on $\hil{H}^{Q}$ can be computed as:
\begin{eqnarray}
\langle A \rangle & = & \trace{\ket{\Psi}\bra{\Psi} A \otimes I} \nonumber \\
& = & \ptrace{Q}{\ptrace{E}{\ket{\Psi}\bra{\Psi} A}} \nonumber \\
& = & \ptrace{Q}{\sum_{i} p_{i}\ket{i}\bra{i} A} \nonumber \\
& = & \ptrace{Q}{\rho A}      \label{eq:trace}
\end{eqnarray}
where $\ptrace{Q}{} = \sum_{q}\bra{q_{i}}\cdot\ket{q_{i}}$, for any orthonormal basis in $\hil{H}^{Q}$,  is the partial trace over $\hil{H}^{Q}$, $\ptrace{E}{}$ is the partial trace over $\hil{H}^{E}$, and $\rho = \sum_{i}p_{i}\ket{i}\bra{i} \in \hil{H}^{Q}$ is the \textit{reduced density operator}\index{reduced density operator} of the open system $Q$, a positive operator with unit trace. Since the density operator $\rho$ yields the statistics of all $Q$-observables via Eq. (\ref{eq:trace}), $\rho$ is taken as representing the quantum state of the system $Q$.  

If $QE$ is an entangled pure state, then the open system $Q$ is in a \emph{mixed state}\index{mixed state} $\rho$, i.e., $\rho \neq \rho^{2}$; for pure states, $\rho$ is a projection operator onto a ray and $\rho = \rho^{2}$. A mixed state represented by a density operator $\rho = \sum\rho_{i}\ket{i}\bra{i}$ can be regarded as a mixture of pure states $\ket{i}$ prepared with prior probabilities $p_{i}$, but this representation is not unique---not even if the states combined in the mixture are orthogonal. For example, the equal-weight mixture of orthonormal states $\ket{0}, \ket{1}$ in a 2-dimensional Hilbert space $\hil{H}_{2}$ has precisely the same statistical properties, and hence the same density operator $\rho = I/2$, as the equal weight mixture of any pair of orthonormal states, e.g., the states $\frac{1}{\sqrt{2}}(\ket{0} + \ket{1}), \frac{1}{\sqrt{2}}(\ket{0} - \ket{1})$, or the equal-weight mixture of nonorthogonal states $\ket{0}, \frac{1}{2}\ket{0} + \frac{\sqrt{3}}{2}\ket{1}, \frac{1}{2}\ket{0} - \frac{\sqrt{3}}{2}\ket{1}$ $120^{\circ}$ degrees apart, or the uniform continuous distribution over all possible states in $\hil{H}_{2}$. 

More generally, for any basis of orthonormal states $\ket{e_{i}} \in \hil{H}^{E}$, the entangled state $\ket{\Psi}$ can be expressed as:
\begin{equation}
\ket{\Psi} =  \sum_{ij}c_{ij}\ket{q_{i}}\ket{e_{j}} 
=  \sum_{j}\sqrt{w_{j}} \ket{r_{j}}\ket{e_{j}}
\end{equation}
where the normalized states $\ket{r_{j}} = \sum_{i}\frac{c_{ij}}{\sqrt{w_{j}}}\ket{q_{i}}$ are \textit{relative states}\index{relative states} to the $\ket{e_{j}}$ ($\sqrt{w_{j}} = \sum_{j}\abs{c_{ij}}^{2}$). Note that the states $\ket{r_{j}}$ are not in general orthogonal. Since the $\ket{e_{j}}$ are orthogonal, we can express the density operator representing the state of $Q$ as:
\begin{equation}
\rho = \sum_{i}w_{i}\ket{r_{i}}\bra{r_{i}}.
\end{equation}

In effect, a measurement of an $E$-observable with eigenstates $\ket{e_{i}}$ will leave the composite system $QE$ in one of the states $\ket{r_{i}}\ket{e_{i}}$ with probability $w_{i}$, and a measurement of an $E$-observable with eigenstates $\ket{i}$ (the orthogonal states of the Schmidt decomposition in (\ref{eq:Schmidt}) above) will leave the system $QE$ in one of the states $\ket{i}\ket{i}$ with probability $p_{i}$. Since $Q$ and $E$ could be widely separated from each other in space, no measurement at $E$ could affect the statistics of any $Q$-observable; or else measurements at $E$ would allow superluminal signaling between $Q$ and $E$. It follows that the mixed state $\rho$ can  be realized as a mixture of orthogonal states $\ket{i}$ (the eigenstates of $\rho$) with weights $p_{i}$, or as a mixture of non-orthogonal relative states $\ket{r_{j}}$ with weights $w_{j}$ in infinitely many ways, depending on the choice of basis in $\hil{H}^{E}$:
\begin{equation}
\rho = \sum_{i}p_{i}\ket{i}\bra{i} = \sum_{j}w_{j}\ket{r_{j}}\bra{r_{j}}
\end{equation}
and all these different mixtures with the same density operator $\rho$ must be physically indistinguishable.

Note that any mixed state density operator $\rho \in \hil{H}^{Q}$ can be `purified'\index{purification} by adding a suitable ancilla system $E$, in the sense that $\rho$ is the partial trace of a pure state $\ket{\Psi} \in \hil{H}^{Q} \otimes \hil{H}^{E}$ over $\hil{H}^{E}$. A purification\index{purification} of a mixed state is, clearly, not unique, but depends on the choice of $\ket{\Psi}$ in $\hil{H}^{E}$. The Hughston-Jozsa-Wootters\index{Hughston-Jozsa-Wootters theorem} theorem \cite{HJW} shows that for \emph{any} mixture of pure states $\ket{r_{i}}$ with weights $w_{i}$, where 
$\rho = \sum_{j}w_{j}\ket{r_{j}}\bra{r_{j}}$,
there is a purification\index{purification} of $\rho$ and a suitable measurement on the system $E$ that will leave $Q$ in the mixture $\rho$.  So an observer at $E$ can remotely prepare $Q$ in any mixture that corresponds to the density operator $\rho$ (and of course all these different mixtures are physically indistinguishable). Similar results were proved earlier by \index{Schr\"{o}dinger}Schr\"{o}dinger \shortcite{Schr2},
Jaynes \shortcite{Jaynes}\index{Jaynes} and Gisin \shortcite{Gisin}\index{Gisin}. See Halvorson \shortcite{Halvorson1}\index{Halvorson} for a generalization to hyperfinite von Neuman algebras\index{hyperfinite von Neuman algebras}.

\subsubsection{Measurement}

\label{sec:measurement}

A standard von Neumann `yes-no' measurement is associated with a projection operator; so a standard observable is represented in the spectral representation as a sum of projection operators, with coefficients representing the eigenvalues of the observable. Such a measurement is the quantum analogue of the measurement of a property of a system in classical physics. Classically, we think of a property of a system as being associated with a subset in the state space (phase space) of the system, and determining whether the system has the property amounts to determining whether the state of the system lies in the corresponding subset. In quantum mechanics, the counterpart of a subset in phase space is a closed linear subspace in Hilbert space. Just as the different possible values of an observable (dynamical quantity) of a classical system correspond to the subsets in a mutually exclusive and collectively exhaustive set of subsets covering the classical state space, so the different values of a quantum observable correspond to the subspaces in a mutually exclusive (i.e., orthogonal) and collectively exhaustive set of subspaces spanning the quantum state space. (For further discussion, see Dickson, this vol., ch. 4, \cite{Mackey}, and \cite{Bubbook}.)

In quantum mechanics, and especially in the theory of quantum information (where any read-out of the quantum information encoded in a quantum state requires a quantum measurement), it is useful to consider a more general class of measurements than the projective measurements\index{measurement!measurement} associated with the determination of the value of an observable. It is common to speak of generalized measurements\index{measurement!generalized} and generalized observables\index{observable!generalized}. But in fact this terminology is more misleading than illuminating, because a generalized measurement is not a procedure that reveals whether or not a quantum system has some sort of generalized property. Rather, the point of the generalization is to exploit the difference between quantum and classical states for new possibilities in the representation and manipulation of information.

To clarify the idea, I will follow the excellent discussion by Nielsen and Chuang 
\shortcite[\S 2.2.3--2.2.6]{NielsenChuang}. A quantum measurement can be characterized, completely generally, as a certain sort of interaction between two quantum systems, $Q$ (the measured system) and $M$ (the measuring system). We suppose that $Q$ is initially in a state $\ket{\psi}$ and that $M$ is initially in some standard state $\ket{0}$, where $\ket{m}$ is an orthonormal basis of `pointer' eigenstates in $\hil{H}^{M}$. The interaction is defined by a unitary transformation $U$ on the Hilbert space $\hil{H}^{Q}\otimes\hil{H}^{M}$ that yields the transition:
\begin{equation}
\ket{\psi}\ket{0} \stackrel{U}{\longrightarrow} \sum_{m}M_{m}\ket{\psi}\ket{m} \label{eq:evol}
\end{equation}
where  $\{M_{m}\}$ is a set of linear operators (the Kraus operators\index{Kraus operator}) defined on $\hil{H}^{Q}$ satisfying the \emph{completeness condition}\index{completeness condition}:
\begin{equation}
\sum_{m}M_{m}^{\dagger}M_{m} = I.
\end{equation}
(The symbol $\dagger$ denotes the adjoint or Hermitian conjugate.) The completeness condition\index{completeness condition} guarantees that this evolution is unitary, because it guarantees that $U$ preserves inner products, i.e.
\begin{eqnarray}
\bra{\phi}\bra{0}U^{\dagger}U\ket{\psi}\ket{0} & =  & \sum_{m,m^{\prime}}\bra{m}\bra{\phi}M_{m}^{\dagger}M_{m^{\prime}}\ket{\psi}\ket{m^{\prime}}  \nonumber \\
& = & \sum_{m}\bra{\phi}M^{\dagger}M\ket{\psi} \nonumber \\
& = & \braket{\phi}{\psi}
\end{eqnarray}
from which it follows that $U$, defined as above by Eq. (\ref{eq:evol}) for any product state $\ket{\psi}\ket{0}$ (for any $\ket{\psi} \in \hil{H}^{Q}$) can be extended to a unitary operator on the Hilbert space $\hil{H}^{Q} \otimes\hil{H}^{M}$. Accordingly, any set of linear operators $\{M_{m}\}$ defined on the Hilbert space of the system $Q$ satisfying the completeness condition\index{completeness condition} defines a measurement in this general sense, with the index $m$ labeling the possible outcomes of the measurement, and any such set is referred to as a set of measurement operators\index{measurement!operator}.

If we now perform a standard projective measurement\index{measurement!projective} on $M$ to determine the value $m$ of the pointer observable, defined by the projection operator
\[
P_{m} = I_{Q}\otimes \ket{m}\bra{m}
\]
then the probability of obtaining the outcome $m$ is, by  (\ref{eq:trace})\footnote{The expected value of a projection operator, which is an idempotent observable with eigenvalues 0 and 1, is equal to the probability of obtaining the eigenvalue 1. Here the eigenvalue 1 corresponds to the outcome $m$.}:
\begin{eqnarray}
p(m) & = & \bra{0}\bra{\psi}U^{\dagger}P_{m}U\ket{\psi}\ket{0} \nonumber \\
& =  & \sum_{m^{\prime}m^{\prime\prime}}\bra{m^{\prime}}\bra{\psi}
M_{m^{\prime}}^{\dagger}(I_{Q}\otimes \ket{m}\bra{m})M_{m^{\prime\prime}}
\ket{\psi}\ket{m^{\prime\prime}} \nonumber \\
& = & \sum_{m^{\prime}m^{\prime\prime}}\bra{\psi}M_{m^{\prime}}^{\dagger}\braket{m^{\prime}}{m}\braket{m}
{m^{\prime\prime}}M_{m^{\prime\prime}}\ket{\psi} \nonumber \\
& = & \bra{\psi}M_{m}^{\dagger}M_{m}\ket{\psi};
\end{eqnarray}
and, more generally, if the initial state of $Q$ is a mixed state $\rho$, then
\begin{equation}
p(m) = \ptrace{Q}{M\rho M^{\dagger}}.
\end{equation}
The final state of $QM$ after the projective measurement\index{measurement!projective} on $M$ yielding the outcome $m$ is:
\begin{equation}
\frac{P_{m}U\ket{\psi}\ket{0}}{\sqrt{\bra{\psi}U^{\dagger}PU\ket{\psi}}} = \frac{M_{m}\ket{\psi}\ket{m}}
{\sqrt{\bra{\psi}M_{m}^{\dagger}M_{m}\ket{\psi}}}.
\end{equation}
So the final state of $M$ is $\ket{m}$ and the final state of $Q$ is:
\[
\frac{M_{m}\ket{\psi}}{\sqrt{\bra{\psi}M_{m}^{\dagger}M_{m}\ket{\psi}}};
\]
and, more generally, if the initial state of $Q$ is a mixed state $\rho$, then the final state of $Q$ is:
\[
\frac{M_{m}\rho M_{m}^{\dagger}}{\ptrace{Q}{M_{m}\rho M_{m}^{\dagger}}}.
\]

Note that this general notion of measurement covers the case of standard projective measurements\index{measurement!projective}. In this case $\{M_{m}\} = \{P_{m}\}$, where $\{P_{m}\}$ is the set of projection operators defined by the spectral measure of a standard quantum observable represented by a self-adjoint operator. It also covers the measurement 
of `generalized observables'\index{observable!generalized} associated with positive operator valued measures (POVMs)\index{positive operator valued measure (POVM)}. Let
\begin{equation}
E_{m}
= M_{m}^{\dagger}M_{m}
\end{equation}
then the set $\{E_{m}\}$ defines a set of positive operators (`effects')\index{effect} such that 
\begin{equation}
\sum E_{m} = I  \label{eq:resid}
\end{equation}
A POVM can be regarded as a generalization of a projection valued measure (PVM)\index{projection valued measure (PVM)}, in the sense that Eq. (\ref{eq:resid}) defines a `resolution of the identity' without requiring the PVM orthogonality condition:
\begin{equation}
P_{m}P_{m^{\prime}} = \delta_{mm^{\prime}}P_{m}.
\end{equation}
Note that for a POVM:
\begin{equation}
p(m) = \bra{\psi}E_{m}\ket{\psi}.
\end{equation}

Given a set of positive operators $\{E_{m}\}$ such that $\sum E_{m} = I$, measurement operators\index{measurement!operator} $M_{m}$ can be defined via
\begin{equation}
M_{m} = U\sqrt{E_{m}},
\end{equation}
where $U$ is a unitary operator, from which it follows that
\begin{equation}
\sum_{m}M_{m}^{\dagger}M_{m} = \sum E_{m} = I
\end{equation}
As a special case, of course, we can take $U = 1$ and $M_{m} = \sqrt{E_{m}}$. Conversely, given a set of measurement operators $\{M_{m}\}$, there exist unitary operators $U_{m}$ such that $M_{m} = U_{m}\sqrt{E_{m}}$, where $\{E_{m}\}$ is a POVM. (This follows immediately from \cite[Theorem 2.3, p. 78]{NielsenChuang}; see \cite[Exercise 2.63, p. 92]{NielsenChuang}.)

Except for the standard case of projective measurements, one might wonder why it might be useful to single out such unitary transformations, and why in the general case such a process should be called a \emph{measurement} of $Q$. The following example, taken from \cite[p. 92]{NielsenChuang}, is illuminating. Suppose we know that a system with a 2-dimensional Hilbert space is in one of two nonorthogonal states:
\begin{eqnarray*}
\ket{\psi_{1}} & = & \ket{0} \\
\ket{\psi_{2}} & = & \frac{1}{\sqrt{2}}(\ket{0} + \ket{1})
\end{eqnarray*}
It is impossible to \textit{reliably} distinguish these states by a quantum measurement, even in the above generalized sense. Here `reliably' means that the state is identified correctly with zero probability of error. 

To see this, suppose there is such a measurement, defined by two measurement operators\index{measurement!operator} $M_{1}, M_{2}$ satisfying the completeness condition. Then we require
\begin{equation}
p(1) = \bra{\psi_{1}}M_{1}^{\dagger}M_{1}\ket{\psi_{1}} = 1,
\end{equation}
to represent reliability if the state is $\ket{\psi_{1}}$; and 
\begin{equation}
p(2) = \bra{\psi_{2}}M_{2}^{\dagger}M_{2}\ket{\psi_{2}} = 1 \label{eq:prob1}
\end{equation}
to represent reliability if the state is $\ket{\psi_{2}}$. By the completeness condition we must have
\begin{equation}
\bra{\psi_{1}}M_{1}^{\dagger}M_{1} + M_{2}^{\dagger}M_{2}\ket{\psi_{1}} = 1
\end{equation}
from which it follows that $\bra{\psi_{1}}M_{2}^{\dagger}M_{2}\ket{\psi_{1}} = 0$, i.e., $M_{2}\ket{\psi_{1}} = M_{2}\ket{0} = 0$. Hence
\begin{equation}
M_{2}\ket{\psi_{2}} = M_{2}\frac{1}{\sqrt{2}}(\ket{0} + \ket{1}) = \frac{1}{\sqrt{2}}M_{2}\ket{1}
\end{equation}
and so 
\begin{equation}
p(2) = \bra{\psi_{2}}M_{2}^{\dagger}M_{2}\ket{\psi_{2}} = \frac{1}{2}\bra{1}M_{2}^{\dagger}M_{2}\ket{1}.
\end{equation}
But by the completeness condition we also have
\begin{equation}
\bra{1}M_{2}^{\dagger}M_{2}\ket{1} \leq \bra{1}M_{1}^{\dagger}M_{1} + M_{2}^{\dagger}M_{2}\ket{1} = \braket{1}{1} = 1
\end{equation}
from which it follows that
\begin{equation}
p(2) \leq \frac{1}{2}
\end{equation}
which contradicts Eq. (\ref{eq:prob1}).

However, it is possible to perform a measurement in the generalized 
sense\index{measurement!generalized}, with \textit{three} possible outcomes, that will allow us to correctly identify the state some of the time, i.e., for two of the possible outcomes, while nothing about the identity of the state can be inferred from the third outcome. 

Here's how: The three operators
\begin{eqnarray}
E_{1} & = & \frac{\sqrt{2}}{1+ \sqrt{2}}\frac{(\ket{0} - \ket{1})(\bra{0} - \bra{1})}{2} \nonumber \\
E_{2} & = & \frac{\sqrt{2}}{1+ \sqrt{2}}\ket{1}\bra{1} \nonumber \\
E_{3} & = & I - E_{1} - E_{2}
\end{eqnarray}
are all positive operators and $E_{1}+E_{2}+E_{3} = I$, so they define a POVM. In fact, $E_{1}, E_{2}, E_{3}$ are each multiples of projection operators onto the states 
\begin{eqnarray}
\ket{\phi_{1}} & = & \ket{\psi_{2}}^{\perp} \nonumber \\
\ket{\phi_{2}} & = & \ket{\psi_{1}}^{\perp} \nonumber \\
\ket{\phi_{3}} & = & \frac{(1+ \sqrt{2})\ket{0} + \ket{1}}{\sqrt{2\sqrt{2}(1+\sqrt{2})}}
\end{eqnarray}
with coefficients $\frac{\sqrt{2}}{1+\sqrt{2}}, \frac{\sqrt{2}}{1+\sqrt{2}}, \frac{1}{1+\sqrt{2}}$ respectively. The measurement involves a system $M$ with three orthogonal pointer states $\ket{1}, \ket{2}, \ket{3}$. The appropriate unitary interaction $U$ results in the transition, for an input state $\ket{\psi}$:
\begin{equation}
\ket{\psi}\ket{0} \stackrel{U}{\longrightarrow}\sum_{m}M_{m}\ket{\psi}\ket{m}
\end{equation}
where $M_{m} = \sqrt{E_{m}}$.

If the input state is $\ket{\psi_{1}} = \ket{0}$, we have the transition:
\begin{eqnarray}
\ket{\psi_{1}}\ket{0} & \stackrel{U}{\longrightarrow} &\sqrt{E}_{1}\ket{0}\ket{1} + \sqrt{E}_{3}\ket{0}\ket{3} \nonumber \\
& = & \alpha\ket{\phi_{1}}\ket{1} + \beta\ket{\phi_{3}}\ket{3}
\end{eqnarray}
(because $\sqrt{E}_{2}\ket{\psi_{1}} = \sqrt{E}_{2}\ket{0} = 0$). And if the input state is $\ket{\psi_{2}} = \frac{1}{\sqrt{2}}(\ket{0} + \ket{1})$, we have the transition:
\begin{eqnarray}
\ket{\psi_{2}}\ket{0} & \stackrel{U}{\longrightarrow} & \sqrt{E}_{2}\frac{\ket{0} + \ket{1}}{\sqrt{2}}\ket{2} + \sqrt{E}_{3}\frac{\ket{0} + \ket{1}}{\sqrt{2}}\ket{3} \nonumber \\
& = & \gamma\ket{\phi_{2}}\ket{2} + \delta\ket{\phi_{3}}\ket{3}
\end{eqnarray}
(because $\sqrt{E}_{1}\ket{\psi_{2}} = \sqrt{E}_{1}\frac{\ket{0} + \ket{1}}{\sqrt{2}} = 0$), where $\alpha, \beta, \gamma, \delta$ are real numerical coefficients. 

We see that a projective measurement of the pointer of $M$ that yields the outcome $m = 1$ indicates, with certainty, that the input state was $\ket{\psi_{1}} = \ket{0}$. In this case, the measurement leaves the system $Q$ in the state $\ket{\phi_{1}}$. A measurement outcome $m = 2$ indicates, with certainty, that the input state was $\ket{\psi_{2}} = \frac{1}{\sqrt{2}}(\ket{0} + \ket{1})$, and in this case the measurement leaves the system Q in the state $\ket{\phi_{2}}$. If the outcome is $m = 3$, the input state could have been either $\ket{\psi_{1}} = \ket{0}$ or $\ket{\psi_{2}} = \frac{1}{\sqrt{2}}(\ket{0} + \ket{1})$, and $Q$ is left in the state $\ket{\phi_{3}}$.

\subsubsection{Quantum Operations\index{quantum!operation}}

\label{sec:operations}

When a closed system $QE$ evolves under a unitary transformation, $Q$ can be shown to evolve under a \emph{quantum operation}\index{quantum!operation}, i.e., a completely positive linear map:
\begin{equation}
\hil{E}: \rho \rightarrow \rho'
\end{equation}
where
\begin{equation}
\hil{E}(\rho) = \ptrace{E}{U\rho\otimes\rho_{E}U^{\dagger}}
\end{equation}
(See \cite[p. 356 ff]{NielsenChuang}.) The map $\hil{E}$ is linear (or convex-linear) in the sense that $\hil{E}(\sum_{i}p_{i}\rho_{i}) = \sum_{i}p_{i}\hil{E}(p_{i})$, positive in the sense that $\hil{E}$ maps positive operators to positive operators, and completely positive in the sense that $\hil{E} \otimes I$ is a positive map on the extension of $\hil{H}^{Q}$ to a Hilbert space $\hil{H}^{Q} \otimes \hil{H}^{E}$, associated with the addition of any ancilla system $E$ to $Q$.

Every quantum operation\index{quantum!operation} (i.e., completely positive linear map) on a Hilbert space $\hil{H}^{Q}$ has a (non-unique) representation as a unitary evolution on an extended Hilbert space $\hil{H}^{Q} \otimes \hil{H}^{E}$, i.e.,
\begin{equation}
\hil{E}(\rho) = \ptrace{E}{U(\rho \otimes \rho_{E})U^{\dagger}} \label{eq:extended}
\end{equation}
where $\rho_{E}$ is an appropriately chosen initial state of an ancilla system $E$ (which we can think of as the environment of $Q$). It turns out that it suffices to take $\rho_{E}$  as a pure state, i.e., $\ket{0}\bra{0}$, since a mixed state of $E$ can always be purified\index{purification} by enlarging the Hilbert space (i.e., adding a further ancilla system). So the evolution of a system $Q$ described by a quantum operation can always be modeled as the unitary evolution of a system $QE$, for an initial pure state of $E$. 

Also, every quantum operation on a Hilbert space $\hil{H}^{Q}$ has a (non-unique) operator sum representation intrinsic to $\hil{H}^{Q}$:
\begin{equation}
\hil{E}(\rho) = \sum_{i}E_{i}\rho E^{\dagger}_{i}
\end{equation}
where $E_{i} = \bra{i}U\ket{0}$ for some orthonormal basis $\{\ket{i}\}$ of $E$. (See \cite[Theorem 8.1, p. 368]{NielsenChuang}.) If the operation is trace-preserving (or nonselective), then $\sum_{i}E^{\dagger}_{i}E_{i} = I$. For operations that are not trace-preserving (or selective),  $\sum_{i}E^{\dagger}_{i}E_{i} \leq I$. This corresponds to the case where the outcome of a measurement on $QE$ is taken into account (selected) in the transition $\hil{E} \rightarrow \hil{E}(\rho)$. 

If there is no interaction between $Q$ and $E$, then $\epsilon(\rho) = U_{Q} \rho U^{\dagger}_{Q}$, $U_{Q}U^{\dagger}_{Q} = I$, i.e., there is only one operator in the sum. In this case, $U = U_{Q} \otimes U_{E}$ and 
\begin{eqnarray}
\hil{E}(\rho) & = & \ptrace{E}{U_{Q }\otimes U_{E}(\rho \otimes \ket{0}\bra{0})U^{\dagger}_{Q} \otimes U^{\dagger}_{E}} \\
& = & U_{Q} \rho U^{\dagger}_{Q}.
\end{eqnarray}
So unitary evolution is a special case of the operator sum representation of a quantum operation and, of course, another special case  is  the transition $\hil{E} \rightarrow \hil{E}(\rho)$ that occurs in a quantum measurement process, where $E_{i} = M_{i}$. A trace-preserving operation corresponds to a non-selective measurement\index{non-selective measurement}: 
\begin{equation}
\hil{E}(\rho) = \sum_{i}M_{i}\rho M^{\dagger}_{i};
\end{equation}
while an operation that is not trace-preserving corresponds to a selective measurement\index{selective measurement}, where the state `collapses' onto the corresponding measurement outcome: 
\begin{equation}
M_{i}\rho M^{\dagger}_{i}/\trace{M_{i}\rho M^{\dagger}_{i}}.
\end{equation}

The operator sum representation applies  to quantum operations between possibly different input and output Hilbert spaces, and characterizes the following general situation: a quantum system in an unknown initial state $\rho$ is allowed to interact unitarily with other systems prepared in standard states, after which some part of the composite system is discarded, leaving the final system in a state $\rho'$. The transition $\rho \rightarrow \rho'$ is defined by a quantum operation. So a quantum operation represents, quite generally, the unitary evolution of a closed quantum system, the nonunitary evolution of an open quantum system in interaction with its environment, and  evolutions that result from a combination of unitary interactions and selective or nonselective measurements.

As we have seen, the creed of the Church of the Larger Hilbert Space\index{Church of the Larger Hilbert Space} is that every state can be made pure, every measurement can be made ideal, and every evolution can be made unitary -- on a larger Hilbert space.\footnote{The Creed originates with John Smolin\index{Smolin}. I owe this formulation to Ben Schumacher. See his \emph{Lecture Notes on Quantum Information Theory} \shortcite{Schumacher98}.}

\subsection{Von Neumann Entropy\index{von Neumann entropy}\index{entropy!von Neumann}}
\label{sec:Neumann}

In this section,
I define the von Neumann entropy\index{von Neumann entropy}\index{entropy!von Neumann} of a mixture of quantum states (von Neumann's generalization of the Shannon entropy of a classical probability distribution characterizing a classical information source) and the corresponding notions of conditional entropy\index{entropy!conditional} and mutual information\index{mutual information}\index{information!mutual}.

Information\index{information!Shannon's sense} in Shannon's sense is a quantifiable resource associated with the output of a (suitably idealized) stochastic source of symbolic states, where the physical nature of the systems embodying these states is irrelevant to the amount of classical information associated with the source. The quantity of information associated with a stochastic source is defined by its optimal compressibility, and this is given by the Shannon entropy\index{Shannon entropy}\index{entropy!Shannon}. The fact that some feature of the output of a stochastic source can be optimally compressed is, ultimately, what justifies the attribution of a quantifiable resource to the source. 

Information\index{information!represented physically} is represented physically in the states of physical systems. The essential difference between classical\index{information!classical} and quantum information\index{information!quantum}\index{quantum!information} arises because of the different distinguishability\index{distinguishability} properties of classical and quantum states. As we will see below, only sets of orthogonal quantum states are reliably distinguishable (i.e., with zero probability of error), as are sets of different classical states (which are represented by disjoint singleton subsets in a phase space, and so are orthogonal as subsets of phase space in a sense analogous to orthogonal subspaces of a Hilbert space). 

Classical information\index{information!classical} is that sort of information represented in a set of distinguishable states---states of classical systems, or orthogonal quantum states---and so can be regarded as a subcategory of quantum information, where the states may or may not be distinguishable. The idea behind quantum information\index{information!quantum}\index{quantum!information} is to extend Shannon's notion of compressibility\index{information!compressibility} to a stochastic source of quantum states, which may or may not be distinguishable. For this we need to define a suitable measure of information for probability distributions of quantum states---mixtures---as a generalization of the notion of Shannon entropy\index{Shannon entropy}\index{entropy!Shannon}.

Consider a system $QE$ in an entangled state $\ket{\Psi}$. Then the subsystem $Q$ is in a mixed state $\rho$, which can always be expressed as:
\begin{equation}
\rho = \sum_{i}p_{i}\ket{i}\bra{i}
\end{equation}
where the $p_{i}$ are the eigenvalues of $\rho$ and the pure states $\ket{i}$ are orthonormal eigenstates of $\rho$. This is the spectral representation of $\rho$, and any density operator---a positive (hence Hermitian) operator---can be expressed in this way. The representation is unique if and only if the $p_{i}$ are all distinct. If some of the $p_{i}$ are equal, there is a unique representation of $\rho$ as a sum of projection operators with the distinct values of the $p_{i}$ as coefficients, but some of the projection operators will project onto multi-dimensional subspaces. 

Since $\rho$ has unit trace, $\sum p_{i} = 1$, and so the spectral representation of $\rho$ represents a classical probability distribution of orthogonal, and hence distinguishable, pure states. If we measure a $Q$-observable with eigenstates $\ket{i}$, then the outcomes can be associated with the values of a random variable $X$, where $\mbox{Pr}(X = i) = p_{i}$. Then 
\begin{equation}
H(X) = -\sum p_{i}\log{p_{i}} 
\end{equation}
is the Shannon entropy\index{Shannon entropy}\index{entropy!Shannon} of the probability distribution of measurement outcomes.

Now,
\begin{equation}
-\trace{\rho\log{\rho}} = -\sum p_{i}\log{p_{i}} 
\end{equation}
(because the eigenvalues of $\rho\log{\rho}$ are $p_{i}\log{p_{i}}$ and the trace of an operator is the sum of the eigenvalues), so a natural generalization of Shannon entropy for any mixture of quantum states with density operator $\rho$ is the \emph{von Neumann entropy\index{von Neumann entropy}\index{entropy!von Neumann}}\footnote{Von Neumann first defined this quantity on the basis of a thermodynamic argument in \shortcite[p. 379]{Neumann}.}:
\begin{equation}
S := -\trace{\rho\log{\rho}}
\end{equation}
which coincides with the Shannon entropy\index{Shannon entropy}\index{entropy!Shannon} for measurements in the eigenbasis of $\rho$. For a completely mixed state $\rho = I/d$, where $\dimn{\hil{H}^{Q}} = d$, the $d$ eigenvalues of $\rho$ are all equal to $1/d$ and $S = \log{d}$. This is the maximum value of $S$ in a $d$-dimensional Hilbert space. The von Neumann entropy\index{von Neumann entropy}\index{entropy!von Neumann} $S$ is zero, the minimum value, if and only if $\rho$ is a pure state, where the eigenvalues of $\rho$ are 1 and 0. So $0 \leq S \leq \log{d}$, where $d$ is the dimension of $\hil{H}^{Q}$.

Recall that we can think of the Shannon entropy\index{Shannon entropy}\index{entropy!Shannon} as a measure of the average amount of information gained by identifying the state produced by a known stochastic source. Alternatively, the Shannon entropy\index{Shannon entropy}\index{entropy!Shannon} represents the optimal compressibility of the information produced by an information source. The von Neumann entropy\index{von Neumann entropy}\index{entropy!von Neumann} does \emph{not}, in general, represent the amount of information gained by identifying the quantum state produced by a stochastic source characterized as a mixed state, because nonorthogonal quantum states in a mixture cannot be reliably identified. However, as we will see in \S\ref{sec:quantcompress}, the von Neumann entropy can be interpreted in terms of compressibility\index{information!compressibility} via Schumacher's source coding theorem{ Schumacher's source coding theorem} for quantum information \cite{Schumacher95}, a generalization of Shannon's source coding theorem\index{Shannon's source coding theorem (noiseless channel coding theorem)} for classical information. For an elementary two-state quantum system with a 2-dimensional Hilbert space considered as representing the output of an elementary quantum information source, $S = 1$ for an equal weight distribution over two orthogonal states (i.e., for the density operator $\rho = I/2$), so Schumacher takes the basic unit of quantum information as the \index{qubit}`qubit.' By analogy with the term `bit,' the term\index{qubit} `qubit' refers to the basic unit of quantum information in terms of the von Neumann entropy, and to an elementary two-state quantum system considered as representing the possible outputs of an elementary quantum information source.

The difference between quantum information\index{quantum!information}\index{information!quantum} as measured by von Neumann entropy\index{von Neumann entropy}\index{entropy!von Neumann} $S$ and classical information\index{information!classical} as measured by Shannon entropy\index{Shannon entropy}\index{entropy!Shannon} $H$ can be brought out by considering the quantum notions of conditional entropy and mutual information (cf. \S \ref{sec:entropy}), and in particular the peculiar feature of \emph{inaccessibility}\index{information!inaccessible} associated with quantum information. 

For a composite system $AB$, conditional von Neumann entropy\index{entropy!conditional} and mutual information\index{mutual information}\index{information!mutual} are defined in terms of the joint entropy\index{joint entropy}\index{entropy!joint} $S(AB) = -\trace{\rho^{AB}\log{\rho^{AB}}}$ by analogy with the corresponding notions for Shannon entropy (cf. Eqs. (\ref{eq:chainrule}), (\ref{eq:mutinfo2}), (\ref{eq:mutinfo3})):
\begin{eqnarray}
S(A|B) & = & S(A,B) - S(B) \\
S(A\!:\!B) & = & S(A) - S(A|B) \label{eq:qmutinfo1} \\
& = & S(B) - S(B|A) \label{eq:qmutinfo2} \\
& = & S(A) + S(B) - S(A,B) \label{eq:qmutinfo3}
\end{eqnarray}
The joint entropy\index{joint entropy} satisfies the subadditivity inequality\index{subadditivity inequality}:
\begin{equation}
S(A,B) \leq S(A) + S(B)
\end{equation}
with equality if and only if $A$ and $B$ are uncorrelated, i.e., $\rho^{AB} = \rho^{A}\otimes\rho^{B}$.

Now, $S(A|B)$ can be negative, while the conditional Shannon entropy is always positive or zero. Consider, for example, the entangled state $\ket{\Psi} = (\ket{00} + \ket{11})/\sqrt{2}$. Since $\ket{\Psi}$ is a pure state, $S(A,B) = 0$. But $S(A) = S(B) = 1$. So $S(A|B) = S(A,B) - S(A) = -1$. In fact, for a pure state $\ket{\Psi}$ of a composite system $AB$, $S(A|B) < 0$ if and only if $\ket{\Psi}$ is entangled.

For a composite system $AB$ in a product state $\rho\otimes\sigma$, it follows from the definition of joint entropy\index{joint entropy} that:
\begin{equation}
 S(A,B) = S(\rho\otimes\sigma) = S(\rho) + S(\sigma) = S(A) + S(B). \label{eq:prodentropy}
\end{equation}
If $AB$ is in a pure state $\ket{\Psi}$, it follows from the Schmidt decomposition theorem\index{Schmidt decomposition} that $\ket{\Psi}$ can be expressed as 
\begin{equation}
\ket{\Psi} = \sum_{i}\sqrt{p_{i}}\ket{i}\bra{i}
\end{equation}
from which it follows that
\begin{equation}
\begin{array}{lllll}
\rho_{A}  & = & \ptrace{B}{\ket{\psi}\bra{\psi}} & = & \sum_{i}\ket{i}\bra{i} \nonumber  \\
 \rho_{B} & = &  \ptrace{A}{\ket{\psi}\bra{\psi}} & = & \sum_{i}\ket{i}\bra{i};
\end{array}
\end{equation}
and so:
\begin{equation}
S(A) = S(B) = -\sum_{i}p_{i}\log{p_{i}}.
\end{equation}

Consider a mixed state prepared as a mixture of states $\rho_{i}$ with weights $p_{i}$. It can be shown that
\begin{equation}
S(\sum_{i}p_{i}\rho_{i}) \leq H(p_{i}) + \sum_{i}p_{i}S(\rho_{i}) \label{eq:ineqmixture}
\end{equation}
with equality if and only if the states $\rho_{i}$ have support on orthogonal subspaces (see \cite[Theorem 11.10, p. 518]{NielsenChuang}). The entropy $H(p_{i})$ is referred to as the \emph{entropy of preparation} of the mixture $\rho$. 

If the states $\rho_{i}$ are pure states, then $S(\rho) \leq H(p_{i})$. For example, suppose $\hil{H}^{Q}$ is 2-dimensional  and $p_{1} = p_{2} = 1/2$, then $H(p_{i}) = 1$. So if we had a classical information source producing the symbols 1 and 2 with equal probabilities, no compression\index{information!compression} of the information would be possible. However, if the symbols 1 and 2 are encoded as nonorthogonal quantum states $\ket{r_{1}}$ and $\ket{r_{2}}$, then $S(\rho) < 1$. As we will see in \S\ref{sec:quantcompress}, according to Schumacher's source coding theorem\index{Schumacher's quantum source coding theorem (noiseless channel coding theorem)}, since $S(\rho) < 1$, quantum compression is possible, i.e., we can transmit long sequences of qubits reliably using $S < 1$ qubits per quantum state produced by the source.

Note that if $AB$ is prepared in a mixture of states $\rho_{i}\otimes\ket{i}\bra{i}$ with weights $p_{i}$, where the $\rho_{i}$ are any density operators, not necessarily orthogonal, then it follows from (\ref{eq:ineqmixture}), (\ref{eq:prodentropy}), and the fact that $S(\ket{i}\bra{i}) = 0$ that
\begin{eqnarray}
S(\sum_{i}p_{i}\rho_{i}\otimes\ket{i}\bra{i}) & = & H(p_{i}) + \sum_{i}p_{i}S(\rho_{i}\otimes\ket{i}\bra{i}) \nonumber \\
& = & H(p_{i}) + \sum_{i}p_{i}S(\rho_{i}). \label{eq:entropysum}
\end{eqnarray}

The von Neumann entropy\index{von Neumann entropy}\index{entropy!von Neumann} of a mixture of states $\rho_{i}$ with weights $p_{i}$, $\sum p_{i}\rho_{i}$, is a concave function of the states in the distribution, i.e., 
\begin{equation}
S(\sum_{i}p_{i}\rho_{i}) \geq \sum_{i}p_{i}S(\rho_{i}).
\end{equation}
To see this, consider a composite system $AB$ in the state
\begin{equation}
\rho^{AB} = \sum p_{i}\rho_{i}\otimes\ket{i}\bra{i}.
\end{equation}
We have
\begin{equation}
S(A) = S(\sum_{i}p_{i}\rho_{i})
\end{equation}
\begin{equation}
S(B) = S(\sum_{i}p_{i}\ket{i}\bra{i}) = H(p_{i})
\end{equation}
and
\begin{equation}
S(A,B) = H(p_{i}) + \sum_{i}p_{i}S(\rho_{i})
\end{equation}
by equation (\ref{eq:entropysum}). By subadditivity $S(A) +S(B) \geq S(A,B)$, so:
\begin{equation}
S(\sum_{i}p_{i}\rho_{i}) \geq \sum_{i}p_{i}S(\rho_{i}).
\end{equation}

It turns out that projective measurements\index{measurement!projective!and entropy} always increase entropy, i.e., if $\rho' = \sum_{i}P_{i}\rho P_{i}$, then $S(\rho') \geq S(\rho)$, but generalized measurements\index{measurement!generalized} can \emph{decrease} entropy. Consider, for example, the generalized measurement on a qubit in the initial state $\rho$ defined by the measurement operators $M_{1} = \ket{0}\bra{0}$ and $M_{2} = \ket{0}\bra{1}$. (Note that these operators \emph{do} define a generalized measurement because $M_{1}^{\dagger}M_{1} + M_{2}^{\dagger}M_{2} = \ket{0}\bra{0} + \ket{1}\bra{1} = I$.) After the measurement
\begin{eqnarray}
\rho' & = & \ket{0}\bra{0}\rho\ket{0}\bra{0} + \ket{0}\bra{1}\rho\ket{1}\bra{0}  \nonumber \\
& = & \trace{\rho}\ket{1}\bra{1}  \nonumber \\
& = & \ket{1}\bra{1}.
\end{eqnarray}
So $S(\rho^{\prime}) = 0 \leq S(\rho)$.

\subsection{The `No Cloning' Theorem\index{cloning!`no cloning' theorem}}

\label{sec:nocloning}

In \S~\ref{sec:quantum} we saw that two nonorthogonal quantum states cannot be reliably distinguished\index{distinguishability} by any measurement. A \emph{`no cloning' theorem}\index{cloning!`no cloning' theorem} establishes that nonorthogonal quantum states cannot be copied. To see this, suppose there were a device $D$ that could copy any input quantum state of a system $Q$ with states in $\hil{H}^{Q}$. Suppose the initial ready state of the device $D$ is $\ket{0} \in \hil{H}^{D}$. Then we require, for any orthonormal set of input states $\{\ket{i}\}$:
\begin{equation}
\ket{i}\ket{0} \stackrel{U}{\longrightarrow} \ket{i}\ket{i}
\end{equation}
where $U$ is the unitary transformation that implements the copying process. By linearity, it then follows that for any input state $\sum_{i}c_{i}\ket{i}$:
\begin{equation}
(\sum_{i}c_{i}\ket{i})\ket{0} \stackrel{U}{\longrightarrow} \sum_{i}c_{i}\ket{i}\ket{i}
\end{equation}
But for copying we require that:
\begin{equation}
(\sum_{i}c_{i}\ket{i})\ket{0} \stackrel{U}{\longrightarrow} (\sum_{i}c_{i}\ket{i})(\sum_{i}c_{i}\ket{i})
\end{equation}
and
\begin{equation}
\sum_{i}c_{i}\ket{i}\ket{i}  \neq  (\sum_{i}c_{i}\ket{i})(\sum_{i}c_{i}\ket{i}) 
 =  \sum_{ij}c_{i}c_{j}\ket{i}\ket{j}
\end{equation}
unless $c_{i}c_{j} = \delta_{ij}$, which means that the device could not copy any states that are not in the orthonormal set $\ket{i}$. 

Alternatively, one might note that if two states $\ket{\psi}$ and $\ket{\phi}$ could be copied, then
\begin{eqnarray}
\ket{\psi}\ket{0} & \stackrel{U}{\longrightarrow} & \ket{\psi}\ket{\psi} \label{eq:copy1}\\
\ket{\phi}\ket{0} & \stackrel{U}{\longrightarrow} & \ket{\phi}\ket{\phi} \label{eq:copy2}
\end{eqnarray}
Since unitary transformations preserve inner produces, we require that
\begin{equation}
\braket{\psi}{\phi} = \braket{\psi}{\phi}\braket{\psi}{\phi} \label{eq:copy3}
\end{equation}
which is possible if and only if if$\braket{\psi}{\phi} = 1$ or 0. That is: for cloning to be possible, either the states are identical, or they are orthogonal.

The `no cloning' theorem was proved independently by Dieks\index{Dieks} \shortcite{Dieks} and Wootters and Zurek\index{Wootters}\index{Zurek} \shortcite{WoottersZurek}. An important extension of this result to mixtures is due to Barnum, Caves, Fuchs, Jozsa, and Schumacher \shortcite{BCFJS96b}. In a cloning process, a ready state $\sigma$ of a system $B$ and the state to be cloned $\rho$ of a system $A$ are transformed into two copies of $\rho$. In a more general \emph{broadcasting}\index{broadcasting} process, a ready state $\sigma$ and the state to be broadcast $\rho$ are transformed to a new state $\omega$ of $AB$, where the marginal state $\omega$ with respect to both $A$ and $B$ is $\rho$, i.e., 
\begin{equation}
\rho_{A} = \ptrace{B}{\omega} = \rho \nonumber
\end{equation}
\begin{equation}
\rho_{B} = \ptrace{A}{\omega} = \rho 
\end{equation}
The `no cloning' theorem\index{cloning!`no cloning' theorem} states that a set of pure states can be cloned if and only if the states are mutually orthogonal. The \emph{`no broadcasting' theorem}\index{broadcasting!`no broadcasting' theorem} states that an arbitrary set of states can be broadcast if and only if they are represented by mutually commuting density operators. Classically, since all pure states are, in a formal sense, orthogonal and all operators (representing real-valued functions on phase space) commute, both cloning and broadcasting are possible. Note that broadcasting reduces to cloning for pure states.

Of course, it is always possible to build a special-purpose device to clone a given (known) quantum state $\ket{\psi}$, because this would simply be a device that prepares the state $\ket{\psi}$. The `no cloning' theorem\index{cloning!`no cloning' theorem}, from another perspective, is just the statement of the quantum measurement problem (see \S \ref{sec:foundations}): measurements, in the classical sense of reproducing in a second system a copy of the state of the first system (or, more generally, a `pointer state' that represents the state of the first system), are impossible in quantum mechanics, except for measurements restricted to orthogonal sets of input states.

A modification of the argument leading to Eqs. (\ref{eq:copy1})--(\ref{eq:copy3}) shows that no information gain about the identity of nonorthogonal states is possible without disturbing the states. Suppose the device $D$ acts as a measuring device that records some information about the identity of the input state, i.e., the output state of the device is different for different input states $\ket{\psi}, \ket{\phi}$; and that the device does not disturb the input states. Then
\begin{eqnarray}
\ket{\psi}\ket{0} & \stackrel{U}{\longrightarrow} & \ket{\psi}\ket{\psi'} \\
\ket{\phi}\ket{0} & \stackrel{U}{\longrightarrow} & \ket{\phi}\ket{\phi'}
\end{eqnarray}
from which it follows that
\begin{equation}
\braket{\psi}{\phi} = \braket{\psi}{\phi}\braket{\psi'}{\phi'}
\end{equation}
and so
\begin{equation}
\braket{\psi'}{\phi'} = 1
\end{equation}
since $\braket{\psi}{\phi} \neq 0$ if $\ket{\psi}$ and $\ket{\phi}$ are nonorthogonal. In other words, if there is no disturbance to the nonorthogonal input states, there can be no information gain about the identity of the states. So, for example, an eavesdropper, Eve, could gain no information about the identity of nonorthogonal quantum states communicated between Alice and Bob without disturbing the states, which means that passive eavesdropping is impossible for quantum information.

The observation that a set of pure states can be cloned if and only if they are mutually orthogonal is equivalent to the observation that a set of pure states can  be reliably distinguished\index{distinguishability!states} if and only if they are mutually orthogonal. For if we could distinguish a pair of states $\ket{\psi}$ and $\ket{\phi}$, then we could copy them by simply preparing the states with special-purpose preparation devices for $\ket{\psi}$ and $\ket{\phi}$. And if we could copy the states, then we could prepare as many copies as we liked of each state. Because the product states $\ket{\psi}^{\otimes n}$ and $\ket{\phi}^{\otimes n}$ become orthogonal in the limit as $n \rightarrow \infty$, these states are certainly distinguishable, and so the possibility of cloning the states $\ket{\psi}$ and $\ket{\phi}$ would provide a means of distinguishing them.

Note also that, by a similar argument, cloning\index{cloning} would allow different mixtures associated with the same density operator to be distinguished\index{distinguishability!mixtures}. The equal-weight mixture of qubit states $\ket{\!\uparrow_{z}} = \ket{0}, 
\ket{\!\downarrow_{z}} = \ket{1}$ (the eigenstates of the spin observable $Z = \sigma_{z}$) has the same density operator, $I/2$, as the equal-weight mixture of states $\ket{\!\uparrow_{x}} = \frac{1}{\sqrt{2}}(\ket{0} + \ket{1})$, $\ket{\!\downarrow_{x}} = \frac{1}{\sqrt{2}}(\ket{0} - \ket{1})$ (the eigenstates of $X = \sigma_{x}$). Since the cloned states $\ket{\!\uparrow_{x}}^{\otimes n}$, $\ket{\!\downarrow_{x}}^{\otimes n}$ become distinguishable from the cloned states $\ket{\!\uparrow_{z}}^{\otimes n}$, $\ket{\!\downarrow_{z}}^{\otimes n}$, cloning would allow the two mixtures to be distinguished. 

This possibility would also allow superluminal signalling\index{superluminal signalling}. For suppose Alice and Bob shared the entangled state $\frac{1}{\sqrt{2}}(\ket{0}\ket{1} -\ket{1}\ket{0})$. If Alice measured $X$ or $Z$ on her qubit, she would steer Bob's qubit into the mixture $\frac{1}{2}\ket{\!\uparrow_{x}}\bra{\!\uparrow_{x}} + \frac{1}{2}\ket{\!\downarrow_{x}}\bra{\!\downarrow_{x}}$ or the mixture 
$\frac{1}{2}\ket{\!\uparrow_{z}}\bra{\!\uparrow_{z}} + 
\frac{1}{2}\ket{\!\downarrow_{z}}\bra{\!\downarrow_{z}}$. If Bob could distinguish these mixtures by cloning, in a shorter time than the time taken for light to travel between Alice and Bob, he would be able to ascertain whether Alice measured $X$ or $Z$, so 1 bit of information would be transferred from Alice to Bob superluminally.

\subsection{Accessible Information\index{information!accessible}}

\label{sec:accessible}

The ability to exploit quantum states to perform new sorts of information-processing tasks arises because quantum states have different distinguishability properties than classical states. Of course, it is not the mere lack of distinguishability of quantum states that is relevant here, but the different sort of distinguishability enjoyed by quantum states. This indistinguishability is reflected in the limited \emph{accessibility} of quantum information. 

To get a precise handle on this notion of accessibility, consider a classical information source in Shannon's sense, with Shannon entropy $H(X)$. Suppose the source produces symbols represented as the values $x$ (in an alphabet $\hil{X})$ of a random variable $X$, with probabilities $p_{x}$, and that the symbols are encoded as quantum states $\rho_{x}$, $x \in X$. The mutual information $H(X:Y)$ (as defined by Eqs. (\ref{eq:mutinfo1}),  (\ref{eq:mutinfo2}),  (\ref{eq:mutinfo3})) is a measure of how much information one gains, on average, about the value of the random variable $X$ on the basis of the outcome $Y$ of a measurement on a given quantum state. The \emph{accessible information}\index{information!accessible} is defined as:
\begin{equation}
\mbox{Sup }H(X\!:\!Y)
\end{equation}
over all possible measurements. 

The \emph{Holevo bound}\index{Holevo bound} on mutual information provides an important upper bound to accessible information:
\begin{equation}
H(X\!:\!Y) \leq S(\rho) - \sum_{x}p_{x}S(\rho_{x}) \label{eq:Hol}
\end{equation}
where $\rho = \sum_{x}p_{x}\rho_{x}$ and the measurement outcome $Y$ is obtained from a measurement defined by a POVM $\{E_{y}\}$. Since $S(\rho) - \sum_{x}p_{x}S(\rho_{x}) \leq H(X)$ by Eq. (\ref{eq:ineqmixture}), with equality if and only if the states $\rho_{x}$ have orthogonal support, we have:
\begin{equation}
H(X\!:\!Y) \leq H(X)
\end{equation}
Note that $X$ can be distinguished from $Y$ if and only if $H(X\!:\!Y) = H(X)$. If the states $\rho_{x}$ are orthogonal pure states, then in principle there exists a measurement that will distinguish the states, and for such a measurement $H(X\!:\!Y) = H(X)$. In this case, the accessible information is the same as the entropy of preparation of the quantum states, $H(X)$. But if the states are nonorthogonal, then $H(X\!:\!Y) < H(X)$ and there is no measurement, even in the generalized sense, that will enable the reliable identification of $X$. 

Note, in particular, that if the values of $X$ are encoded as the pure states of a qubit, then $H(X\!:\!Y) \leq S(\rho)$ and $S(\rho) \leq 1$. It follows that at most 1 bit of information can be extracted from a qubit by measurement. If $X$ has $k$ equiprobable values, $H(X) = \log{k}$. Alice could encode these $k$ values into a qubit by preparing it in an equal-weight mixture of $k$ nonorthogonal pure states, but Bob could only extract at most 1 bit of information about the value of $X$. For an $n$-state quantum system associated with an $n$-dimensional Hilbert space, $S(\rho) \leq \log{n}$. So even though Alice could encode any amount of information into such an $n$-state quantum system (by preparing the state as a mixture of nonorthogonal states), the most information that Bob could extract from the state by measurement is $\log{n}$, which is the same as the maximum amount of information that could be encoded into and extracted from an $n$-state classical system. It might seem, then, that the inaccessibility of quantum information as quantified by the Holevo bound would thwart any attempt to exploit quantum information to perform nonclassical information-processing tasks. In the following sections, we shall see that this is not the case: surprisingly, the  inaccessibility of quantum information can actually be  exploited in information-processing tasks that transcend the scope of classical information.

For an insightful derivation of the Holevo bound\index{Holevo bound!derivation} (essentially reproduced below), see \cite[Theorem 12.1, p. 531]{NielsenChuang}. The basic idea is the following: Suppose Alice encodes the distinguishable symbols of a classical information source with entropy $H(X)$ as quantum states $\rho_{x}$ (not necessarily orthogonal). That is, Alice has a quantum system $P$, the preparation device, with an orthonormal pointer basis $\ket{x}$ corresponding to the values of the random variable $X$, which are produced by the source with probabilities $p_{x}$. The preparation interaction correlates the pointer states $\ket{x}$ with the states $\rho_{x}$ of a quantum system $Q$, so that the final state of $P$ and $Q$ after the preparation interaction is:
\begin{equation}
\rho^{PQ} = \sum_{x}p_{x}\ket{x}\bra{x}\otimes\rho_{x}. \label{eq:initial}
\end{equation}
Alice sends the system $Q$ to Bob, who attempts to determine the value of the random variable $X$ by measuring the state of $Q$. The initial state of $P, Q,$ and Bob's measuring instrument $M$ is:
\begin{equation}
\rho^{PQM} = \sum_{x}p_{x}\ket{x}\bra{x}\otimes\rho_{x}\otimes\ket{0}\bra{0} 
\end{equation}
where $\ket{0}\bra{0}$ is the initial ready state of $M$. Bob's measurement can be described by a quantum operation $\hil{E}$ on the Hilbert space $\hil{H}^{Q}\otimes\hil{H}^{M}$ that stores a value of $y$, associated with a POVM $\{E_{y}\}$ on $\hil{H}^{Q}$, in the pointer state $\ket{y}$ of $M$, i.e., $\hil{E}$ is defined for any state $\sigma \in \hil{H}^{Q}$ and initial ready state $\ket{0} \in \hil{H}^{M}$ by:
\begin{equation}
\sigma\otimes\ket{0}\bra{0} \stackrel{\hil{E}}{\longrightarrow} \sum_{y}\sqrt{E_{y}}\sigma\sqrt{E_{y}}\otimes\ket{y}\bra{y}.
\end{equation}

We have (recall the definition of quantum mutual information in Eqs. (\ref{eq:qmutinfo1})--(\ref{eq:qmutinfo3})):
\begin{equation}
S(P\!:\!Q) = S(P\!:\!Q,M)\footnote{The notation $S(P\!:\!Q,M)$ refers to the mutual information between the system $P$ and the composite system consisting of the system $Q$ and the measuring device $M$, in the initial state (\ref{eq:initial}). That is, the comma notation refers to the joint system (cf. Eqs. (\ref{eq:qmutinfo1}), (\ref{eq:qmutinfo3})): $S(P\!:\!Q,M) = S(P)-S(P|Q,M) = S(P) + S(Q,M) - S(P,Q,M)$}
\end{equation}
because $M$ is initially uncorrelated with $PQ$ and
\begin{equation}
S(P'\!:\!Q',M') \leq S(P\!:\!Q,M)
\end{equation}
because it can be shown (\cite[Theorem 11.15, p. 522]{NielsenChuang}) that quantum operations never increase mutual information (primes here indicate states after the application of $\hil{E}$). Finally:
\begin{equation}
S(P'\!:\!Q^{\prime},M')
\end{equation}
because discarding systems never increases mutual information (\cite[Theorem 11.15, p. 522]{NielsenChuang}), and so:
\begin{equation}
S(P'\!:\!M') \leq S(P\!:\!Q) \label{eq:hol}
\end{equation}
which (following some algebraic manipulation) is the statement of the Holevo bound, i.e., (\ref{eq:hol}) reduces to (\ref{eq:Hol}).

To see this, note (from (\ref{eq:initial})) that
\begin{equation}
\rho^{PQ} = \sum_{x}p_{x}\ket{x}\bra{x}\otimes\rho_{x}
\end{equation}
So $S(P) = H(p_{x)}), S(Q) = S(\sum_{x}p_{x}\rho_{x}) = S(\rho)$ and, by (\ref{eq:entropysum}), 
\begin{equation}
S(P,Q) = H(p_{x)} + \sum_{x}p_{x}S(\rho_{x})
\end{equation}
since the states $\ket{x}\bra{x}\otimes\rho_{x}$ have support on orthogonal subspaces in $\hil{H}^{P}\otimes\hil{H}^{Q}$. It follows that 
\begin{eqnarray}
S(P\!:\!Q) & = & S(P) + S(Q) -S(P,Q) \nonumber \\
& = & S(\rho) - \sum_{x}p_{x}S(\rho_{x})
\end{eqnarray}
which is the right hand side of the Holevo bound.

For the left hand side:
\begin{eqnarray}
\rho^{P'M'} & = & \ptrace{Q'}{\rho^{P'Q'M'}} \\
& = & \ptrace{Q'}{\sum_{xy}p_{x}\ket{x}\bra{x}\otimes\sqrt{E_{y}}\rho_{x}\sqrt{E_{y}}\otimes\ket{y}\bra{y}} \\
& = & \sum_{xy}p_{x}\trace{E_{y}\rho_{x}E_{y}}\ket{x}\bra{x}\otimes\ket{y}\bra{y} \\
& = & \sum_{xy}p(x,y)\ket{x}\bra{x} \otimes\ket{y}\bra{y},
\end{eqnarray}
since $p(x,y) = p_{x}p(y\mid x) = p_{x}\trace{\rho_{x}E_{y}} = p_{x}\trace{\sqrt{E_{y}}\rho_{x}\sqrt{E_{y}}}$, and so $S(P':M') = H(X:Y)$.

The Holevo bound\index{Holevo bound} limits the representation of classical bits by qubits. Putting it another way, the Holevo bound characterizes the resource cost of encoding classical bits as qubits: one qubit is necessary and sufficient. Can we represent qubits by bits? If so, what is the cost of a qubit in terms of bits? This question is answered by the following result \cite{BHJW2001}: A quantum source of nonorthogonal signal states can be compressed with arbitarily high fidelity to $\alpha$ qubits per signal plus any number of classical bits per signal if and only if $\alpha$ is at least as large as the von Neumann entropy $S$ of the source. This means that a generic quantum source cannot be separated into a classical and quantum part: quantum information cannot be traded for any amount of classical information.

\subsection{Quantum Information Compression\index{information!compression!quantum}}
\label{sec:quantcompress}

As pointed out in \S \ref{sec:Neumann}, Shannon's source coding theorem (noiseless channel coding theorem) and the core notion of a typical sequence can be generalized for quantum sources. This was first shown by Jozsa\index{Jozsa} and Schumacher \shortcite{JS94} and Schumacher\index{Schumacher} \shortcite{Schumacher95}. See also \cite{BFJS96}.

For a classical information bit source, where the output of the source is given by a random variable $X$ with two possible values $x_{1}, x_{2}$ with probabilities $p_{1}, p_{2}$, the Shannon entropy of the information produced by the source is $H(X) = H(p_{1},p_{2})$. So by Shannon's source coding theorem the information can be compressed and communicated to a receiver with arbitrarily low probability of error by using $H(X)$ bits per signal, which is less than one bit if $p_{1} \neq p_{2}$.

Now suppose the source produces qubit states $\ket{\psi_{1}}, \ket{\psi_{2}}$ with probabilities $p_{1}, p_{2}$. The Shannon entropy of the mixture $\rho = p_{1}\ket{\psi_{1}}\bra{\psi_{1}} + p_{2}\ket{\psi_{2}}\bra{\psi_{2}}$ is $S(\rho)$. Schumacher's generalization of Shannon's source coding theorem shows that the quantum information encoded in the mixture $\rho$ can be compressed and communicated to a receiver with arbitrarily low probability of error by using $S(\rho)$ qubits per signal, and $S(\rho) < 1$ if the qubit states are nonorthogonal.

Note that the signals considered here are \emph{qubit states}. What Schumacher's theorem\index{Schumacher's quantum source coding theorem (noiseless channel coding theorem)} shows is that we can reliably communicate the sequence of qubit states produced by the source by sending less than one qubit per signal. Note also that since $S(\rho) < H(p_{1},p_{2})$ if the qubit states are nonorthogonal, the quantum information represented by the sequence of qubit states can be compressed beyond the classical limit of the classical information associated with the entropy of preparation of $\rho$ (i.e., the Shannon entropy of the random variable whose values are the labels of the qubit states). 

Since the individual states in a mixture are not in general distinguishable, there are two distinct sorts of compression applicable to quantum information that do not apply to classical information. In \emph{blind compression}\index{information!compression!blind}, the sequence of quantum states produced by a source is compressed via a compression scheme that depends only on the identities of the quantum states and their probabilities, i.e., the input to the compression scheme is the density operator associated with the distribution. In \emph{visible compression}\index{information!compression!visible}, the identity of each individual quantum state produced by the source is assumed to be known, i.e., the input to the compression scheme is an individual quantum state in the sequence produced by the source, and the compression of the state is based on the probability distribution of such states. 

An example: an inefficient visible compression scheme of the above qubit source ($\ket{\psi_{1}},\ket{\psi_{2}}$ with probabilities $p_{1},p_{2}$) would simply involve sending the classical information of the quantum state labels, compressed to $H(p_{1}, p_{2})$ bits per signal, to the receiver, where the original qubit states could then be prepared after decompression of the classical information. This scheme is not optimal by Schumacher's theorem (for nonorthogonal qubit states) because $S(\rho) < H(p_{1}, p_{2})$. Of course, Schumacher's theorem refers to a compression rate\index{information!compression rate} of $S(\rho)$ \emph{qubits} per quantum signal, while the application of Shannon's theorem here refers to $H(p_{2}, p_{2})$ \emph{bits} per classical signal. But note that the communication of one classical bit requires the same physical resource as the communication of one qubit, prepared in one of two orthogonal basis states. Note also that sending the (nonorthogonal) qubit states themselves, which would require one qubit per signal, would not convey the identity of the states in the sequence to the transmitter. So the classical information about the individual state labels in the sequence (which would be bounded by $\log{n}$ per signal if we considered a source producing $n$ qubit states) is really redundant if the aim is to communicate the quantum information associated with the sequence of qubit states. 

Remarkably, Schumacher's theorem\index{Schumacher's quantum source coding theorem (noiseless channel coding theorem)}  shows that the optimal compressibility of the quantum information associated with a sequence of quantum pure states is $S(\rho)$ qubits per signal, for blind \emph{or} visible compression. 

To see the general idea, consider a source of (possibly nonorthogonal) qubits $\ket{\psi_{1}}, \ket{\psi_{2}}$ with probabilities $p_{1}, p_{2}$. The density operator of the probability distribution is $\rho = p_{1}\ket{\psi_{1}}\bra{\psi_{1}} + p_{2}\ket{\psi_{2}}\bra{\psi_{2}}$.

An $n$-sequence of states produced by the source is represented by a state
\begin{equation}
\ket{\Psi_{i_{1} \ldots i_{n}}} = \ket{\psi_{i_{1}}} \ldots \ket{\psi_{i_{n}}}
\end{equation}
 in $\hil{H}^{\otimes n}_{2}$. Each such state has a probability $p_{i_{1} \ldots i_{n}} = p_{i_{1}} \ldots p_{i_{n}}$. The $n$-sequences span the $2^{n}$-dimensional Hilbert space $\hil{H}^{\otimes n}_{2}$, but as $n \rightarrow \infty$ it turns out that the probability of finding an $n$-sequence in a `typical subspace' (in a measurement, on an $n$-sequence produced by the source,
  of the projection operator onto the subspace ) tends to 1.  That is, for any $\epsilon, \delta > 0$, there is a subspace $\hil{T}^{(n)}_{\delta}$ of dimension between $2^{n(S(\rho) - \delta)}$ and $2^{n(S(\rho) + \delta)}$, with projection operator $P^{(n)}_{\delta}$, such that:
\begin{equation}
\sum_{\mbox{all sequences}}p_{i_{1} \ldots i_{n}}\trace{\ket{\Psi_{i_{1} \ldots i_{n}}}\bra{\Psi_{i_{1} \ldots i_{n}}}P^{(n)}_{\delta}} 
 =  \trace{\rho^{\otimes n}P^{(n)}_{\delta}} \geq 1- \epsilon.
\end{equation}
Here $\rho^{\otimes n} = \rho \otimes \rho \ldots \rho$, the $n$-fold tensor product of $\rho$ with itself, is the density operator of $n$-sequences of states produced by the source:
\begin{eqnarray}
\rho^{\otimes n} & = & \sum_{\mbox{all }n\mbox{-sequences}}p_{i_{1}\ldots i_{n}}\ket{\Psi_{i_{1} \ldots i_{n}}}\bra{\Psi_{i_{1} \ldots i_{n}}} \\
& = & \sum_{\mbox{all }n\mbox{-sequences}}p_{i_{1}} \ldots p_{i_{n}}\ket{\psi_{i_{1}}}\bra{\psi_{i_{1}}}\otimes \ldots \otimes \ket{\psi_{i_{n}}}\bra{\psi_{i_{n}}}
\end{eqnarray}
where each state $\ket{\psi_{i_{j}}}$ is one of $k$ possible states in a $d$-dimensional Hilbert space. Recall that the statistical properties of such  $n$-sequences of states, for all possible measurements, is given by $\rho^{\otimes n}$ and does not depend on the representation of $\rho^{\otimes n}$ as a particular mixture of states. Since $S(\rho) \leq 1$ for a qubit source, the dimension of $\hil{T}^{(n)}_{\delta}$ decreases exponentially in $\hil{H}^{\otimes n}_{2}$ as $n \rightarrow \infty$, i.e., the typical subspace is exponentially small in $\hil{H}^{\otimes n}_{2}$ for large $n$. 

Note that this does \textit{not} mean that almost all $n$-sequences of states produced by the source lie in the typical subspace. Rather, almost all $n$-sequences produced by the source are such that a measurement of $P^{(n)}_{\delta}$ on the sequence will yield the value 1, i.e., almost all $n$-sequences produced by the source will answer `yes' in a measurement of the projection operator onto the typical subspace. So, in this sense, most sequences produced by the source will be found to lie in the typical subspace on measurement, and for any subspace \hil{V} of dimension less than $2^{n(S(\rho) - \delta)}$ it can be shown that the average probability of finding an $n$-sequence produced by the source in $\hil{V}$ is less than  any pre-assigned $\epsilon$ for sufficiently large $n$.

Consider now the general case where the source produces $k$ states $\ket{\psi_{1}}. \ldots, \ket{\psi_{k}} \in \hil{H}_{d}$ (not necessarily orthogonal) with probabilities $p_{1}. \ldots, p_{k}$. Here the density operator associated with the source is $\rho = \sum_{i=1}^{k}p_{i}\ket{\psi_{i}}\bra{\psi_{i}}$. Sequences of length $n$ span a subspace of $d^{n} = 2^{n\log d}$ dimensions and the typical subspace $\hil{T}^{(n)}_{\delta}$ has dimension between $2^{n(s(\rho) - \delta)}$ and $2^{n(s(\rho) + \delta)}$, which is again exponentially small in $\hil{H}^{\otimes n}_{d}$ because $S(\rho) \leq \log d$.

For comparison with Shannon's theorem, we write $\rho$ in the spectral representation as:
\begin{equation}
\rho = \sum_{x}p(x)\ket{x}\bra{x}
\end{equation}
where $\{p(x)\}$ is the set of non-zero eigenvalues of $\rho$ and $\{\ket{x}\}$ is an orthonormal set of eigenstates of $\rho$. If $\rho$ has eigenvalues $p(x)$ and eigenstates $\ket{x}$, then $\rho^{\otimes n}$ has eigenvalues $p(x_{1})p(x_{2}) \ldots p(x_{n})$ and eigenstates $\ket{x_{1}}\ket{x_{2}} \ldots \ket{x_{n}}$.

A $\delta$-typical state is defined as a state $\ket{x_{1}}\ket{x_{2}} \ldots \ket{x_{n}}$ for which the sequence $x_{1},x_{2},\ldots,x_{n}$ is a $\delta$-typical sequence, in the sense that (cf. Eq. (\ref{eq:typical})):
\begin{equation}
2^{-n(S(\rho)+\delta)} < p(x_{1}\ldots x_{n}) < 2^{-n(S(\rho)-\delta)}.
\end{equation}
The $\delta$-typical subspace $\hil{T}^{(n)}_{\delta}$ is the subspace spanned by all the $\delta$-typical states. Denote the projection operator onto $\hil{T}^{(n)}_{\delta}$ by:
\begin{equation}
P^{(n)}_{\delta} = \sum_{\delta-\mbox{typical states}}\ket{x_{1}}\bra{x_{1}}\otimes\ket{x_{2}}\bra{x_{2}}\ldots\ket{x_{n}}\bra{x_{n}}
\end{equation}
Then, for a fixed $\delta > 0$, it can be shown that for any $\epsilon > 0$ and sufficiently large $n$
\begin{equation} 
\trace{P^{(n)}_{\delta}\rho^{\otimes n}}  \geq 1-\epsilon;
\end{equation}
and the dimension of $\hil{T}^{(n)}_{\delta}$ ($= \trace{P^{(n)}_{\delta}}$) satisfies
\begin{equation}
(1-\epsilon)2^{n(S(\rho)-\delta)} \leq \dim\hil{T}^{(n)}_{\delta}  \leq 2^{n(S(\rho)+\delta)}.
\end{equation}
That is, the dimension of $\hil{T}^{(n)}_{\delta}$ is roughly $2^{nS(\rho)}$, which is exponentially smaller than the dimension of $\hil{H}^{\otimes n}$, as $n \rightarrow \infty$.

It follows that the density operator $\rho^{\otimes n}$  can be replaced with a density operator $\tilde{\rho}^{\otimes n}$ with support on the typical subspace (take $\rho^{\otimes n}$ in the spectral representation, where the matrix is diagonal with $2^{n\log d}$ eigenvalues $p(x_{1} \ldots x_{n}) = p(x_{1}) \ldots p(x_{n})$, and replace all $p(x_{1} \ldots x_{n})$ that do not correspond to typical sequences with zeros).

Before considering a compression/decompression scheme for quantum information, we need a measure of the reliability of such a scheme in terms of the fidelity\index{fidelity}, as in the case of classical information. The following definition generalizes the classical notion of fidelity in \S\ref{sec:Shannon} (see  \cite[p. 70]{Jozsa98}): If $\ket{\psi}$ is any pure quantum state and $\rho$ any mixed state, the fidelity\index{fidelity} between $\rho$ and $\ket{\psi}$ is:
\begin{equation}
F(\rho,\ket{\psi}) = \trace{(\rho\ket{\psi}\bra{\psi})} = \bra{\psi}\rho\ket{\psi}
\end{equation}
which is the probability that a measurement of the projection operator $\ket{\psi}\bra{\psi}$ in the state $\rho$ yields the outcome 1, i.e., it is the probability that $\rho$ passes a test of being found to be $\ket{\psi}$ on measurement. Note that for a pure state $\rho = \ket{\psi}\bra{\psi}$, $F(\ket{\phi},\ket{\psi}) = \abs{\braket{\psi}{\phi}}^{2}$. The fidelity\index{fidelity} between two mixed states $\rho$ and $\sigma$ is defined as:\footnote{Note that Nielsen and Chuang \shortcite[p. 409]{NielsenChuang} define the fidelity\index{fidelity} $F(\rho,\sigma)$ as the square root of the quantity defined here. If $\rho$ and $\sigma$ commute, they can be diagonalized in the same basis. The definition then reduces to their definition of the classical fidelity between two probablity distributions defined by the eigenvalues of $\rho$ and $\sigma$ in footnote \ref{foot:NielsenChuang} in \S \ref{sec:classinfocomp}.}
\begin{equation}
F(\rho,\sigma) = \mbox{max}\abs{\braket{\psi}{\phi}}^{2} = (\trace{\sqrt{\rho^{1/2}\sigma\rho^{1/2}}})^{2}.
\end{equation}
for all purifications $\ket{\psi}$ of $\rho$ and $\ket{\phi}$ of $\sigma$. Note that in spite of appearances, $F(\rho,\sigma)$ is symmetric in $\rho$ and $\sigma$. 

In the case of a source of $n$-sequences of quantum states $\ket{\Psi_{i_{1}\ldots i_{n}}} = \ket{\psi_{i_{1}}}\ldots \ket{\psi_{i_{n}}}$ with prior probabilities $p_{i_{1}\ldots i_{n}} = p_{i_{1}}\ldots p_{i_{n}}$, a compression/decompression scheme will in general yield a mixed state 
$\rho_{i_{1}\ldots i_{n}}$. The average fidelity of a compression-decompression scheme for an $n$-sequence of quantum states is defined as:
\begin{equation}
F_{n} = \sum_{\mbox{all }n\mbox{-sequences}}p_{i_{1}\ldots i_{n}}\trace{\rho_{i_{1}\ldots i_{n}}\ket{\Psi_{i_{1}\ldots i_{n}}}\bra{\Psi_{i_{1}\ldots i_{n}}}}
\end{equation}

Schumacher's quantum source coding theorem\index{Schumacher's quantum source coding theorem (noiseless channel coding theorem)} (or quantum noiseless channel coding theorem) for a quantum source that produces quantum states $\ket{\psi_{1}}\ldots \ket{\psi_{n}} \in \hil{H}_{d}$ with probabilities $p_{1}\ldots p_{n}$ (so the density operator corresponding to the output of the source is 
$\rho = \sum p_{i}\ket{\psi_{i}}\bra{\psi_{i}}$), states that 
\begin{quote}
for any $\epsilon, \delta > 0$: (i) there exists a compression/decompression scheme\index{compression/decompression scheme} using $S(\rho) + \delta$ qubits per state for $n$-length sequences produced by the source that can be decompressed by the receiver with fidelity $F_{n} > 1 - \epsilon$, for sufficiently large $n$, and (ii) any compression/decompression scheme using $S(\rho) - \delta$ qubits per state for $n$-length sequences will have a fidelity $F_{n} < \epsilon$, for sufficiently large $n$.
\end{quote}

A compression/decompression scheme\index{compression/decompression scheme} for such a quantum source would go as follows: The transmitter applies a unitary transformation $U$ in $\hil{H}^{\otimes n}_{d}$ (dimension = $d^{n} = 2^{n\log d}$) which maps any state in the typical subspace onto a linear superposition of sequences of $n\log d$ qubits, where all but the first $nS(\rho)$ qubits are in the state $\ket{0}$, and then transmits the first $nS(\rho)$ qubits to the receiver. So the transmitter compresses $n\log d$ qubits to $nS(\rho)$ qubits. The receiver adds $n\log d - nS(\rho)$ qubits in the state $\ket{0}$ and applies the unitary transformation $U^{-1}$. Since the initial $nS(\rho)$ qubits will in general be slightly entangled with the remaining $n\log d - nS(\rho)$ qubits, discarding these qubits amounts to tracing over the associated dimensions, so replacing these qubits with the state $\ket{0}$ will produce a mixed state $\tilde{\rho_{n}}$. The state $U^{-1}\tilde{\rho_{n}}$ will pass a test of being found to be the original state $\ket{\Psi_{i_{1}\ldots i_{n}}}$ with fidelity greater than $1 - \epsilon$.

\section{Entanglement Assisted Quantum Communication\index{quantum!communication}}

In this section I show how entanglement\index{entanglement} can be exploited as a channel for the reliable transmission of quantum information. I discuss two related forms of entanglement assisted communication: quantum teleportation in \S \ref{sec:teleportation} and quantum dense coding in \S \ref{sec:densecoding}.

\subsection{Quantum Teleportation\index{quantum!teleportation}}

\label{sec:teleportation}

As mentioned in \S1, Schr\"{o}dinger\index{Schr\"{o}dinger} introduced the term `entanglement'\index{entanglement} to describe the peculiar
nonlocal correlations of the EPR-state in an  extended two-part
commentary \shortcite{Schr1,Schr2} on the
Einstein-Podolsky-Rosen argument \cite{EPR}. Schr\"{o}\-ding\-er regarded entangled states as problematic because they allow the possibility of what he called `remote steering,'\index{remote steering} which he regarded as a mathematical artefact of the Hilbert space theory and discounted as a physical possibility.
As it turns out, quantum teleportation is an experimentally confirmed application of remote steering between two separated systems. This was first pointed out in a paper by Bennett\index{Bennett}, Brassard\index{Brassard}, Cr\'{e}peau\index{Cr\'{e}peau}, Jozsa\index{Jozsa}, Peres\index{Peres}, and Wootters\index{Wootters} \shortcite{BBC+93} and later experimentally confirmed by several groups using a variety of different techniques \cite{BPM+97,BBM+98,FSB+98,NKL98}.

In the
1935 paper, Schr\"{o}dinger\index{Schr\"{o}dinger} considered pure entangled states with a unique biorthogonal
decomposition, as well as cases like the EPR-state, where
a biorthogonal decomposition is non-unique. He showed that suitable measurements on one system can fix the (pure) state of the
entangled distant system, and that this state depends on what
observable one chooses to measure, not merely on the outcome of that
measurement. In the second paper, he showed that a
`sophisticated experimenter,' by performing a suitable local
measurement on
one system, can `steer'\index{remote steering} the
distant system into any mixture of pure states represented by
its reduced density
operator. So the distant
system can be steered (probabilistically, depending on the outcome of the local measurement) into any pure state in the support of the
reduced density operator, with a nonzero probability that depends
only
on the pure state. For a mixture of linearly
independent states of the distant system, the steering can be done by performing a local standard
 projection-valued measurement in a
suitable basis. If the states are linearly dependent, the
experimenter
performs a generalized measurement (associated with a POVM), which amounts to enlarging the
experimenter's Hilbert space by adding an ancilla, so that the
dimension of the enlarged Hilbert space is equal to the number of
linearly independent states. As indicated in \S\ref{sec:entanglement}, Schr\"{o}dinger's analysis
anticipated the later result by Hughston, Jozsa, and Wootters
 \shortcite{HJW}.

Suppose Alice and Bob,  the traditional protagonists in any two-party communication protocol, each holds one of a pair of qubits in the entangled state:
\begin{equation}
\ket{\Psi} = \frac{1}{\sqrt{2}}(\ket{0}_{A}\ket{1}_{B} - \ket{1}_{A}\ket{0}_{B}) \label{eq:EPR}
\end{equation}
Bob's qubit separately is in the mixed state $\rho_{B} = I/2$, which can be interpreted as an equal weight  mixture of the orthogonal states
$\ket{0}_{B}$, $\ket{1}_{B}$,
or, equivalently, as an infinity of other
mixtures including, to take a specific example,
the equal weight mixture of the four
nonorthogonal normalized states:
\begin{eqnarray*}
\ket{\phi_{1}}_{B} &=& \alpha\ket{0}_{B} + \beta\ket{1}_{B}  \\
\ket{\phi_{2}}_{B} &=& \alpha\ket{0}_{B} - \beta\ket{1}_{B}  \\
\ket{\phi_{3}}_{B} &=& \beta\ket{0}_{B} + \alpha\ket{1}_{B}  \\
\ket{\phi_{4}}_{B} &=& \beta\ket{0}_{B} - \alpha\ket{1}_{B} 
\end{eqnarray*}
That is:
\begin{equation}
\rho_{B} = I/2 = \frac{1}{4}(\ket{\phi_{1}}\bra{\phi_{1}} + \ket{\phi_{2}}\bra{\phi_{2}}  + \ket{\phi_{3}}\bra{\phi_{3}} + \ket{\phi_{4}}\bra{\phi_{4}})
\end{equation}

If Alice measures the observable with eigenstates $\ket{0}_{A}$,
$\ket{1}_{A}$ on her qubit A, and Bob measures the corresponding
observable on his qubit B, Alice's outcomes will be oppositely
correlated with Bob's outcomes (0 with 1, and 1 with 0). If,
instead, Alice prepares an ancilla qubit
$A^{\prime}$ in the state
$\ket{\phi_{1}}_{A^{\prime}} = 
\alpha\ket{0}_{A^{\prime}} + \beta\ket{1}_{A^{\prime}}$ 
and measures an observable on the pair of
qubits $A'+A$ in her possession with eigenstates:
\begin{eqnarray}
\ket{1} &=& (\ket{0}_{A^{\prime}}\ket{1}_{A} -
  \ket{1}_{A^{\prime}}\ket{0}_{A})/\sqrt{2} \\
  \ket{2} &=& (\ket{0}_{A^{\prime}}\ket{1}_{A} +
  \ket{1}_{A^{\prime}}\ket{0}_{A})/\sqrt{2}  \\
  \ket{3} &=& (\ket{0}_{A^{\prime}}\ket{0}_{A} -
  \ket{1}_{A^{\prime}}\ket{1}_{A})/\sqrt{2} \\
  \ket{4} &=& (\ket{0}_{A^{\prime}}\ket{0}_{A} +
  \ket{1}_{A^{\prime}}\ket{1}_{A})/\sqrt{2} 
  \end{eqnarray}
(the Bell states defining the Bell basis in $\hil{H}^{A'}\otimes\hil{H}^{A}$), she will obtain the outcomes 1, 2, 3, 4
with equal probability of 1/4, and these outcomes will be correlated with Bob's
states $\ket{\phi_{1}}_{B}$, $\ket{\phi_{2}}_{B}$,
$\ket{\phi_{3}}_{B}$, $\ket{\phi_{4}}_{B}$. That is, if Bob
checks to see whether his particle is in the state
$\ket{\phi_{i}}_{B}$ when Alice reports that she obtained the outcome
$i$, he will find that this is always in fact the case. This follows
because
\begin{equation}
\ket{\phi_{1}}_{A^{\prime}}\ket{\Psi} =
\frac{1}{2}(-\ket{1}\ket{\phi_{1}}_{B} -
\ket{2}\ket{\phi_{2}}_{B} + \ket{3}\ket{\phi_{3}}_{B} -
\ket{4}\ket{\phi_{4}}_{B})  
\end{equation}
In this sense, Alice can steer Bob's particle into any  mixture
compatible with his density operator $\rho_{B} = I/2$ by an
appropriate local measurement.

What Schr\"{o}dinger\index{Schr\"{o}dinger} found problematic about entanglement\index{entanglement} was the possibility of remote steering in the above sense \shortcite[p. 556]{Schr1}:
\begin{quote}
It is rather discomforting that the theory should allow a system to
be steered or piloted into one or the other type of state at the
experimenter's mercy in spite of his having no access to it.
\end{quote}
Now, remote steering\index{remote steering}  in this probabilistic sense is precisely
what makes quantum teleportation\index{quantum!teleportatioon} possible. Suppose Alice and Bob share
a pair of qubits in the entangled state (\ref{eq:EPR}) and Alice is given a
qubit $A^{\prime}$ in an \textit{unknown} state $\ket{\phi_{1}}$ that she would like to send to Bob. There is no procedure by which Alice can determine the identity of the unknown state, but even if she could, the amount of classical information that Alice would have to send to Bob in order for him to prepare the state $\ket{\phi_{1}}$ is potentially infinite, since the precise specification of a general normalized qubit state
$\alpha\ket{0} + \beta\ket{1}$ requires two real parameters ( the number of independent parameters is reduced from four to two because $\abs{\alpha}^{2} + \abs{\beta}^{2} = 1$ and the overall phase is irrelevant). Alice could send the qubit itself to Bob, but  the quantum information in the qubit state might be corrupted by transmission through a possibly noisy environment. 

Instead, for the cost of just two bits of classical information, Alice can succeed in communicating the unknown quantum state $\ket{\phi_{1}}$ to Bob with perfect reliability.
What Alice does is to measure the 2-qubit system $A'+A$ in her possession
in the Bell basis. Depending on the outcome of her measurement, $i = 1, 2, 3$, or 4 with equal probability, Bob's qubit will be steered into
one of the states $\ket{\phi_{1}}_{B}$, $\ket{\phi_{2}}_{B}$,
$\ket{\phi_{3}}_{B}$, $\ket{\phi_{4}}_{B}$. If
Alice communicates the outcome of her measurement to Bob (requiring the transmission of two bits of classical information), Bob can apply one of four local unitary transformations in his Hilbert space to obtain the state $\ket{\phi_{1}}_{B}$:
\begin{itemize}
\item[] $i=1$: do nothing, i.e., apply the identity transformation $I$ 
\item[] $i=2$: apply the transformation $\sigma_{z}$ 
\item[] $i=3$: apply the transformation $\sigma_{x}$
\item[] $i=4$: apply the transformation $i\sigma_{y}$
\end{itemize}
where $\sigma_{x}, \sigma_{y}, \sigma_{z}$ are the Pauli spin matrices.

The trick that results in the communication of the state $\ket{\phi_{1}}$ from Alice to Bob, without the qubit $A^{\prime}$ literally traveling from Alice to Bob, is the ability afforded Alice by the shared entangled state to correlate one of four measurement outcomes (each occurring with probability 1/4) with one of four states that together represent a particular decomposition of Bob's mixed state. The communication of the state of $A^{\prime}$ is completed by Bob's operation, which requires that Alice sends the two bits of classical information about her measurement outcome to Bob.  In the teleportation protocol, the state of the particle $A^{\prime}$ is destroyed by Alice's measurement and re-created as the state of Bob's particle by Bob's operation---in fact, the systems $A$ and $A^{\prime}$ end up in an entangled state as the result of Alice's measurement. Note that if the state $\ket{\phi_{1}}$ of $A^{\prime}$ were not destroyed there would be two copies of the state, which would violate the quantum `no cloning' theorem. So neither Alice nor Bob, nor any other party, can gain any information about the identity of the teleported state, because the recording of such information in the state of another quantum system would amount to a partial copying of the information in the teleported state.

Shared entanglement\index{entanglement} provides a secure and reliable channel for quantum communication\index{quantum!communication}. This might be useful for the communication of quantum information between parties in a cryptographic protocol, or for the transmission of quantum information between the processing components of a quantum computer. It is a feature of an entangled state shared by two parties that the entanglement is not affected by noise in the environment between them. So the reliability of the communication of  quantum information by teleportation depends on the reliability of the required classical communication, which can be protected against noise by well-known techniques of error-correcting codes. An entangled state shared by two parties is also unaffected by changes in their relative spatial location. So Alice could teleport a quantum state to Bob without even knowing Bob's location, by broadcasting the two bits of information.

\subsection{Quantum Dense Coding\index{quantum!dense coding}}

\label{sec:densecoding}

We know from the Holevo bound\index{Holevo bound} (see \S\ref{sec:accessible}) that the maximum amount of classical information that can be reliably communicated by encoding the information in the quantum state of  a qubit is one bit, even though an arbitrarily large amount of classical information can be encoded in the state of a qubit (by encoding symbols as nonorthogonal quantum states). Quantum dense coding\index{quantum!dense coding} is a procedure, first pointed out by Bennett\index{Bennett} and Wiesner\index{Wiesner} \shortcite{BennettWiesner}, for exploiting entanglement\index{entanglement} to double the amount of classical information that can be communicated by a qubit.

Consider again the Bell states:
\begin{eqnarray}
\ket{1} &=& (\ket{0}\ket{1} -
  \ket{1}\ket{0})/\sqrt{2} \\
  \ket{2} &=& (\ket{0}\ket{1} +
  \ket{1}\ket{0})/\sqrt{2}  \\
  \ket{3} &=& (\ket{0}\ket{0} -
  \ket{1}\ket{1})/\sqrt{2} \\
  \ket{4} &=& (\ket{0}\ket{0} +
  \ket{1}\ket{1})/\sqrt{2} 
  \end{eqnarray}

Suppose Alice and Bob share a pair of qubits in the state
\begin{equation}
\ket{1} =  (\ket{0}_{A}\ket{1}_{B} -
  \ket{1}_{A}\ket{0}_{B})/\sqrt{2}
\end{equation}
By performing one of four \emph{local} operations on the qubit in her possession defined by the unitary transformations in $\hil{H}^{A}$: 
\begin{eqnarray}
U_{1} & = & I \\
U_{2} & = & \sigma_{x} \\
U_{3} & = & \sigma_{x} \\
U_{4} & = & i\sigma_{y}
\end{eqnarray}
Alice can transform the state $\ket{1}$ of the qubit pair into any Bell state. For example: 
\begin{eqnarray}
I\ket{1} & = & \ket{1} \\
\sigma_{z}\ket{1} & = &  \ket{2} \\
\sigma_{x}\ket{1} & = & \ket{3} \\
i\sigma_{y}\ket{1} & = & \ket{4}
\end{eqnarray}

So to communicate two classical bits to Bob, Alice applies one of the four operations above to her qubit and sends the qubit to Bob. Bob then performs a measurement on the two qubits in the Bell basis. Since these are orthogonal states, he can distinguish the states and identify Alice's operation. 

\section{Quantum Cryptography\index{quantum!cryptography}}

Over the past few years, quantum cryptography as emerged as perhaps the most successful
area of application of quantum information theoretic ideas. The main results have been a variety of provably secure 
protocols for key distribution, following an original proposal by Bennett\index{Bennett} and 
Brassard\index{Brassard} \shortcite{BB84}, and an important `no go' theorem\index{quantum!bit commitment!`no go' theorem}\index{bit commitment!`no go' theorem} 
by Mayers\index{Mayers} \shortcite{Mayers2,Mayers3} and Lo\index{Lo} and Chau\index{Chau} \shortcite{LoChau98}: the 
impossibility of unconditionally secure two-party quantum bit commitment\index{quantum!bit commitment}\index{bit commitment}. The 
quantum bit commitment theorem generalizes previous results restricted 
to one-way communication protocols by Mayers \shortcite{Mayers1} 
and by Lo and 
Chau \shortcite{LoChau97} and applies to quantum, classical, and 
quantum-classical hybrid schemes (since classical information, as we have seen, can be regarded as quantum information subject to certain constraints). The 
restriction to two-party schemes excludes schemes that involve a trusted 
third-party or trusted channel properties, and the restriction to 
schemes based solely on the principles of quantum mechanics excludes 
schemes that exploit special relativistic signalling 
constraints, or schemes that might involve time 
machines or the thermodynamics of black holes, etc.

In \S\ref{sec:key}, I show how the security of quantum key distribution depends on features of quantum information---no cloning, no information gain without disturbance, entanglement---that prevent an eavesdropper from secretly gaining information about the quantum communication between two parties, i.e., completely undetectable eavesdropping is in principle impossible for quantum communication. In \S\ref{sec:bit}, I dscuss quantum bit commitment and show why unconditionally secure quantum bit commitment is impossible.

\subsection{Key Distribution\index{key distribution}}

\label{sec:key}

\subsubsection{Quantum Key Distribution Protocols}

\label{sec:protocols}

In a quantum key distribution protocol, the object is for two parties, Alice 
and Bob, who initially share no information, to exchange information 
via quantum and classical channels, so as to end up sharing a secret 
key which they can then use for encryption, 
in such a way as to ensure that any attempt by an 
eavesdropper, Eve, to gain information about the secret key will be 
detected with non-zero probability.

The one-time pad\index{one-time pad} provides a perfectly secure way for Alice and Bob to communicate classical information, but this is also the \emph{only} way that two parties can achieve perfectly security classical communication. The one-time pad is, essentially, a random sequence of bits. If Alice and Bob both have a copy of the one-time pad, Alice can communicate a message to Bob securely by converting the message to an $n$-bit binary number (according to some scheme known to both Alice and Bob), and adding (bitwise, modulo 2) the sequence of bits in the binary number to an $n$-length sequence of bits from the top of the one-time pad. Alice sends the encrypted sequence to Bob, which Bob can then decrypt using the same sequence of bits from his copy of the one-time pad. Since the encrypted message is random, it is impossible for Eve to decrypt the message without a copy of the one-time pad. It is essential to the security of the scheme that the $n$ random bits used to encrypt the message are discarded once the message is transmitted and decrypted, and that a unique random sequence is used for each distinct message---hence the term `one-time pad.' 

This procedure guarantees perfect privacy, so long as Alice and Bob, and only Alice and Bob, can each be assumed to possess a copy of an arbitrarily long one-time pad. But this means that in order for two parties to communicate secretly, they must already share a secret: the random key. The \emph{key distribution problem}\index{key distribution problem} is the problem of how to distribute the key securely in the first place without the key being secretly intercepted during transmission and copied, and the \emph{key storage problem}\index{key storage problem} is the problem of how to store the key securely without it being secretly copied. We would like a procedure that can be guaranteed to be secure against passive eavesdropping, so that Alice and Bob can be confident that their communications are in fact private.

The key idea in \emph{quantum} cryptography\index{quantum!cryptography} is to exploit the indistinguishability\index{distinguishability!states} of nonorthogonal quantum states, which we saw in \S\ref{sec:nocloning} entails that any information gained by Eve about the identity of such states will introduce some disturbance of the states that can be detected by Alice and Bob, and the `no cloning' theorem, which makes it impossible for Eve to copy quantum communications between Alice and Bob and store them for later analysis (perhaps using, in addition, intercepted classical communications between Alice and Bob).

A large variety of quantum key distribution schemes\index{key distribution} have been 
proposed following the original Bennett and 
Brassard protocol~\shortcite{BB84}, now known as BB84\index{BB84}. The core idea there was for Alice to send Bob a sequence of qubits, prepared with equal probability in one of the states $\ket{0}, \ket{1}, \ket{+}, \ket{-}$, where the pair of orthogonal states  $\ket{0}, \ket{1}$ are nonorthogonal to the pair of orthogonal states $\ket{+}, \ket{-}$. Bob measures each qubit randomly in either the basis $\ket{0}, \ket{1}$ or the basis $\ket{+}, \ket{-}$. Following his measurements, he publicly broadcasts the basis he used for each qubit in the sequence, and Alice publicly broadcasts which of these bases is the same as the basis she used to prepare the qubit. Alice and Bob then discard the qubits for which their bases disagree. Since the outcome states of Bob's measurements are the same as the states Alice prepared, Alice and Bob share a random key on the remaining qubits. They can then sacrifice a portion of these qubits to detect eavesdropping. Alice publicly announces the qubit state she prepared and Bob checks his measurement outcome to confirm this. If they agree on a sufficient number of qubit states (depending on the expected error rate), they conclude that there has been no eavesdropping and use the remaining portion as the secret key. If they don't agree, they conclude that the qubits have been disturbed by eavesdropping, in which case they discard all the qubits and begin the procedure again. The actual protocol involves further subtleties in which a perfectly secure secret key is distilled from the `raw key' obtained in this way by techniques of error correction and privacy amplification.

The BB84\index{BB84} scheme solves the key distribution problem, in the sense that Alice and Bob, who initially share no secrets, can end up sharing a secret key via a key distribution protocol that excludes the possibility of eavesdropping, with arbitrarily high reliability(since the length of the sequence of qubits sacrificed to detect eavesdropping can be arbitrarily long).  Clearly, it does not solve the key storage problem, since the output of the key distribution protocol is stored as classical information, which is subject to passive eavesdropping. 

A scheme proposed by Ekert~\shortcite{Ekert}\index{Ekert} allows Alice and Bob to create a shared random key by performing measurements on two entangled qubits. Suppose Alice and Bob share many copies of an entangled pure state of two qubits, say the Bell state $\frac{1}{\sqrt{2}}(\ket{0}\ket{1} - \ket{1}\ket{0})$ (perhaps emitted by a common source of entangled pairs between Alice and Bob). Alice and Bob agree on three observables that they each measure on their qubits, where the measurements are chosen 
randomly and independently for each qubit. After a sequence of 
measurements on an appropriate number of pairs, Alice and Bob 
announce the directions of their measurements publicly and divide the 
measurements into two groups: those in which they measured the spin 
in different directions, and those in which they measured the spin in 
the same direction. They publicly reveal the outcomes of the first group of 
measurements and use these to 
check that the singlet states have not been disturbed by eavesdropping. Essentially, they calculate a correlation 
coefficient: any attempt by an eavesdropper,
Eve, to monitor the particles will disturb the entangled state  and 
result in a correlation coefficient 
that is bounded by Bell's inequality\index{Bell's inequality} and is therefore distinguishable 
from the correlation 
coefficient for the entangled state. If 
Alice and Bob are satisfied that no eavesdropping has occurred, they 
use the second group of oppositely correlated measurement outcomes as 
the key.

\subsubsection{Quantum Key Distribution via Pre- and Post-Selection\index{pre- and post-selected states}}

\label{sec:prepost}

The Ekert scheme solves the key distribution problem\index{key distribution problem} as well as the 
key storage problem\index{key storage problem}, because a new key is generated for each message from the stored entangled states, and there is no information about the key in the entangled states. Here I describe a  key distribution protocol that  also involves entangled states (see \cite{BubKey}), but with a different type of
 test for eavesdropping. Instead of a statistical test 
based on Bell's theorem, the test exploits conditional statements about 
measurement outcomes generated by pre- and 
post-selected quantum states.

The peculiar features of pre- and post-selected\index{pre- and post-selected states} quantum states 
were first pointed out by 
Aharonov\index{Aharonov}, Bergmann\index{Bergmann}, and 
Lebowitz\index{Lebowitz}~\shortcite{ABL}. If:
\begin{enumerate}
\item[(i)] Alice prepares a system in a 
certain state $|\mbox{pre}\rangle$ 
at time $t_{1}$,
\item[(ii)] Bob measures some observable $M$ on the system 
at time $t_{2}$,
\item[(iii)] Alice measures an observable of which 
$|\mbox{post}\rangle$ is an eigenstate at time $t_{3}$, 
and post-selects for $|\mbox{post}\rangle$,
\end{enumerate}
then Alice can assign 
probabilities to the outcomes of Bob's $M$-measurement at $t_{2}$, 
conditional on the states $|\mbox{pre}\rangle$ and $|\mbox{post}\rangle$ at times 
$t_{1}$ and $t_{3}$, respectively, as follows \cite{ABL,VAA}:
\begin{equation}
    \mbox{prob}(q_{k}) =
    \frac{|\langle \mbox{pre}|P_{k}| \mbox{post}\rangle|^{2}}
    {\sum_{i} |\langle \mbox{pre} |P_{i}|\mbox{post}\rangle|^{2}}
    \label{eq:ABL}
\end{equation}
where $P_{i}$ is the projection operator onto the $i$'th eigenspace 
of $M$. Notice that (\ref{eq:ABL})---referred to as the `ABL-rule'\index{ABL-rule} 
(Aharonov-Bergmann-Lebowitz rule) in the following---is 
time-symmetric, in the sense that the states $|\mbox{pre}\rangle$ and 
$|\mbox{post}\rangle$ can be interchanged.

If $M$ is unknown to Alice, she can use the ABL-rule\index{ABL-rule} to assign 
probabilities to the outcomes of various hypothetical 
$M$-measurements. The interesting peculiarity of the ABL-rule, by 
contrast with the usual Born rule for pre-selected states, is that it 
is possible---for an appropriate choice of observables $M$, $M'$, 
\ldots, and states $|\mbox{pre}\rangle$ and $|\mbox{post}\rangle$---to 
assign unit probability to the outcomes of a set of mutually 
\textit{noncommuting} observables. That is, Alice can be in a 
position to assert a conjunction of conditional statements of the 
form: `If Bob measured $M$, then the outcome must have been $m_{i}$, 
with certainty, and if Bob measured $M'$, then the outcome must have been 
$m'_{j}$, with certainty, \ldots,' where $M, M', \ldots$ are mutually 
noncommuting observables. Since Bob could only have measured at most 
one of these noncommuting observables, Alice's conditional information 
does not, of course, contradict quantum mechanics: she only knows the 
eigenvalue $m_{i}$ of an observable $M$ if she knows that Bob in fact 
measured $M$. 

Vaidman\index{Vaidman}, Aharonov\index{Aharonov}, and Albert\index{Albert}~\shortcite{VAA} discuss a case of this sort, 
where the outcome of a 
measurement of any of the three spin observables $X = \sigma_{x}$, 
$Y = \sigma_{y}$, $Z = \sigma_{z}$ of a spin-$\frac{1}{2}$ particle can be 
inferred from an appropriate pre- and post-selection. Alice prepares 
the Bell state
\begin{equation}
    |\mbox{pre}\rangle = 
    \frac{1}{\sqrt{2}}(|\uparrow_{z}\rangle_{A}|\uparrow_{z}\rangle_{C} + 
    |\downarrow_{z}\rangle_{A}|\downarrow_{z}\rangle_{C}
    \label{eq:Bell}
\end{equation}
where $|\uparrow_{z}\rangle$ and $|\downarrow_{z}\rangle$  
denote the $\sigma_{z}$-eigenstates. Alice sends one of the 
particles---the channel particle, denoted by the subscript $C$---to Bob 
and keeps the ancilla, denoted by $A$. Bob measures either 
$X, Y$, or $Z$ on the channel 
particle and returns the channel particle to Alice. Alice then 
measures an observable $R$ on the pair of particles, where $R$ has 
the eigenstates (the subscripts $A$ and $C$ are suppressed):
\begin{eqnarray}
    |r_{1}\rangle & = & 
    \frac{1}{\sqrt{2}}|\uparrow_{z}\rangle|\uparrow_{z}\rangle + 
    \frac{1}{2}(|\uparrow_{z}\rangle|\downarrow_{z}\rangle 
    e^{i\pi/4} + |\downarrow_{z}\rangle|\uparrow_{z}\rangle 
    e^{-i\pi/4}) \\
    |r_{2}\rangle & = & 
    \frac{1}{\sqrt{2}}|\uparrow_{z}\rangle|\uparrow_{z}\rangle - 
    \frac{1}{2}(|\uparrow_{z}\rangle|\downarrow_{z}\rangle 
    e^{i\pi/4} + |\downarrow_{z}\rangle|\uparrow_{z}\rangle 
    e^{-i\pi/4}) \\
    |r_{3}\rangle & = & 
    \frac{1}{\sqrt{2}}|\downarrow_{z}\rangle|\downarrow_{z}\rangle + 
    \frac{1}{2}(|\uparrow_{z}\rangle|\downarrow_{z}\rangle 
    e^{-i\pi/4} + |\downarrow_{z}\rangle|\uparrow_{z}\rangle 
    e^{i\pi/4}) \\
    |r_{4}\rangle & = & 
    \frac{1}{\sqrt{2}}|\downarrow_{z}\rangle|\downarrow_{z}\rangle - 
    \frac{1}{2}(|\uparrow_{z}\rangle|\downarrow_{z}\rangle 
    e^{-i\pi/4} + |\downarrow_{z}\rangle|\uparrow_{z}\rangle 
    e^{i\pi/4})
\end{eqnarray}

Note that:
\begin{eqnarray}
    |\mbox{pre}\rangle & = & 
    \frac{1}{\sqrt{2}}(|\uparrow_{z}\rangle|\uparrow_{z}\rangle + 
    |\downarrow_{z}\rangle|\downarrow_{z}\rangle \\ \label{eq:R1}
                       & = & 
    \frac{1}{\sqrt{2}}(|\uparrow_{x}\rangle|\uparrow_{x}\rangle + 
    |\downarrow_{x}\rangle|\downarrow_{x}\rangle \\
                       & = & 
    \frac{1}{\sqrt{2}}(|\uparrow_{y}\rangle|\downarrow_{y}\rangle + 
    |\downarrow_{y}\rangle|\uparrow_{y}\rangle \\
                       & = & 
    \frac{1}{2}(|r_{1}\rangle + |r_{2}\rangle + |r_{3}\rangle + 
    |r_{4}\rangle) \label{eq:R4}
\end{eqnarray}
    
Alice can now assign values to the outcomes of Bob's spin measurements  
via the ABL-rule,
whether Bob measured $X, Y$, or $Z$,
based on the post-selections $|r_{1}\rangle$, $|r_{2}\rangle$, 
$|r_{3}\rangle$, or $|r_{4}\rangle$, according to Table~\ref{table:xyz}  
(where 0 represents the outcome $\uparrow$ and 1 represents the 
outcome $\downarrow$) \cite{VAA}:
\begin{table}[ht]
    \begin{center}
  $
    \begin{array}{r|ccc}
	      & \sigma_{x} & \sigma_{y} & \sigma_{z} \\ \hline
        r_{1} & 0 & 0 & 0 \\ 
        r_{2} & 1 & 1 & 0 \\ 
        r_{3} & 0 & 1 & 1 \\ 
        r_{4} & 1 & 0 & 1 
     \end{array}
  $
  \end{center}
    \caption{\protect $\sigma_{x}$, \protect $\sigma_{y}$, \protect 
    $\sigma_{z}$ measurement outcomes correlated with eigenvalues of R}
    \label{table:xyz}
\end{table}

This case can be exploited to enable Alice and Bob to share a private 
random key in the following way: Alice prepares a certain number of 
copies (depending on the length of the key and the level of privacy 
desired) of the Bell state $\ket{\mbox{pre}}$ in Eq. (\ref{eq:Bell}).
She sends the channel particles to Bob in sequence and keeps the 
ancillas. Bob measures $X$ or $Z$ randomly on the 
channel particles and returns the particles, in sequence, to Alice. 
Alice then measures the observable $R$ on the ancilla and channel 
pairs and divides the sequence into two subsequences: the 
subsequence $S_{14}$ 
for which she obtained the outcomes $r_{1}$ or $r_{4}$, and the 
subsequence $S_{23}$ for which she obtained the outcomes $r_{2}$ or 
$r_{3}$. The sequence of operations can be implemented on a 
quantum circuit; see \cite{Metzger}. 

To check that the channel particles have not been monitored by Eve, 
Alice now publicly 
announces (broadcasts) the indices of the subsequence $S_{23}$. As is 
evident from Table~\ref{table:xyz}, 
for this subsequence she can make conditional 
statements of the form: `For channel particle $i$, if $X$ 
was measured, the outcome was 1 (0), and if $Z$ 
was measured, the outcome was 0 (1),' depending on whether the outcome of her 
$R$-measurement was $r_{2}$ or $r_{3}$. She publicly announces these 
statements as well. If one of these statements, for some index $i$, 
does not agree with Bob's records, Eve must have monitored the $i$'th 
channel particle. (Of course, agreement does not entail that the 
particle was \textit{not} monitored.)

For suppose Eve 
measures a different spin component observable than Bob on a channel particle 
and Alice subsequently obtains one of the eigenvalues $r_{2}$ or $r_{3}$ 
when she measures $R$. 
Bob's measurement outcome, either 0 or 1, 
will be compatible with just one of these eigenvalues, assuming no 
intervention by Eve. 
But after Eve's measurement, both of these eigenvalues will 
be 
possible outcomes of Alice's measurement. So Alice's retrodictions of 
Bob's measurement outcomes for the subsequence $S_{23}$ 
will not necessarily correspond to Bob's 
records. In fact, one can show that if Eve measures $X$ 
or $Z$ randomly on the channel particles, or if she measures 
a particular one of the observables $X$, $Y$, or 
$Z$ on the channel particles (the same observable on each 
particle), the probability of detection in the subsequence $S_{23}$ 
is 3/8.   

In the subsequence $S_{14}$, the 0 and 1 outcomes of Bob's measurements 
correspond to the outcomes $r_{1}$ and $r_{4}$ of Alice's 
$R$-measurements. If, following their public communication about the 
subsequence $S_{23}$, Alice and Bob agree that there has been no 
monitoring of the 
channel particles by Eve, they use the subsequence $S_{14}$ to define 
a shared raw key.

Note that even a single disagreement between Alice's 
retrodictions and Bob's records is sufficient to reveal that the channel 
particles have been monitored by Eve. This differs from the 
eavesdropping test in the Ekert protocol. 
Note also that Eve only has access to the channel particles, not 
the particle pairs. So no strategy is possible in which Eve replaces all the 
channel particles with her own particles and 
entangles the original channel particles, treated as a single system, 
with an ancilla by some unitary transformation, and then delays any
measurements until after Alice and Bob have communicated publicly. 
There is no way that Eve can ensure agreement between Alice and Bob 
without having access to the particle pairs, or without information 
about Bob's measurements.

The key distribution protocol as outlined above solves the key 
distribution problem\index{key distribution problem} but not the key storage problem\index{key storage problem}. If Bob actually 
makes the random choices, measures $X$ or $Z$, and 
records definite outcomes for the spin measurements 
before Alice measures $R$, as required by the protocol, Bob's 
measurement records---stored as classical information---could in 
principle be 
copied by Eve without detection. In that case, 
Eve would know the raw key (which is 
contained in this information), following the public communication
between Alice and 
Bob to verify the integrity of the quantum communication channel. 

To solve the key storage problem\index{key storage problem}, the protocol is modified in the 
following way: Instead of actually making the random choice for each 
channel particle,  
measuring one of the spin observables, and recording the outcome of the 
measurement, Bob keeps the random choices and the spin 
measurements `at the quantum level' until after Alice announces the 
indices of the subsequence $S_{23}$ of her $R$ measurements. To do 
this, Bob enlarges the Hilbert space by entangling the 
quantum state of the channel particle via a unitary transformation with 
the states of two ancilla 
particles that he introduces. One particle is associated with a 
Hilbert space spanned by two 
eigenstates, $|d_{X}\rangle$ and $|d_{Z}\rangle$, of a choice
observable or  `quantum die' observable $D$. The 
other particle is associated with a 
Hilbert space spanned by two eigenstates, $|p_{\uparrow}\rangle$ and 
$|p_{\downarrow}\rangle$, of a pointer 
observable $P$. (See \S\ref{sec:bittheoremidea} for a discussion of how to implement the 
unitary transformation on the enlarged Hilbert space.) 

On the modified protocol (assuming the ability 
to store entangled states indefinitely), Alice and Bob share a large 
number of copies of an entangled 4-particle state. When they wish to establish a 
random key of a certain length, Alice measures $R$ on an appropriate 
number of particle pairs in her possession and announces the indices of the 
subsequence $S_{23}$. Before Alice announces the indices of the 
subsequence $S_{23}$, neither Alice nor Bob have stored any classical information. So there is nothing for Eve to copy.
After Alice announces the 
indices of the subsequence $S_{23}$, Bob measures the observables $D$ 
and $P$ on his ancillas
with these indices and announces 
the eigenvalue $|p_{\uparrow}\rangle$ or $|p_{\downarrow}\rangle$ as the
outcome of his $X$ or $Z$ measurement, depending on 
the eigenvalue of $D$. If Alice and 
Bob decide that there has been no eavesdropping by Eve, Bob measures 
$C$ and $P$ on his ancillas in the 
subsequence $S_{14}$. It is easy 
to see that the ABL-rule applies in this case, just as it applies 
in the case where Bob actually makes the random choice and actually 
records definite outcomes of his $X$ or $Z$ 
measurements before Alice measures $R$. In 
fact, if the two cases were not equivalent for Alice---if Alice could 
tell from her $R$-measurements 
whether Bob had actually made the random choice and actually performed 
the spin measurements, or had merely implemented these actions 
`at the quantum level'---the difference could be 
exploited to signal superluminally.

\subsection{Bit Commitment\index{bit commitment}\index{quantum!bit commitment}}

\label{sec:bit}

\subsubsection{Some History}

\label{sec:bithistory}

In a bit commitment protocol, one party, Alice, supplies an encrypted bit 
to a second party, Bob. The information available in the encrypted bit
should be insufficient for Bob to ascertain the value of the bit, but 
sufficient, together with further information supplied by Alice at a 
subsequent stage when she is supposed to reveal the 
value of the bit, for Bob to be convinced that the protocol does not 
allow Alice to cheat by encrypting the bit in a way that leaves her free 
to reveal either 0 or 1 at will. 

To illustrate the idea, suppose Alice 
claims the ability to predict the outcomes of elections. To substantiate her claim without 
revealing valuable information (perhaps to a potential 
employer, Bob) she suggests the following demonstration: She proposes 
to record her prediction about whether a certain candidate will win or lose 
by writing a 0 (for `lose') or a 1 (for 
`win') on a note a month before the election. She will then lock the note in a safe and hand the safe to Bob, but keep the key. 
After the election, she will announce the bit she chose and prove  
that she in fact made the commitment at the earlier time by handing 
Bob the key. Bob can then open the safe and read the note. 

Obviously, the security of this procedure depends on the strength of the safe walls or the ingenuity of the locksmith. More generally, 
 Alice can send (encrypted) information to Bob that guarantees the
 truth of an exclusive classical disjunction (equivalent
to her commitment to a 0 or a 1) only if the information is biased
towards one of the alternative disjuncts (because a classical exclusive
disjunction is true
if and only if one of the disjuncts is true and the other false). No
principle of classical mechanics precludes Bob from extracting this
information,
so the security of a classical bit commitment protocol can only be a matter of
computational complexity.

The question is whether there exists a quantum analogue 
of this procedure that is \emph{unconditionally secure}\index{bit commitment!unconditionally secure}\index{quantum!bit commitment!unconditionally secure}: 
provably secure as a matter of physical law (according to quantum theory) against cheating by either Alice or 
Bob. Note that Bob can cheat if he can obtain \textit{some} 
information about Alice's commitment before she reveals it 
(which would give him an advantage in repetitions of the protocol with 
Alice). Alice can cheat if she can delay actually making a commitment 
until the final stage when she is required to reveal her commitment, 
or if she can change her commitment at the final stage with a very low 
probability of detection.

Bennett\index{Bennett} and Brassard\index{Brassard} originally proposed a quantum bit 
commitment protocol in \shortcite{BB84}. The basic idea was to associate the 0 and 1 
commitments with two different mixtures 
represented by the same density operator. As they showed in the 
same paper, Alice can cheat by adopting an `EPR attack' or cheating strategy\index{EPR! cheating strategy (attack)}: she prepares entangled pairs of 
qubits, keeps one of each pair (the ancilla) and sends the second 
qubit (the channel particle) to Bob. In this way she can fake 
sending one of two equivalent mixtures to Bob 
and reveal either bit at will at the opening stage by 
effectively steering Bob's particle into
the desired mixture by an appropriate measurement. Bob 
cannot detect this cheating strategy. 

In a later paper, 
Brassard, Cr\'{e}peau, Josza, and Langlois \shortcite{BCJL} proposed a quantum bit commitment 
protocol that they 
claimed to be unconditionally secure. The BCJL scheme was first shown 
to be insecure by Mayers\index{Mayers}~\shortcite{Mayers1}. Subsequently, 
Mayers~\shortcite{Mayers2,Mayers3} and Lo\index{Lo} and 
Chau\index{Chau}~\shortcite{LoChau97,LoChau98} independently showed that  
the insight of 
Bennett and Brassard
in~\shortcite{BB84} can be extended to a proof that a generalized version of 
the EPR cheating strategy\index{EPR! cheating strategy (attack)} can always be applied, if the 
Hilbert space is enlarged in a suitable way by introducing additional 
ancilla particles. 

The impossibility of unconditionally  secure quantum bit commitment came as something of a surprise to the community of quantum cryptologists and has profound consequences. Indeed, it would not 
be an exaggeration to say that the significance of the quantum bit commitment 
theorem for our understanding of quantum mechanics is comparable to Bell's theorem \cite{BellEPR}. Brassard\index{Brassard} and Fuchs\index{Fuchs} have 
speculated (\cite{Brassard,Fuchs1,Fuchs2,FuchsJacobs}) 
that quantum mechanics can be derived from two 
postulates about quantum information: the possibility of secure 
key distribution and the impossibility of secure bit commitment. 
We shall see in \S\ref{sec:foundations} what this means for the foundations of quantum mechanics.

Perhaps because of the simplicity of the proof and the 
universality of the claim, the quantum bit commitment theorem is 
continually challenged in the literature, 
on the basis that the proof does not cover all possible procedures 
that might be exploited to implement quantum bit commitment (see, e.g., Yuen
\shortcite{Yuen2005}). 
There seems to be a 
general feeling that the theorem is `too good to be true' and that 
there must be a loophole. 

In fact, there is no loophole. While Kent\index{Kent} \shortcite{Kent1,Kent2} has shown 
how to implement a secure classical 
bit commitment protocol by exploiting relativistic signalling 
constraints in a timed sequence of communications between verifiably 
separated sites for both Alice and Bob, 
and Hardy and Kent~\shortcite{HardyKent} and Aharonov, Ta-Shma, Vazirani, 
and Yao~\shortcite{ATVY} have investigated the security of 
`cheat-sensitive' or `weak' versions of quantum bit commitment, 
these results are not in conflict with the quantum bit commitment theorem. 
In a bit commitment protocol as usually understood, there is a time interval 
of arbitrary 
length, where no information is exchanged, 
between the end of the commitment stage of the protocol and 
the opening or unveiling stage, when Alice reveals the value of the 
bit. Kent's\index{Kent} ingenious scheme effectively involves a third stage between the 
commitment state and the unveiling stage, in which information is 
exchanged between Bob's sites and Alice's sites at regular intervals 
until one of Alice's sites 
chooses to unveil the originally committed bit. At this moment of 
unveiling the protocol is not yet complete, because a further sequence of 
unveilings is required between Alice's sites and corresponding sites 
of Bob before Bob has all the information required to verify the 
commitment at a single site. If a bit commitment protocol 
is understood to 
require an arbitrary amount of free time between the end of the 
commitment stage and the opening stage (in which no step is to be 
executed in the protocol), then the quantum bit commitment theorem 
covers protocols that exploit special relativistic signalling 
constraints.\footnote{I am indebted to Dominic Mayers for clarifying this 
point.}

\subsubsection{A Key Observation}

\label{sec:bittheoremidea}

The crucial insight underlying the proof of the quantum bit commitment theorem is that any 
step in a quantum bit commitment protocol that requires Alice or Bob to make a 
definite choice (whether to perform one of a number of alternative 
measurements, or whether to implement one of a number of alternative 
unitary transformations) can always be replaced by an EPR cheating 
strategy in the generalized sense, assuming that Alice and Bob are 
both equipped with quantum computers. That is, a classical 
disjunction over definite possibilities---this operation 
\textit{or} that operation---can always be replaced by a quantum 
entanglement and a subsequent measurement (perhaps at a more 
convenient time for the cheater) in which one of the possibilities 
becomes definite. Essentially, the classical disjunction is 
replaced by a quantum disjunction. This cheating strategy cannot be 
detected. Similarly, a measurement can be `held at the quantum level' 
without detection: instead of performing the measurement and obtaining 
a definite outcome as one of a number of possible outcomes, a 
suitable unitary transformation can be performed on an enlarged 
Hilbert space, in which the system is entangled with a `pointer' 
ancilla in an appropriate way, and the procedure of obtaining a 
definite outcome can 
be delayed. The key point is the possibility of keeping the series of transactions 
between Alice and Bob at the quantum level by enlarging the 
Hilbert space, until the final exchange of classical information when 
Alice reveals her commitment.

Any quantum bit commitment scheme will involve a series of transactions 
between Alice and Bob, where a certain number, $n$, of quantum 
systems---the `channel particles'---are passed between them and
subjected to various quantum operations (unitary transformations, 
measurements, etc.), possibly chosen randomly. These 
operations can always be replaced, without detection, 
by entangling a channel particle with one or more ancilla particles 
that function as `pointer' particles for measurements or `die' 
particles for random choices. In effect, this is the (generalized) EPR cheating strategy\index{EPR! cheating strategy (attack)}.

To illustrate: Suppose, at a certain stage of a quantum bit commitment protocol, 
that Bob is required to make a random choice between 
measuring one of two observables, $X$ or $Y$, on each channel particle he 
receives from Alice. For simplicity, assume that $X$ and $Y$ each have two 
eigenvalues, $x_{1}$, $x_{2}$ and $y_{1}$, $y_{2}$.
After recording the outcome of the measurement, 
Bob is required to return the channel particle to Alice. When Alice 
receives the $i$'th channel particle she sends Bob the next channel particle 
in the sequence. We may suppose that the measurement outcomes that Bob 
records form part of the information that enables him to confirm 
Alice's commitment, once she discloses it (together with further 
information), so he is not required to report his measurement outcomes 
to Alice until the final stage of the protocol when she reveals her commitment.

Instead of following the protocol, Bob can construct a device that 
entangles the input state $|\psi\rangle_{C}$ of a 
channel particle with the initial 
states, $|d_{0}\rangle_{B}$ and $|p_{0}\rangle_{B}$, of two ancilla 
particles that he introduces, 
the first of which functions as a `quantum die' for the 
random choice and the 
second as a `quantum pointer' for the measurement. It is assumed that 
Bob's ability to construct such a device---in effect, a special purpose quantum 
computer---is restricted only by the laws of quantum mechanics.

The entanglement  
is implemented by a unitary transformation in the following way:\footnote{Note 
that there is 
no loss of generality in assuming that the channel particle is in a 
pure state. If the channel particle is entangled with Alice's 
ancillas, the device implements the entanglement via the 
transformation $I\otimes \cdots$, 
where $I$ is the identity operator in the Hilbert space of Alice's 
ancillas.} 
Define two unitary transformations, $U_{X}$ and $U_{Y}$, that 
implement the $X$ and $Y$ measurements `at the quantum level' on the 
tensor product of the Hilbert space of the channel particle, 
$\hil{H}_{C}$, and the Hilbert space of 
Bob's pointer ancilla, $\hil{H}_{B_{P}}$: 
\begin{eqnarray}
   \ket{x_{1}}_{C}\ket{p_{0}}_{B} \stackrel{U_{X}}{\longrightarrow} 
    \ket{x_{1}}_{C}\ket{p_{1}}_{B}
    \nonumber \\
    \ket{x_{2}}_{C}\ket{p_{0}}_{B} \stackrel{U_{X}}{\longrightarrow} 
    \ket{x_{2}}_{C}\ket{p_{2}}_{B} \label{eq:bc1}
\end{eqnarray}
and
\begin{eqnarray}
    \ket{y_{1}}_{C}\ket{p_{0}}_{B} \stackrel{U_{Y}}{\longrightarrow} 
    \ket{y_{1}}_{C}\ket{p_{1}}_{B}
    \nonumber \\
    \ket{y_{2}}_{C}|\ket{p_{0}}_{B} \stackrel{U_{Y}}{\longrightarrow} 
    \ket{y_{2}}_{C}\ket{p_{2}}_{B}
\end{eqnarray}
so that 
\begin{equation}
    \ket{\psi}_{C}\ket{p_{0}}_{B} 
    \stackrel{U_{X}}{\longrightarrow}
   \braket{x_{1}}{\psi}\ket{x_{1}}_{C} \ket{p_{1}}_{B}
    + \braket{x_{2}}{\psi}\ket{x_{2}}_{C} \ket{p_{2}}_{B}
\end{equation}
and
\begin{equation}
    \ket{\psi}_{C}\ket{p_{0}}_{B} 
    \stackrel{U_{Y}}{\longrightarrow}
    \braket{y_{1}}{\psi}\ket{y_{1}}_{C}\ket{p_{1}}_{B}
    + \braket{y_{2}}{\psi}\ket{y_{2}}_{C}\ket{p_{2}}_{B}
\end{equation}

The random choice is defined similarly by a unitary transformation 
$V$ on 
the tensor product of the Hilbert space of Bob's die ancilla, 
$\hil{H}_{B_{D}}$, and the Hilbert space 
$\hil{H}_{C}\otimes\hil{H}_{B_{P}}$. Suppose $\ket{d_{X}}$ 
and $\ket{d_{Y}}$ are two orthogonal states in $\hil{H}_{B_{D}}$ 
and that $\ket{d_{0}} = \frac{1}{\sqrt{2}}\ket{d_{X}} + 
\frac{1}{\sqrt{2}}\ket{d_{Y}}$. Then (suppressing 
the obvious subscripts) $V$ is defined by:
\begin{eqnarray}
    \ket{d_{X}}\otimes \ket{\psi}\ket{p_{0}} &
    \stackrel{V}{\longrightarrow} & \ket{d_{X}} \otimes 
    U_{X}\ket{\psi}\ket{p_{0}} \nonumber \\
    \ket{d_{Y}}\otimes \ket{\psi}\ket{p_{0}} &
    \stackrel{V}{\longrightarrow} & \ket{d_{Y}} \otimes 
    U_{Y}\ket{\psi}\ket{p_{0}}
\end{eqnarray}
so that
\begin{eqnarray}
    \lefteqn{\ket{d_{0}} \otimes\ket{\psi}\ket{p_{0}}
    \stackrel{V}{\longrightarrow}} \nonumber \\
    & & \frac{1}{\sqrt{2}}\ket{d_{X}} \otimes 
    U_{X}\ket{\psi}\ket{p_{0}}
      + \frac{1}{\sqrt{2}}\ket{d_{Y}} \otimes
     U_{Y}\ket{\psi}\ket{p_{0}}
     \label{eq:V}
\end{eqnarray}
where the tensor product symbol has been introduced selectively to indicate 
that $U_{x}$ and $U_{y}$ are defined on 
$\hil{H}_{C}\otimes\hil{H}_{B_{P}}$.

If Bob were to actually 
choose the observable $X$ or $Y$ randomly, and actually perform the 
measurement and obtain a particular eigenvalue, Alice's density operator for 
the channel particle would be:
\begin{eqnarray}
    \lefteqn{\frac{1}{2}(\mid\braket{x_{1}}{\psi}\mid^{2}\ket{x_{1}}\bra{x_{1}}
    + \mid\braket{x_{2}}{\psi}\mid^{2}\ket{x_{2}}\bra{x_{2}})} 
    \nonumber \\
    & & +\frac{1}{2}(\mid\braket{y_{1}}{\psi}\mid^{2}\ket{y_{1}}\bra{y_{1}}
    + \mid\braket{y_{2}}{\psi}\mid^{2}\ket{y_{2}}\bra{y_{2}})
    \label{eq:density1}
\end{eqnarray}
assuming that Alice does not know what observable Bob chose to 
measure, nor what outcome he obtained.
But this is precisely the same density operator generated by tracing 
over Bob's ancilla particles for the state produced in (\ref{eq:V}). 
In other words, the density operator for 
the channel particle is the same for Alice, whether Bob 
randomly chooses which observable to measure and actually performs 
the measurement, or whether he implements an EPR cheating strategy 
\index{EPR! cheating strategy (attack)}with his two ancillas that produces the transition (\ref{eq:V}) on the 
enlarged Hilbert space. 

If Bob is required to eventually report what measurement he performed 
and what outcome he obtained, he can at that stage measure the die 
ancilla for the eigenstate $\ket{d_{X}}$ or $\ket{d_{Y}}$, and then 
measure the pointer ancilla for the eigenstate $\ket{p_{1}}$ or 
$\ket{p_{2}}$. In effect, if we consider the ensemble of possible 
outcomes for the two measurements, Bob will have converted the 
`improper' mixture generated by tracing over his ancillas to a `proper' 
mixture. But the difference between a proper and improper mixture 
is undetectable by Alice since she has no access to Bob's ancillas, 
and it is only by measuring the composite system consisting of the 
channel particle together with Bob's ancillas that Alice could 
ascertain that the channel particle is entangled with the ancillas. 

In fact, if it were possible to distinguish between a proper and 
improper mixture, it would be possible to signal superluminally: 
Alice could know instantaneously whether or not Bob performed a 
measurement on his ancillas by monitoring the channel particles in her 
possession. Note that it makes no difference whether Bob or Alice 
measures first, since the measurements are of observables in 
different Hilbert spaces, which therefore commute. 

Clearly, a similar argument applies if Bob is required to choose between 
alternative unitary operations at some stage of a bit commitment 
protocol. Perhaps less obviously, an EPR cheating strategy\index{EPR! cheating strategy (attack)} is also possible if 
Bob is required to perform a measurement or choose between alternative operations
on channel particle $i+1$, 
conditional on the outcome 
of a prior measurement on channel particle $i$, or conditional on a 
prior choice of some operation from among a set of alternative 
operations. Of course, if Bob is in possession of all the channel particles 
at the same time, he can perform an entanglement with ancillas on the entire 
sequence, considered as a single composite system. But even if Bob 
only has access to one channel particle at a time (which he is 
required to return to Alice after performing a measurement or other operation 
before 
she sends him the next channel particle), he can always entangle 
channel particle $i+1$ with the 
ancillas he used to entangle channel 
particle $i$. 

For example, suppose Bob is presented with two channel particles in 
sequence. He is supposed to decide randomly whether to measure $X$ or 
$Y$ on the first particle, perform the measurement, and return the 
particle to Alice. After Alice receives the first particle, she sends 
Bob the second particle. If Bob measured $X$ on the first particle and 
obtained the outcome $x_{1}$, he 
is supposed to measure $X$ on the second particle; if he obtained 
the outcome $x_{2}$, he is supposed to measure $Y$ on the second 
particle. If he measured $Y$ on the first particle and obtained the 
outcome $y_{1}$, he is supposed to apply the unitary transformation 
$U_{1}$ to the second particle; if he obtained the outcome $y_{2}$, he 
is supposed to apply the unitary transformation $U_{2}$. After 
performing the required operation, he is supposed to return the 
second particle to Alice. 

It would seem at first sight that Bob has to 
actually perform a measurement on the first channel particle and obtain a 
particular outcome before he can apply the protocol to the second 
particle, given that he only has access to one channel particle at a time, 
so an EPR cheating strategy\index{EPR! cheating strategy (attack)} is excluded. But this is not so. Bob's strategy is 
the following: He applies the EPR strategy discussed above for two 
alternative measurements to the first 
channel particle. 
For the second channel particle, he applies the following unitary 
transformation on the tensor product of the Hilbert spaces of his 
ancillas and the channel particle, where the state of the second channel 
particle is denoted by $|\phi\rangle$, 
and the state of the pointer ancilla for the second channel particle 
is denoted by $|q_{0}\rangle$ (a second die particle is not required):

\begin{eqnarray}
    \ket{d_{X}}\ket{p_{1}}\ket{\phi}\ket{q_{0}} 
    \stackrel{U_{C}}{\longrightarrow}
    \ket{d_{X}}\ket{p_{1}} \otimes U_{X} \ket{\phi}\ket{q_{0}}
    \nonumber \\
    \ket{d_{X}}\ket{p_{2}}\ket{\phi}\ket{q_{0}} 
    \stackrel{U_{C}}{\longrightarrow}
    \ket{d_{X}}\ket{p_{2}} \otimes U_{Y} \ket{\phi}\ket{q_{0}}    \nonumber \\
   \ket{d_{Y}}\ket{p_{1}}\ket{\phi}\ket{q_{0}} 
    \stackrel{U_{C}}{\longrightarrow}
    \ket{d_{Y}}\ket{p_{1}} \otimes U_{1} \ket{\phi}\ket{q_{0}}
    \nonumber \\
   \ket{d_{Y}}\ket{p_{2}}\ket{\phi}\ket{q_{0}} 
    \stackrel{U_{C}}{\longrightarrow}
    \ket{d_{Y}}\ket{p_{2}} \otimes U_{2} \ket{\phi}\ket{q_{0}}
\end{eqnarray}

\subsubsection{Proof of the Quantum Bit Commitment Theorem\index{bit commitment!`no go' theorem}\index{quantum!bit commitment!`no go' theorem}}

\label{sec:bittheoremproof}

Since an EPR cheating strategy\index{EPR! cheating strategy (attack)} can always be applied without detection, the 
proof of the quantum bit commitment theorem assumes that at the end of the 
commitment stage the composite system consisting of Alice's ancillas, 
the $n$ channel particles, and Bob's
ancillas will be represented by some composite entangled 
state $\ket{0}$ or $\ket{1}$, depending on Alice's 
commitment,\footnote{More precisely, depending on whether Alice intends to reveal 0 or 1---since we are assuming that Alice will apply an EPR cheating strategy whenever this is relevant.} on a Hilbert space 
$\hil{H}_{A}\otimes \hil{H}_{B}$, where 
$\hil{H}_{A}$ is the Hilbert space of the particles in Alice's 
possession at that stage (Alice's ancillas and the channel particles 
retained by Alice, if any), and 
$\hil{H}_{B}$ is the Hilbert space of the particles in Bob's 
possession at that stage (Bob's ancillas and the channel particles 
retained by Bob, if any). 

Now, the density operators $W_{B}(0)$ and 
$W_{B}(1)$, characterizing the information available to 
Bob for the two alternative commitments,
are obtained by tracing the states $|0\rangle$ and $|1\rangle$ 
over $\mathcal{H}_{A}$. If these density operators are the same, 
then Bob will be unable to distinguish the 
0-state from the 1-state without further information from 
Alice. In this case, the protocol is said to be `concealing.'\index{bit commitment!concealing protocol} What the proof 
establishes, by an application of the biorthogonal decomposition 
theorem, is that if $W_{B}(0) = W_{B}(1)$ then there exists a unitary 
transformation in $\mathcal{H}_{A}$ that will transform $|0\rangle$ 
to $|1\rangle$. That is, if the protocol is `concealing' then 
it cannot be `binding'\index{bit commitment!binding protocol} on Alice: she can always follow the protocol (with appropriate substitutions of an EPR  
strategy) to establish the state $|0\rangle$. At the 
final stage when she is required to reveal her commitment, she can 
choose to reveal the alternative commitment, depending on circumstances, by applying a 
suitable unitary transformation in her own Hilbert space to transform 
$|0\rangle$ to $|1\rangle$ without Bob being able to 
detect this move. So 
either Bob can cheat by obtaining some information about Alice's 
choice before she reveals her commitment, or Alice can cheat. 

The essentials of the proof can be sketched as follows: In the 
Schmidt decomposition, the states $\ket{0}$ and 
$\ket{1}$ can be expressed as:
\begin{eqnarray}
    \ket{0} & = & \sum_{i}\sqrt{p_{i}}\ket{a_{i}}\ket{b_{i}}
    \nonumber \\
    \ket{1} & = & \sum_{j}\sqrt{p'_{j}}\ket{a'_{j}}\ket{b'_{j}}
\end{eqnarray}
where $\{\ket{a_{i}}\}, \{\ket{a'_{j}}\}$ are two 
orthonormal sets of states in $\hil{H}_{A}$, and 
$\{\ket{b_{i}}\}, \{\ket{b'_{j}}\}$ are two orthonormal 
sets in $\hil{H}_{B}$.

The density operators $W_{B}(0)$ and $W_{B}(1)$ are defined by:
\begin{eqnarray}
    W_{B}(0) = Tr_{A}\ket{0}\bra{0} & = & 
    \sum_{i}p_{i}\ket{b_{i}}\bra{b_{i}}
    \nonumber \\
     W_{B}(1) = Tr_{A}\ket{1}\bra{1} & = & 
    \sum_{j}p'_{j}\ket{b'_{j}}\bra{b'_{j}}
\end{eqnarray}

Bob can't cheat if and only if $W_{B}(0) = W_{B}(1)$. Now, by the spectral 
theorem, the decompositions:
\begin{eqnarray*}
    W_{B}(0) & = & \sum_{i}p_{i}\ket{b_{i}}\bra{b_{i}}
    \nonumber \\
    W_{B}(1) & = & \sum_{j}p'_{j}\ket{b'_{j}}\bra{b'_{j}}
\end{eqnarray*}
are unique for the nondegenerate case, where the $p_{i}$ are all 
distinct and the $p'_{j}$ are all distinct.  The condition $W_{B}(0) = 
W_{B}(1)$ implies that for all $k$:
\begin{eqnarray}
    p_{i} & = & p'_{i} 
    \nonumber \\
    \ket{b_{i}} & = & \ket{b'_{i}}
\end{eqnarray}
and so
\begin{eqnarray}
    \ket{0} & = & \sum_{i}\sqrt{p_{i}}\ket{a_{i}}\ket{b_{i}}    \nonumber \\
    \ket{1} & = & \sum_{i}\sqrt{p_{i}}\ket{a'_{i}}\ket{b_{i}}
\end{eqnarray}
It follows that there exists a unitary transformation $U\in \mathcal{H}_{A}$ 
such that
\begin{equation}
    \{\ket{a_{k}}\} \stackrel{U}{\longrightarrow} \{\ket{a'_{k}}\}
\end{equation}
and hence
\begin{equation}
    \ket{0} \stackrel{U}{\longrightarrow} \ket{1}
\end{equation}

As we shall see in \S \ref{sec:bitexample}, instead of transforming $\ket{0}$ to $\ket{1}$ by a unitary transformation, Alice could achieve the same effect by preparing the state $\ket{0}$ and measuring in either of two bases, depending on whether she intends to reveal 0 or 1.
    
The degenerate case can be handled in a similar way. Suppose that 
$p_{1} = p_{2} = p'_{1} = p'_{2} = p$. Then $\ket{b_{1}}, 
\ket{b_{2}}$ and $\ket{b'_{1}}, 
\ket{b'_{2}}$ span the same subspace $\hil{H}$ in 
$\hil{H}_{B}$, and hence (assuming the coefficients are distinct 
for $k>2$):
\begin{eqnarray}
    \ket{0} & = &  
    \sqrt{p}(\ket{a_{1}}\ket{b_{1}} + \ket{a_{2}}\ket{b_{2}})
    + \sum_{k>2}\sqrt{p_{k}}\ket{a_{k}}\ket{b_{k}}
    \nonumber \\
    \ket{1} & = &  
    \sqrt{p}(\ket{a'_{1}}\ket{b'_{1}} + \ket{a'_{2}}\ket{b'_{2}})
    + \sum_{k>2}\sqrt{p_{k}}\ket{a'_{k}}\ket{b_{k}}
    \nonumber \\
    & = & 
    \sqrt{p}(\ket{a''_{1}}\ket{b_{1}} + \ket{a''_{2}}\ket{b_{2}})
    + \sum_{k>2}\sqrt{p_{k}}\ket{a'_{k}}\ket{b_{k}}
\end{eqnarray}
where $\ket{a''_{1}}, \ket{a''_{2}}$ are orthonormal states 
spanning $\hil{H}$. Since $\{\ket{a''_{1}}, 
\ket{a''_{2}}, \ket{a_{3}}, \ldots \}$ is an orthonormal set in 
$\hil{H}_{A}$, there exists a unitary transformation in 
$\hil{H}_{A}$ that transforms $\{\ket{a_{k}}; k = 1, 2, 3, \ldots\}$ to 
$\{\ket{a''_{1}}, \ket{a''_{2}}, \ket{a'_{3}}, \dots\}$, 
and hence $\ket{0}$ to $\ket{1}$.

The extension of the theorem to the nonideal case, where $W_{B}(0) 
\approx W_{B}(1)$, so that there is a small probability that Bob could distinguish the 
alternative commitments, shows that Alice has a correspondingly large 
probability of cheating successfully: there exists a unitary transformation $U$ in 
$\hil{H}_{A}$  that will 
transform $W_{B}(0)$ sufficiently close to $W_{B}(1)$so that Alice can reveal whichever commitment she chooses, with a corresponding small probability of Bob being able to detect this move.  

\subsubsection{How the Theorem Works: An Example}

\label{sec:bitexample}

The following example by Asher Peres\index{Peres} (private communication) is a beautiful illustration of how the theorem works. (My analysis of the example owes much to correspondence with Adrian Kent\index{Kent} and Dominic Mayers\index{Mayers}.)

Suppose Alice is required to send Bob a channel particle $C$ in an equal weight mixture of the qubit states:
\begin{eqnarray}
\ket{c_{0}} & = & \ket{0} \\
\ket{c_{2}} & = & -\frac{1}{2}\ket{0} + \frac{\sqrt{3}}{2}\ket{1} \\
\ket{c_{4}} & = & -\frac{1}{2}\ket{0} - \frac{\sqrt{3}}{2}\ket{1}
\end{eqnarray}
if she commits to 0, and an equal weight mixture of the qubit states:
\begin{eqnarray}
\ket{c_{1}} & = & \ket{1} \\
\ket{c_{3}} & = & \frac{\sqrt{3}}{2}\ket{0} - \frac{1}{2}\ket{1} \\
\ket{c_{5}} & = & -\frac{\sqrt{3}}{2}\ket{0} - \frac{1}{2}\ket{1}
\end{eqnarray}
if she commits to 1. Note that these two mixtures have the same density operator:
\begin{equation}
\rho_{0} = \rho_{1} = I/2
\end{equation}

Suppose Alice tries to implement an EPR cheating strategy\index{EPR! cheating strategy (attack)} by preparing the entangled state of a system $AC$:
\begin{equation}
\ket{0} = \frac{1}{\sqrt{3}} (\ket{a_{0}}\ket{c_{0}} + \ket{a_{2}}\ket{c_{2}}
 + \ket{a_{4}}\ket{c_{4}})
\end{equation}
where $\{\ket{a_{0}},\ket{a_{2}},\ket{a_{4}}\}$ is an orthonormal basis in the 3-dimensional Hilbert space $\hil{H}^{A}$ of a suitable ancilla system $A$. If Alice could transform the state $\ket{0}$ to the state:
\begin{equation}
\ket{1} = \frac{1}{\sqrt{3}} (\ket{a_{1}}\ket{c_{1}} + \ket{a_{3}}\ket{c_{3}}
 + \ket{a_{5}}\ket{c_{5}})
\end{equation}
where $\{\ket{a_{1}},\ket{a_{3}},\ket{a_{5}}\}$ is another orthonormal basis in $\hil{H}^{A}$,
by a local unitary transformation in $\hil{H}^{A}$, she could delay her commitment to the opening stage. If, at that stage, she decides to commit to 0, she  measures the observable with eigenstates $\{\ket{a_{0}},\ket{a_{2}},\ket{a_{4}}\}$. If she decides to commit to 1, she performs the local unitary transformation  taking the state $\ket{0}$ to the state $\ket{1}$ and measures the observable with eigenstates $\{\ket{a_{1}},\ket{a_{3}},\ket{a_{5}}\}$.

Now, $\ket{0}$ can be expressed as:
\begin{eqnarray}
\ket{0} & = & \frac{1}{\sqrt{3}}\left(\ket{a_{0}}\frac{\ket{c_{3}} - \ket{c_{5}}}{\sqrt{3}} + 
\ket{a_{2}}\frac{\ket{c_{1}} - \ket{c_{3}}}{\sqrt{3}} + \ket{a_{4}}\frac{\ket{c_{5}} - \ket{c_{1}}}{\sqrt{3}}\right) \\
& = & \frac{1}{\sqrt{3}}\left(\frac{\ket{a_{2}} - \ket{a_{4}}}{\sqrt{3}}\ket{c_{1}} + 
\frac{\ket{a_{0}} - \ket{a_{2}}}{\sqrt{3}}\ket{c_{3}} + \frac{\ket{a_{4}} - \ket{a_{0}}}{\sqrt{3}}\ket{c_{5}}\right) 
\end{eqnarray}
In this representation of $\ket{0}$, the factor states $\frac{\ket{a_{2}} - \ket{a_{4}}}{\sqrt{3}}, \frac{\ket{a_{0}} - \ket{a_{2}}}{\sqrt{3}}, \frac{\ket{a_{4}} - \ket{a_{0}}}{\sqrt{3}}$  in $\hil{H}^{A}$ are not orthogonal---in fact, they are coplanar:
\begin{equation}
\ket{a_{0}} - \ket{a_{2}} = -(\ket{a_{2}} - \ket{a_{4}}) - (\ket{a_{4}} - \ket{a_{0}}
\end{equation}
So it seems that there cannot be a suitable unitary transformation that will map $\ket{0}$ to $\ket{1}$ and the EPR strategy\index{EPR! cheating strategy (attack)} is blocked!

Of course, this is not the case. To see that there is such a unitary transformation, note that $\ket{0}$ and $\ket{1}$ can be expressed in the Schmidt decomposition\index{Schmidt decomposition} as:
\begin{eqnarray}
\ket{0} & = & \frac{1}{\sqrt{2}} \left(\frac{2\ket{a_{0}} - \ket{a_{2}} - \ket{a_{4}}}{\sqrt{6}} \ket{c_{0}}
+ \frac{\ket{a_{2}} - \ket{a_{4}}}{\sqrt{2}} \ket{c_{1}}\right) \\
\ket{1} & = & \frac{1}{\sqrt{2}}\left(\frac{\ket{a_{3}} - \ket{a_{5}}}{\sqrt{2}} \ket{c_{0}}
+ \frac{-2\ket{a_{1}} + \ket{a_{3}} + \ket{a_{5}}}{\sqrt{6}} \ket{c_{1}}\right)
\end{eqnarray}
Clearly, now, there exists a unitary transformation $U$ in $\hil{H}^{A}$ such that:
\begin{equation}
\ket{0} \stackrel{U}{\longrightarrow} \ket{1}
\end{equation}
It follows that:
\begin{equation}
\{\ket{a_{0}},\ket{a_{2}},\ket{a_{4}}\} \stackrel{U}{\longrightarrow} \{\ket{a'_{0}},\ket{a'_{2}},\ket{a'_{4}}\}
\end{equation}
where $\{\ket{a'_{0}},\ket{a'_{2}},\ket{a'_{4}}\}$ is a basis in $\hil{H}^{A}$, and so
\begin{eqnarray}
\ket{1} & = & \frac{1}{\sqrt{3}}(\ket{a'_{0}}\ket{c_{0}} + \ket{a'_{2}}\ket{c_{2}} + \ket{a'_{4}}\ket{c_{4}}) \\
& = & \frac{1}{\sqrt{3}}(\ket{a_{1}}\ket{c_{1}} + \ket{a_{3}}\ket{c_{3}} + \ket{a_{5}}\ket{c_{5}})
\end{eqnarray}

So Alice could implement the EPR cheating strategy\index{EPR! cheating strategy (attack)} by preparing  the state $\ket{1}$ and measuring in the basis $\{\ket{a'_{0}},\ket{a'_{2}},\ket{a'_{4}}\}$ for the 0-commitment, or in the basis $\{\ket{a_{1}},\ket{a_{3}},\ket{a_{5}}\}$ for the 1-commitment. Equivalently, of course, she could prepare the state $\ket{0}$ and measure in two different bases, since the unitary transformation that takes $\ket{1}$ to $\ket{0}$ also takes the basis $\{\ket{a_{1}},\ket{a_{3}},\ket{a_{5}}\}$ to the basis $\{\ket{a''_{1}},\ket{a''_{3}},\ket{a''_{5}}\}$, and so:
\begin{eqnarray}
\ket{0} & = & \frac{1}{\sqrt{3}}(\ket{a_{0}}\ket{c_{0}} + \ket{a_{2}}\ket{c_{2}} + \ket{a_{4}}\ket{c_{4}}) \\
& = & \frac{1}{\sqrt{3}}(\ket{a''_{1}}\ket{c_{1}} + \ket{a''_{3}}\ket{c_{3}} + \ket{a''_{5}}\ket{c_{5}})
\end{eqnarray}
A calculation shows that:
\begin{eqnarray}
\ket{a''_{1}} & = & \frac{1}{3} \left(\ket{a_{0}} + (1 + \sqrt{3})\ket{a_{2}} + (1 - \sqrt{3})\ket{a_{4}}\right) \\
\ket{a''_{3}} & = & \frac{1}{3} \left((1 + \sqrt{3})\ket{a_{0}} + (1 - \sqrt{3})\ket{a_{2}} + \ket{a_{4}}\right) \\
\ket{a''_{5}} & = & \frac{1}{3} \left(1 - \sqrt{3})\ket{a_{0}} + \ket{a_{2}} + (1 + \sqrt{3})\ket{a_{4}}\right)
\end{eqnarray}
In effect, if Alice prepares the entangled state $\ket{0}$ and measures the ancilla $A$ in the $\{\ket{a_{0}},\ket{a_{2}},\ket{a_{4}}\}$ basis, she steers the channel particle into a mixture of nonorthogonal states $\{\ket{c_{0}},\ket{c_{2}},\ket{c_{4}}\}$. If she measures in the $\{\ket{a''_{1}},\ket{a''_{3}},\ket{a''_{5}}\}$ basis, she steers the channel particle into a mixture of nonorthogonal states $\{\ket{c_{1}},\ket{c_{3}},\ket{c_{5}}\}$.

It follows that Alice can implement the EPR cheating strategy\index{EPR! cheating strategy (attack)} without performing any unitary transformation---she simply entangles the channel particle with a suitable ancilla particle and performs one of two measurements at the opening stage, depending on her commitment. This shows that the unitary transformation required by the theorem is not in fact required. If a cheating strategy is possible in which Alice, at the opening stage, either makes a measurement on an entangled state for the 0-commitment, or transforms this entangled state to a different state by a local unitary transformation in her Hilbert space and then makes a measurement on the transformed state for the 1-commitment, then an equally good cheating strategy is available in which Alice prepares one entangled state for both commitments, and measures in two alternative bases at the opening stage, depending on her commitment.

\subsubsection{A Final Worry Laid to Rest}

\label{sec:bitworry}

The heart of the mathematical proof is the Schmidt
decomposition theorem\index{Schmidt decomposition}. But the essential conceptual insight is the possibility 
of enlarging the Hilbert space and implementing an EPR 
strategy without detection. 

This raises the 
following question: Suppose Bob cannot 
cheat because $W_{B}(0) = W_{B}(1)$, so by the theorem there exists a 
unitary transformation $U$ in $\hil{H}_{A}$ that will transform
$\ket{0}$ to $\ket{1}$. Could there be a protocol 
in which Alice also cannot cheat because, although there exists a 
suitable unitary transformation $U$, 
she cannot know what unitary 
transformation to apply? This 
is indeed the case, but only if $U$ 
depends on Bob's operations, which are unknown to Alice. But then Bob 
would have to actually make a definite choice or obtain a 
definite outcome in a measurement, and he could always avoid doing 
so without detection by applying an EPR strategy. 

This raises a further question: How do we know that following an EPR strategy is never 
disadvantageous to the cheater? If so, Bob might choose to avoid an EPR 
strategy in a certain situation because it would be disadvantageous 
to him.  Could there be a bit 
commitment protocol where the application of an EPR 
strategy by Bob at a certain stage of the protocol would give 
Alice the advantage, rather than Bob, while conforming to the protocol 
would ensure that neither party could cheat? If there were such a 
protocol, then Bob would, in effect, be forced to conform to the 
protocol and avoid the EPR strategy, and unconditionally 
secure bit commitment would be possible. 

In fact, the impossibility of such a protocol follows from the theorem (see \cite{BubBC}. 
Suppose there were such a protocol. That is, suppose that 
if Bob applies an EPR strategy then $W_{B}(0) = W_{B}(1)$, so 
by the theorem there exists a unitary transformation $U$ in Alice's 
Hilbert space that will transform $\ket{0}$ to $\ket{1}$. Alice 
must know this $U$ because it is uniquely determined by Bob's 
deviation from the protocol according to an EPR strategy 
that keeps all disjunctions at the quantum level as linear 
superpositions. Suppose also that if, instead, Bob is honest and follows the 
protocol (so that there is a definite choice for every disjunction 
over possible operations or possible measurement outcomes), then 
$W_{B}(0) = W_{B}(1)$,
but the unitary transformation in Alice's Hilbert space that allows 
her to transform $\ket{0}$ to $\ket{1}$ depends on Bob's choices 
or measurement outcomes, which are unknown to Alice. 

The point to note is that the information available in Alice's 
Hilbert space must be the same whether Bob follows the protocol and 
makes determinate choices and obtains determinate measurement 
outcomes before Alice applies the unitary transformation $U$ 
that transforms $\ket{0}$ to $\ket{1}$, or whether he 
deviates from the protocol via an EPR strategy 
in which he implements corresponding entanglements with his ancillas 
to keep choices and measurement outcomes at the quantum level before 
Alice applies the transformation $U$, and only makes these choices and 
measurement outcomes definite at the final stage of the protocol by 
measuring his ancillas. There 
can be no difference for Alice because Bob's measurements on his 
ancillas and any measurements or operations 
that Alice might perform take place in 
different Hilbert spaces, so the operations commute. If Alice's 
density operator (obtained by tracing over Bob's ancillas), which 
characterizes the statistics of measurements that Alice can perform in 
her part of the universe, were different depending on whether or not 
Bob actually carried out the required measurements, as opposed to 
keeping the alternatives at the quantum level by implementing 
corresponding 
entanglements with ancillas, then it would be possible to use this 
difference to signal superluminally. Actual measurements by Bob on his 
ancillas 
that selected alternatives in the entanglements 
as determinate would instantaneously alter the information available 
in Alice's part of the universe. 

It follows that in the hypothetical bit commitment protocol we are considering, 
the unitary transformation $U$ in 
Alice's Hilbert space that transforms $\ket{0}$ to $\ket{1}$ 
must be the same transformation in the honest scenario as in the 
cheating scenario. But we 
are assuming that the transformation in the honest scenario is unknown to Alice 
and depends on Bob's measurement outcomes, while the transformation in 
the cheating scenario is unique and known to Alice. So there can be no such 
protocol: the deviation from the protocol by an EPR strategy 
can never place Bob in 
a worse position than following the protocol honestly.  

The argument can be put formally in terms of the theorem as follows: 
The cheating scenario produces one of two alternative pure 
states $\ket{0}_{c}$ or $\ket{1}_{c}$ in 
$\hil{H}_{A}\otimes\hil{H}_{B}$ (`$c$' for `cheating strategy). 
Since the reduced density 
operators in $\hil{H}_{B}$:
\begin{eqnarray}
    W^{(c)}_{B}(0) & = & Tr_{A}\ket{0}\bra{0}_{c}
    \nonumber \\
    W^{(c)}_{B}(1) & = & Tr_{A}\ket{1}\bra{1}_{c}
\end{eqnarray}
are required by assumption to be the same:
\begin{equation}
    W^{(c)}_{B}(0) = W^{(c)}_{B}(1)
\end{equation}
the states $\ket{0}_{c}$ and $\ket{1}_{c}$ can be expressed in
biorthogonal decomposition as:
\begin{eqnarray}
    \ket{0}_{c} & = & \sum_{i}\sqrt{p_{i}}\ket{a_{i}}\bra{b_{i}}
    \nonumber \\
    \ket{1}_{c} & = & \sum_{i}\sqrt{p_{i}}\ket{a'_{i}}\bra{b_{i}}
\end{eqnarray}
where the reduced density operators in $\hil{H}_{A}$:
\begin{eqnarray}
    W^{(c)}_{A}(0) = Tr_{B}\ket{0}\bra{0}_{c}
    & = & \sum_{i}p_{i}\ket{a_{i}}\bra{a_{i}}
    \nonumber \\ 
    W^{(c)}_{A}(1) = Tr_{B}\ket{1}\bra{1}_{c}
    & = & \sum_{i}p_{i}\ket{a'_{i}}\bra{a'_{i}}
\end{eqnarray}
are different:
\begin{equation}
    W^{(c)}_{A}(0) \neq W^{(c)}_{A}(1)
\end{equation} 

It follows that there exists a unitary 
operator $U_{c} \in \hil{H}_{A}$ defined by the spectral 
representations of $W_{A}^{(c)}(0)$ and $W_{A}^{(c)}(1)$: 
\begin{equation}
    \{\ket{a_{i}}\} \stackrel{U_{c}}{\longrightarrow} \{\ket{a'_{i}}\}
\end{equation}
such that:
\begin{equation}
    \ket{0}_{c} \stackrel{U_{c}}{\longrightarrow} 
    \ket{1}_{c}
\end{equation}

The honest scenario produces one of two alternative pure states 
$\ket{0}_{h}$ and $\ket{1}_{h}$ in 
$\hil{H}_{A}\otimes\hil{H}_{B}$ (`$h$' for `honest'), 
where the pair 
$\{\ket{0}_{h}, \ket{1}_{h}\}$ depends on Bob's choices and 
the outcomes of his measurements. 

By assumption, as in the cheating scenario, the reduced density operators 
$W_{B}^{(h)}(0)$ and $W_{B}^{(h)}(1)$ in $\hil{H}_{B}$ are the 
same:
\begin{equation}
    W^{(h)}_{B}(0) = W^{(h)}_{B}(1)
\end{equation}
which entails the existence of a 
unitary operator $U_{h} \in \hil{H}_{A}$ such that:
\begin{equation}
    \ket{0}_{h} \stackrel{U_{h}}{\longrightarrow} 
    \ket{1}_{h}
\end{equation}
where $U_{h}$ depends on Bob's choices and measurement outcomes. 

Now, the difference between the honest scenario and the cheating scenario is 
undetectable in $\mathcal{H}_{A}$, which means that 
the reduced density operators 
in $\mathcal{H}_{A}$ are the same in the honest scenario as in the 
cheating scenario:
\begin{eqnarray}
     W^{(h)}_{A}(0) & = & W^{(c)}_{A}(0)
     \nonumber \\
     W^{(h)}_{A}(1) & = & W^{(c)}_{A}(1)
\end{eqnarray}
Since $U_{h}$ is defined by the spectral representations of 
$W^{(h)}_{A}(0)$ and $W^{(h)}_{A}(1)$, it follows that 
$U_{h} = U_{c}$. But we are assuming that 
$U_{h}$ depends on Bob's 
choices and measurement outcomes, while $U_{c}$ is uniquely defined 
by Bob's EPR strategy, in which there are no determinate 
choices or measurement outcomes. Conclusion: there can be no bit 
commitment protocol in which neither Alice nor Bob can cheat if Bob 
honestly follows the protocol, but Alice can cheat
if Bob deviates from the protocol via an EPR strategy. 
If neither Bob nor Alice can cheat in the honest scenario, then Bob and not 
Alice  must 
be able to cheat in the cheating scenario.

A similar argument rules out a protocol in which neither party 
can cheat if Bob is honest (as above), but if Bob follows an EPR 
strategy, then $W_{B}(0) \approx W_{B}(1)$, so Bob has some probability of 
cheating successfully, but Alice has a 
greater probability of cheating successfully than Bob. Again, the 
unitary transformation $U_{c}$ that would allow Alice to cheat with a certain 
probability of success if Bob 
followed an EPR strategy would also have to allow Alice to 
cheat successfully if Bob were honest. But the supposition is that 
Alice cannot cheat if Bob is honest, because the unitary 
transformation $U_{h}$ in that case depends on Bob's choices and 
measurement outcomes, which are unknown to Alice. It follows that 
there can be no such protocol. 

So there is no loophole in the proof of the theorem.
Unconditionally secure quantum bit commitment (in the sense of the 
theorem) really 
is impossible.

\section{Quantum Computation\index{quantum!computation}}

\subsection{The Church-Turing Thesis\index{Church-Turing thesis} and Computational Complexity\index{computational complexity}}
\label{sec:complexity}

The classical theory of computation concerns the question of what can be computed, and how efficiently\index{efficiency}. 

Various formal notions of computability by Alonzo Church\index{Church}, Kurt G\"{o}del\index{G\"{o}del}, and others can all be shown to be equivalent to Alan Turing's\index{Turing} notion of computability by a Turing machine\index{Turing machine} (see, e.g., \cite{LewisPap}). A Turing machine is an abstract computational device that can be in one of a finite set of possible states. It has a potentially infinite tape of consecutive cells to store information (0, 1, or blank in each cell) and a movable tape head that reads the information in a cell. Depending on the symbol in a cell and the state of the machine, the tape overwrites the symbol, changes the state, and moves one cell to the right or the left until it finally halts at the completion of the computation. A program for a Turing machine\index{Turing machine} $T$ (e.g., a program that executes a particular algorithm for finding the prime factors of an integer) is a finite string of 
symbols---which can be expressed as a binary number $b(T)$---indicating, for each state and each symbol, a new state, new symbol, and head displacement. Turing showed that there exists a
\emph{universal Turing machine}\index{universal Turing machine} $U$ that can simulate the program of any Turing machine $T$ with at most a polynomial slow-down, i.e., if we initialize $U$ with $b(T)$ and the input to $T$, then $U$ performs the same computation as $T$,  where the number of steps taken by $U$ to simulate each step of $T$ is a polynomial function of $b(T)$.  The Church-Turing thesis\index{Church-Turing thesis} is the proposal to identify the class of computable functions with the class of functions computable by a universal Turing machine. Equivalently, one could formulate the Church-Turing thesis\index{Church-Turing thesis} in terms of decision problems, which have yes-or-no answers (e.g., the problem of determining whether a given number is a prime number).

Intuitively, some computations are harder than others, and some algorithms take more time than others. The computational complexity\index{computational complexity} of an algorithm is measured by the number of steps required by a Turing machine to run through the algorithm. A decision problem is said to be in complexity class \textbf{P}\index{complexity class!\textbf{P}}, hence \emph{easy} or \emph{tractable}\index{computational complexity!easy (tractable)} if there exists an algorithm for solving the problem in polynomial time, i.e., in a number of steps that is a polynomial function of the size $n$ of the input (the number of bits required to store the input). A problem is said to be \emph{hard} or \emph{intractable}\index{computational complexity!hard (intractable)} if there does not exist a polynomial-time algorithm for solving the problem. A problem is in complexity class 
\textbf{EXP}\index{complexity class!\textbf{EXP}} if the most efficient algorithm requires a number of steps that is an exponential function of the size $n$ of the input. The number of steps here refers to the worst-case running time, $\tau$, which is  of the order 
$\mathcal{O}(n^{k})$ for a polynomial-time algorithm and of the order $\mathcal{O}(2^{n})$ for an exponential-time algorithm. 

Note that an exponential-time algorithm could be more efficient than a polynomial-time algorithm for some range of input sizes, so the above terminology should be understood with caution. Consider the following example (taken from \cite[p. 145]{Barenco}): $\tau_{P}(n) = 10^{-23}n^{1000} + 10^{23}/n \approx \mathcal{O}(n^{1000})$ because, for sufficiently large $n$, the polynomial term dominates (i.e., $\tau_{P}(n) < cn^{1000}$ for a fixed factor $c$), and $\tau_{E}(n) = 10^{23}n^{1000} + 10^{-23}2^{n} \approx \mathcal{O}(2^{n})$ because, for sufficiently large $n$, the exponential term dominates (i.e., $\tau_{E}(n) < c2^{n}$ for a fixed factor $c$). But for small enough values of $n$, $\tau_{E}(n) < \tau_{P}(n)$.

A Turing machine as defined  above is a deterministic machine. A \emph{nondeterministic} or \emph{probabilistic} Turing machine\index{Turing machine!nondeterministic (probabilistic) } makes a random choice between multiple transitions (to a new symbol, new state, and head displacement) for each symbol and each state. For each sequence of choices, the sequence of transitions corresponds to a sequence of steps executed by a deterministic Turing machine. If any of these machines halts, the computation is regarded as completed. Evidently, a nondeterministic Turing machine cannot compute a function that is not computable by a deterministic Turing machine, but it is believed (but not proved) that certain problems can be solved more efficiently by nondeterministic Turing machines than by any deterministic Turing machine. The complexity class \textbf{NP}\index{complexity class!\textbf{NP}} is the class of problems that can be solved in polynomial time by a nondeterministic Turing machine. This is equivalent to the class of problems for which proposed solutions can be verified in polynomial time by a deterministic Turing machine. For example, it is believed (but not proved) that the problem of factoring an integer into its prime factors is a `hard' problem: there is no known polynomial-time algorithm for this problem. However, the problem of checking whether a candidate factor of an integer is indeed a factor can be solved in polynomial time, so factorizability is an \textbf{NP} problem.

Clearly \textbf{P} $\subseteq$ \textbf{NP}, but it is an open problem in complexity theory whether \textbf{P} = \textbf{NP}. An \textbf{NP} problem is said to be \textbf{NP}-complete\index{NP-complete@\textbf{NP}-complete} if every
 \textbf{NP} problem can be shown to have a solution with a number of steps that is a polynomial function of the number of steps required to solve the   \textbf{NP}-complete problem. So if an \textbf{NP}-complete problem  can be solved in polynomial time, then all \textbf{NP} problems can be solved in polynomial time, and \textbf{P} = \textbf{NP}. The problem of determining whether a Boolean function 
 $f\{0,1\}^{n} \rightarrow \{0,1\}$ is satisfiable (i.e., whether there is a set of input values for which the function takes the value 1, or equivalently whether there is an assignment of truth values to the atomic sentences of a compound sentence of Boolean logic under which the compound sentence comes out true) is an \textbf{NP}-complete\index{NP-complete@\textbf{NP}-complete} problem. Factorizability is an \textbf{NP} problem that is not known to be \textbf{NP}-complete.

Since a Turing machine can simulate any classical computing device with at most a polynomial slow-down, the complexity classes are the same for any model of computation. For example, a circuit computer computes the value of a function by transforming  data stored in an input register, representing the input to the function,  via Boolean circuits constructed of elementary Boolean gates connected by wires, to data in an output register representing the value of the function computed. The elementary Boolean gates are 1-bit gates (such as the NOT gate, which transforms 0 to 1, and conversely) and 2-bit gates (such as the AND gate, which takes two input bits to 1 if and only if they are both 1, otherwise to 0), and it can be shown that a combination of such gates forms a `universal set' that suffices for any transformation of $n$ bits. In fact, it turns out that one of the  sixteen possible 2-bit Boolean gates, the NAND gate (or NOT AND gate), which takes two input bits to 0 of and only if they are both 1, forms a universal set by itself. 

In a circuit model\index{quantum!computer!circuit model} of a quantum computer, the  registers store qubits, which are then manipulated by elementary unitary gates. It can be shown (see \cite[p. 188]{NielsenChuang}) that a set of single-qubit and two-qubit unitary gates---the CNOT gate, the Hadamard gate, the phase gate, and the $\pi/8$ gate---forms a universal set, in the sense that any unitary transformation of $n$ qubits can be approximated to arbitrary accuracy by a quantum circuit consisting of these gates connected in some combination. The CNOT gate\index{CNOT gate} (`C' for `controlled) has two input qubits, a `control' qubit and a `target' qubit. The gate functions so as to flip the target qubit  if and only if the control qubit is $\ket{1}$. The remaining three gates are single-qubit gates. The Hadamard gate\index{Hadamard gate} transforms $\ket{0}$ to $(\ket{0} + \ket{1})/\sqrt{2}$ and $\ket{1}$ to 
$(\ket{0} - \ket{1})/\sqrt{2}$ and  is sometimes referred to as  the `square root of NOT' gate\index{square root of NOT gate@`square root of NOT' gate} because two successive applications transforms $\ket{0}$ to $\ket{1}$, and conversely. The phase  gate\index{phase gate} leaves $\ket{0}$ unchanged and transforms $\ket{1}$ to $i\ket{1}$. The  $\pi/8$ gate\index{pi@$\pi/8$ gate} leaves $\ket{0}$ unchanged and transforms $\ket{1}$  to $e^{i\pi/4}\ket{1}$. (See \cite[p. 174]{NielsenChuang} for a discussion and why the $\pi/8$ gate is so named.)

There are other models of quantum computation. In the `cluster state' or `one-way' quantum computer\index{quantum!computer!cluster state}\index{quantum!computer!one-way} of Raussendorf\index{Raussendortf} and Briegel\index{Briegel} \shortcite{RaussendorfBriegel00,RaussendorfBriegel01}, a fixed multi-qubit entangled state (called a `cluster state'), independent of the computation, is prepared. Then a sequence of single-qubit measurements is  performed on this state, where the choice of what observables to measure depends on the outcomes of the previous measurements. No unitary transformations are involved. Remarkably, it can be shown that any quantum circuit of unitary gates and measurements can be simulated by a cluster state computer with similar resources of qubits and time 
(see \cite{Jozsa05,Nielsen03,Nielsen05}).

The interesting question is whether a quantum computer can perform computational tasks that are not possible for a Turing machine, or perform such tasks more efficiently than any Turing machine\index{efficiency}. Since a Turing machine is defined by its program, and a program can be specified by a finite set of symbols, there are only countably many Turing machines. There are uncountably many functions on the natural numbers, so there are uncountably many uncomputable functions, i.e., functions that are not computable by any Turing machine. A quantum computer cannot compute a function that is not Turing-computable, because a Turing machine can simulate (albeit inefficiently, with an exponential slow-down \cite{Feynman}) the dynamical evolution of any system, classical or quantum, with arbitrary accuracy. But there are computational tasks that a quantum computer can perform by exploiting entanglement that are impossible for any Turing machine. Recall the discussion of Bell's counterargument to the EPR argument in \S \ref{sec:entanglement}: a quantum computer, but no classical computer, can perform the task of rapidly producing pairs of values (0 or 1) for pairs of input angles at different locations, with correlations that violate Bell's inequality\index{Bell's inequality}, where the response time is less than the time taken by light to travel between the locations.

The current interest in quantum computers concerns the question of whether a quantum computer can compute certain Turing-computable functions more \emph{efficiently}\index{efficiency} than any Turing machine. In the following section, I discuss  quantum algorithms that achieve an exponential speed-up over any  classical algorithm, or an exponential speed-up over any \emph{known} classical algorithm. The most spectacular of these is Shor's factorization algorithm\index{Shor's algorithm}, and a related algorithm for solving the discrete log problem.\footnote{The discrete log of $x$ with respect to a given prime integer $p$ and an integer $q$ coprime to $p$ is the integer $r$ such that $q^{r} = x \mbox{ mod $p$}$. See \cite[p. 238]{NielsenChuang} for a discussion.} 

The factorization algorithm  has an important practical application to cryptography. Public-key distribution protocols\index{public-key cryptography} such as RSA\index{RSA}  \cite{RSA} (widely used in commercial transactions over the internet, transactions between banks and financial institutions, etc.) rely on factoring being a `hard' problem. (Preskill \shortcite{Preskill} notes that currently the 65-digit factors of a 130-digit integer can be found in about a month using a network of hundreds of work stations implementing the best known classical factoring algorithm (the `number sieve algorithm'). He estimates that factoring a 400-digit integer would take about $10^{10}$ years, which is the age of the universe.) To see the idea behind the RSA protocol, suppose Alice wishes to send a secret message to Bob. Bob's public key consists of two large integers, $s$ and $c$. Alice encrypts the message $m$ (in the form of a binary number) as $e = m^{s}\mbox{ mod}\,c$ and sends the encrypted message to Bob. Bob decrypts the message as $e^{t}\mbox{ mod}\,c$ where $t$ is an integer known only to Bob. The integer $t$ for which $m = e^{t}\mbox{ mod}\,c$ can easily be determined from $s$ and the factors of $c$, but since $c = pq$ is the product of two large prime numbers known only to Bob, an eavesdropper, Eve, can read the message only if she can factor $c$ into its prime factors. The cleverness of the scheme resides in the fact that no secret key needs to be distributed between Alice and Bob: Bob's key $\{s, c\}$ is public and allows anyone to send encrypted messages to Bob. If a quantum computer could be constructed that implemented Shor's algorithm, key distribution protocols that rely on the difficulty of factoring very large numbers would be insecure.

\subsection{Quantum Algorithms\index{quantum!algorithms}}

\label{sec:algorithms}

In the following three sections, I look at the information-processing involved in Deutsch's XOR algorithm\index{Deutsch's XOR algorithm}\index{Deutsch} \shortcite{Deutsch1985}, Simon's period-finding algorithm\index{Simon's algorithm} \shortcite{Simon94,Simon97}\index{Simon}, and Shor's factorization algorithm\index{Shor}\index{Shor's algorithm} \shortcite{Shor94,Shor97} in terms of the difference between the Boolean logic underlying a classical computation and the non-Boolean logic represented by the projective geometry of Hilbert space, in which the subspace structure of Hilbert space replaces the set-theoretic structure of classical logic. The three algorithms all turn out to involve a similar geometric formulation. 

Basically, all three algorithms involve the determination of a global property of a function, i.e., a disjunctive property. The disjunction is represented as a subspace in an appropriate Hilbert space, and alternative possible disjunctions turn out to be represented as orthogonal subspaces, except for intersections or overlaps. The true disjunction is determined as the subspace containing the state vector via a measurement. The algorithm generally has to be run several times because the state might be found in the overlap region. The essential feature of these quantum computations is that the true disjunction is distinguished from alternative disjunctions without determining the truth values of the disjuncts. In a classical computation, distinguishing the true disjunction would be impossible without the prior determination of the truth values of the disjuncts. More generally, a quantum computer computes a global property of a function without computing information that is redundant quantum mechanically, but essential for a classical computation of the global property.

There are other quantum algorithms besides these three, e.g., Grover's sorting algorithm \shortcite{Grover97} which achieves a quadratic speed-up over any classical algorithm. For a discussion, see \cite{NielsenChuang}, \cite{Jozsa99}.

\subsubsection{Deutsch's  XOR Algorithm\index{Deutsch's  XOR algorithm} and the Deutsch-Jozsa Algorithm\index{Deutsch-Jozsa algorithm}}

Let $B = \{0,1\}$ be a Boolean algebra (or the additive group of integers mod 2). In Deutsch's XOR problem \shortcite{Deutsch1985}, we are given a `black box' or oracle  that computes a function $f: B \rightarrow B$ and we are required to determine whether the function is `constant' (takes the same value for both inputs) or `'balanced' (takes a different value for each input). Classically, the only way to do this would be to consult the oracle twice, for the input values 0 and 1, and compare the outputs. 

In a quantum computation of the Boolean function, a unitary transformation $U_{f}: \ket{x}\ket{y} \rightarrow \ket{x}\ket{y \oplus f(x)}$ corresponding to the `black box' correlates input values with corresponding output values.\footnote{Note that two quantum registers are required to compute functions that are not 1-1 by a unitary transformation. Different input values, $x$ and $y$, to a function $f$ are represented by orthogonal states $\ket{x}, \ket{y}$. So if $f(x) = f(y)$ for some $x \neq y$, the transformation $W_{f}: \ket{x} \rightarrow \ket{f(x)}$ could not be unitary, because the orthogonal states $\ket{x}, \ket{y}$ would have to be mapped onto the same state by $W_{f}$. The ability of unitary transformations, which are reversible, to compute irreversible functions is achieved by keeping a record of the input for each output value of the function.}   The computation proceeds as follows: The input and output registers are 1-qubit registers initialized to the state $\ket{0}\ket{0}$ in a standard basis. A Hadamard transformation is applied to the input register, yielding a linear superposition of states corresponding to the two possible input values 0 and 1, and the transformation $U_{f}$ is then applied to both registers, yielding the transitions:
\begin{eqnarray}
\ket{0}\ket{0} & \stackrel{H}{\rightarrow} & \frac{1}{\sqrt{2}}(\ket{0} + \ket{1})\ket{0} \\ 
& \stackrel{U_{f}}{\rightarrow} & \frac{1}{\sqrt{2}}(\ket{0}\ket{f(0)} + \ket{1}\ket{f(1)})
\end{eqnarray}

If the function is constant, the final composite state of both registers is one of the two orthogonal states:
\begin{eqnarray}
\ket{c_{1}} & = & \frac{1}{\sqrt{2}}(\ket{0}\ket{0} + \ket{1}\ket{0}) \\
\ket{c_{2}} & = & \frac{1}{\sqrt{2}}(\ket{0}\ket{1} + \ket{1}\ket{1})
\end{eqnarray}
If the function is balanced, the final composite state is one of the two orthogonal states:
\begin{eqnarray}
\ket{b_{1}} & = & \frac{1}{\sqrt{2}}(\ket{0}\ket{0} + \ket{1}\ket{1}) \\
\ket{b_{2}} & = & \frac{1}{\sqrt{2}}(\ket{0}\ket{1} + \ket{1}\ket{0})
\end{eqnarray}

The states $\ket{c_{1}}, \ket{c_{2}}$ and $\ket{b_{1}}, \ket{b_{2}}$  span two planes $P_{c}, P_{b}$  in 
$\hil{H}^{2}\otimes\hil{H}^{2}$, represented by the projection operators:
\begin{eqnarray}
P_{c} & = & P_{\ket{c_{1}}} + P_{\ket{c_{2}}} \\
P_{b} & = & P_{\ket{b_{1}}} + P_{\ket{b_{2}}}
\end{eqnarray}

 These planes are orthogonal, except for an intersection, so their projection operators commute. The intersection is the line (ray) spanned by the vector\footnote{Here $\ket{00} = \ket{0}\ket{0}$, etc.}:
 \begin{equation}
\frac{1}{2}(\ket{00} + \ket{01} + \ket{10} + \ket{11}) = \frac{1}{\sqrt{2}}(\ket{c_{1}} + \ket{c_{2}}) = \frac{1}{\sqrt{2}}(\ket{b_{1}} + \ket{b_{2}})
\end{equation}

In the `prime' basis spanned by the states $\ket{0'} = H\ket{0}, \ket{1'} = H\ket{1}$ the intersection is the state $\ket{0'}\ket{0'}$, the `constant' plane is spanned by 
$\ket{0'}\ket{0'}, \ket{0'}\ket{1'}$, and the `balanced' plane is spanned by $\ket{0'}\ket{0'}, \ket{1'}\ket{1'}$. Note that:
\begin{eqnarray}
\ket{0'}\ket{1'} & = & \frac{1}{\sqrt{2}}(\ket{c_{1}} - \ket{c_{2}}) \\
\ket{1'}\ket{1'} & = & \frac{1}{\sqrt{2}}(\ket{b_{1}} - \ket{b_{2}})
\end{eqnarray}

In the usual formulation of the algorithm, to decide whether the function $f$ is constant or balanced we measure the output register in the prime basis. If the outcome is $0'$ (which is obtained with probability 1/2, whether the state ends up in the constant plane or the balanced plane), the computation is inconclusive, yielding no information about the function $f$. If the outcome is $1'$, then we measure the input register. If the outcome of the measurement on the input register is $0'$, the function is constant; if it is $1'$, the function is balanced. 

Alternatively---and this will be relevant for the comparison with Simon's algorithm and Shor's algorithm---we could measure the observable with eigenstates $\ket{0'0'}$, $\ket{0'1'}$, $\ket{1'0'}$, $\ket{1'1'}$. The final state is in the 3-dimensional subspace orthogonal to the vector $\ket{1'0'}$, either in the constant plane or the balanced plane. If the state is in the constant plane,  we will either obtain the outcome $0'0'$ with probability 1/2 (since the final state is at an angle $\pi/4$ to $\ket{0'0'}$), in which case the computation is inconclusive, or the outcome $0'1'$ with probability 1/2. If the state is in the balanced plane, we will again obtain the outcome $0'0'$ with probability 1/2, in which case the computation is inconclusive, or the outcome $1'1'$ with probability 1/2. So in either case, with probability 1/2, we can distinguish in one run of the algorithm between the two quantum disjunctions `constant' and `balanced' represented by the planes:
\begin{eqnarray}
P_{c} & = & P_{\ket{0'0'}} \vee P_{\ket{0'1'}} \\
P_{b} & = & P_{\ket{0'0'}} \vee P_{\ket{1'1'}} 
\end{eqnarray}
without finding out the truth values of the disjuncts in the computation (i.e., whether in the `constant' case the function maps 0 to 0 and 1 to 0 or whether the function maps  0 to 1 and 1 to 1, and similarly in the `balanced' case). Note that we could also apply a Hadamard transformation to the final states of both registers and measure in the computational basis, since $\ket{0'0'} \stackrel{H}{\longrightarrow} \ket{00}$, etc.

Deutsch's XOR algorithm was the first quantum algorithm with a demonstrated speed-up over any classical algorithm performing the same computational task. However, the algorithm has an even probability of failing, so the improvement in efficiency over a classical computation is only achieved if the algorithm succeeds, and even then is rather modest: one run of the quantum algorithm versus two runs of a classical algorithm. The following variation of Deutsch's algorithm avoids this feature \cite{Cleve98}. 

We begin by initializing the two registers to $\ket{0}$ and $\ket{1}$, respectively (instead of to $\ket{0}$ and $\ket{0}$) and apply a Hadamard transformation to both registers, which yields the transition:
\begin{eqnarray}
\ket{0}\ket{1} \stackrel{H}{\rightarrow} \frac{\ket{0} + \ket{1}}{\sqrt{2}}\frac{\ket{0} - \ket{1}}{\sqrt{2}} \label{eqn:init}
\end{eqnarray}
Since
\begin{equation}
U_{f}\ket{x}\ket{y} = \ket{x}\ket{y\oplus f(x)}
\end{equation}
it follows that 
\begin{equation}
U_{f}\ket{x}\frac{\ket{0} - \ket{1}}{\sqrt{2}} = \left \{\begin{array}{c}
 \ket{x}\frac{\ket{0} - \ket{1}}{\sqrt{2}} \mbox{ if $f(x) = 0$} \vspace{.1 in} \\
 - \ket{x}\frac{\ket{0} - \ket{1}}{\sqrt{2}} \mbox{ if $f(x) = 1$}
 \end{array} \right.
\end{equation}
which can be expressed as
\begin{equation}
U_{f}\ket{x}\frac{\ket{0} - \ket{1}}{\sqrt{2}}  = (-1)^{f(x)}\ket{x}\frac{\ket{0} - \ket{1}}{\sqrt{2}} \label{eqn:phase}
\end{equation}
Notice that the value of the function now appears as a phase of the final state of the input register, a feature referred to as `phase kickback.' For the input state $1/\sqrt{2}(\ket{0} + \ket{1})$, we have:
\begin{equation}
U_{f}\frac{\ket{0} + \ket{1}}{\sqrt{2}}\frac{\ket{0} - \ket{1}}{\sqrt{2}}  = 
\frac{(-1)^{f(0)}\ket{0} + (-1)^{f(1)}\ket{1}}{\sqrt{2}} \frac{\ket{0} - \ket{1}}{\sqrt{2}} 
\end{equation}
which can be expressed as:
\begin{equation}
U_{f}\frac{\ket{0} + \ket{1}}{\sqrt{2}}\frac{\ket{0} - \ket{1}}{\sqrt{2}}  = \left \{\begin{array}{c}
 \pm \frac{\ket{0} + \ket{1}}{\sqrt{2}}\frac{\ket{0} - \ket{1}}{\sqrt{2}} = \pm \ket{0'}\ket{1'} \mbox{ if $f(0) = f(1)$} \vspace{.1 in} \\
  \pm \frac{\ket{0} - \ket{1}}{\sqrt{2}}\frac{\ket{0} - \ket{1}}{\sqrt{2}} = \pm\ket{1'}\ket{1'}  \mbox{ if $f(0) \neq f(1)$} 
  \end{array} \right.
\end{equation}

Instead of the final state of the two registers ending up as one of two orthogonal states in the constant plane, or as one of two orthogonal states in the balanced plane, the final state now ends up as $\pm\ket{0'1'}$ in the constant plane, or as $\pm\ket{1'1'}$ in the balanced plane, and these states can be distinguished because they are orthogonal. So we can decide with certainty whether the function is constant or balanced after only one run of the algorithm. In fact, we can distinguish these two possibilities by simply measuring the input register in the prime basis. Note that if we perform a final Hadamard transformation on the input register (which takes $\ket{0'}$ to $\ket{0}$ and $\ket{1'}$ to 
$\ket{1}$), we can distinguish the two possibilities by measuring the input register in the computational basis. Note also that the state of the output register is unchanged: at the end of the process it is in the  state $\ket{1'} = H\ket{1}$ (as in (\ref{eqn:init})) and is not measured. 

Deutsch's XOR problem can be generalized to the problem (`Deutsch's problem'\index{Deutsch's problem}) of determining whether a Boolean function $f:B^{n} \rightarrow B$ is constant or whether it is balanced, where it is promised that the function is either constant or balanced.  `Balanced' here means that the function takes the values 0 and 1 an equal number of times, i.e., $2^{n-1}$ times each. The Deutsch-Jozsa algorithm\index{Deutsch-Jozsa algorithm} \shortcite{DeutschJozsa92} decides whether $f$ is constant or balanced in one run.

We begin by setting the input $n$-qubit register to the state $\ket{0}$ (an abbreviation for the state $\ket{0 \cdots 0} = \ket{0} \cdots \ket{0}$) and the output 1-qubit register to the state $\ket{1}$. We apply an $n$-fold Hadamard transformation to the input register and a Hadamard transformation to the output register, followed by the unitary transformation $U_{f}$ to both registers, and finally an $n$-fold Hadamard transformation to the input register. 

First note that
\begin{equation}
H\ket{x} = \frac{1}{\sqrt{2}}\sum_{y \in \{0,1\}}(-1)^{xy}\ket{y}
\end{equation}
so
\begin{equation}
H^{\otimes n}\ket{x_{1}, \ldots, x_{n}} = \frac{1}{\sqrt{2}}\sum_{y_{1}, \cdots, y_{n}}(-1)^{x_{1}y_{1}+ \cdots + x_{n}y_{n}}\ket{y_{1}, \ldots, y_{n}}
\end{equation}
This can be expressed as:
  \begin{equation}
H^{\otimes n}\ket{x} = \frac{1}{\sqrt{2}}\sum_{y \in \{0,1\}}(-1)^{x \cdot y}\ket{y}
\end{equation}
where $x \cdot y$ is the bitwise inner product of $x$ and $y$, mod 2.

The unitary transformations (Hadamard transformation, $U_{f}$) yield:
\begin{eqnarray}
\ket{0}^{\otimes n}\ket{1} & \stackrel{H}{\longrightarrow} & \sum_{x \in \{0,1\}^{n}}\frac{\ket{x}}{\sqrt{2^{n}}} \frac{\ket{0} - \ket{1}}{\sqrt{2}} \\
& \stackrel{U_{f}}{\longrightarrow} & \sum_{x}\frac{(-1)^{f(x)}}{\sqrt{2^{n}}} \ket{x} \frac{\ket{0} - \ket{1}}{\sqrt{2}} \\
& \stackrel{H}{\longrightarrow} & \sum_{y}\sum_{x} \frac{(-1)^{x\cdot y + f(x)}}{\sqrt{2^{n}}} \ket{y} \frac{\ket{0} - \ket{1}}{\sqrt{2}}
\end{eqnarray}

Now consider the state of the input register:
\begin{equation}
\sum_{y}\sum_{x} \frac{(-1)^{x\cdot y + f(x)}}{\sqrt{2^{n}}} \ket{y} = \sum_{x}\frac{(-1)^{f(x)}}{\sqrt{2^{n}}} \ket{0 \ldots 0} + \ldots \label{eqn:DJ}
\end{equation}
Note that the coefficient (amplitude) of the state $\ket{0 \ldots 0}$ in the linear superposition (\ref{eqn:DJ}) is $\sum_{x}\frac{(-1)^{f(x)}}{\sqrt{2^{n}}}$. If $f$ is constant, this coefficient is $\pm 1$, so the coefficients of the other terms must all be 0. If $f$ is balanced, $f(x) = 0$ for half the values of $x$ and $f(x) = 1$ for the other half, so the positive and negative contributions to the coefficient of $\ket{0 \ldots 0}$ cancel to 0. In other words, if $f$ is constant, the state of the input register is $\pm\ket{0 \ldots 0}$; if $f$ is balanced, the state is in the orthogonal subspace. 

This is the usual way of describing how the algorithm works, which rather obscures the geometric picture. Consider, for simplicity, the case $n = 2$. After the transformation $U_{f}$, but before the final Hadamard transformation, the state of the input register is:
\begin{equation}
\pm\frac{1}{2}(\ket{00} + \ket{01} + \ket{10} + \ket{11})
\end{equation}
if the function is constant, or a state of the form:
\begin{equation}
\frac{1}{2}(\pm\ket{00} \pm \ket{01} \pm \ket{10} \pm \ket{11})
\end{equation}
if the function is balanced, where two of the coefficients are $+1$ and two of the coefficients are $-1$. Evidently, there are six such balanced states, and they are all orthogonal to the constant state. So the six balanced states lie in a 3-dimensional subspace orthogonal to the constant state and can therefore be distinguished from the constant state. The final Hadamard transformation transforms the constant state:
\begin{equation}
\pm\frac{1}{2}(\ket{00} + \ket{01} + \ket{10} + \ket{11}) \stackrel{H}{\longrightarrow} \pm\ket{00}
\end{equation}
 and the six balanced states to states in the 3-dimensional subspace orthogonal to $\ket{00}$. So to decide whether the function is constant or balanced we need only measure the input register and check whether it is in the state $\ket{00}$.

\subsubsection{Simon's Algorithm\index{Simon's algorithm}}

The problem here is to find the period $r$ of a periodic function $f:B^{n} \rightarrow B^{n}$, i.e., a Boolean function for which
\begin{equation}
f(x_{i}) = f(x_{j}) \mbox{ if and only if $x_{j} = x_{i} \oplus r$, for all $x_{i},x_{j} \in B^{n}$}
\end{equation}
Note that since $x\oplus r \oplus r =  x$, the function is 2-to-1.

Simon's algorithm solves the problem efficiently, with an exponential speed-up over any  classical algorithm (see \cite{Simon94,Simon97})\index{Simon}. The algorithm proceeds as in the Deutsch-Jozsa algorithm, starting with the input and output registers in the  state $\ket{0 \ldots 0}\ket{0}$ in the computational basis:
\begin{eqnarray}
\ket{0 \ldots 0}\ket{0} & \stackrel{H}{\longrightarrow} & \frac{1}{\sqrt{2^{n}}} \sum_{x=0}^{2^{n}-1}\ket{x}\ket{0} \\
& \stackrel{U_{f}}{\longrightarrow} &  \frac{1}{\sqrt{2^{n}}} \sum_{x}\ket{x}\ket{f(x)} \\
& & = \frac{1}{\sqrt{2^{n-1}}}\sum_{x_{i}}\frac{\ket{x_{i}} + \ket{x_{i} \oplus r}}{\sqrt{2}} \ket{f(x_{i})}
\end{eqnarray}
where $U_{f}$ is the unitary transformation implementing the Boolean function as:
\begin{equation}
U_{f}: \ket{x}\ket{y} \rightarrow \ket{x}\ket{y \oplus f(x)}
\end{equation}

The usual way to see how the algorithm works is to consider what happens if we measure the output register and keep the state of the input register,\footnote{The measurement of the output register here is a pedagogical device for ease of conceptualization. Only the input register is actually measured. The input register  is in a mixture of states, which we can think of as the mixture associated with the distribution of  outcomes obtained by measuring the output register.} which will have the form:
\begin{equation}
\frac{\ket{x_{i}} + \ket{x_{i}\oplus r}}{\sqrt{2}}
\end{equation}
This state contains the information $r$, but summed with an unwanted randomly chosen offset $x_{i}$ that depends on the measurement outcome. A direct measurement of the state label would  yield any $x \in B^{n}$ equiprobably, providing no information about $r$.

We now apply a Hadamard transform:
\begin{eqnarray}
\frac{\ket{x_{i}} + \ket{x_{i}\oplus r}}{\sqrt{2}} & \stackrel{H}{\longrightarrow} & \sum_{y \in B^{n}}\frac{(-1)^{x_{i}\cdot y} + (-1)^{(x_{i}\oplus r)\cdot y}}{\sqrt{2}}\ket{y} \\
& = & \sum_{y:r\cdot y = 0} \frac{(-1)^{x_{i}\cdot y}}{\sqrt{2}}\ket{y}
\end{eqnarray}
where the last equality follows  because terms interfere destructively if $r\cdot y =1$. Finally, we measure the input register in the computational basis and obtain a value $y$ (equiprobably) such that $r\cdot y = 0$. Then we repeat the algorithm sufficiently many times to find enough values $y_{i}$ so that $r$ can be determined by solving the linear equations $r\cdot y_{1} = 0, \ldots, r\cdot y_{k} = 0$.

To see what is going on geometrically, consider  the case $n = 2$. The possible values of the period $r$ are: 01, 10, 11, and the corresponding states of the input and output registers after the unitary transformation $U_{f}$ are:
\begin{description}\centering
\item[$r = 01:$] $(\ket{00} + \ket{01})\ket{f(00)} + (\ket{10} + \ket{11})\ket{f(10)}$
\item[$r = 10:$] $(\ket{00} + \ket{10})\ket{f(00)} + (\ket{01} + \ket{11})\ket{f(01)}$
\item[$r = 11:$] $(\ket{00} + \ket{11})\ket{f(00)} + (\ket{01} + \ket{10})\ket{f(01)}$
\end{description}

Notice that this case reduces to the same geometric construction as in Deutsch's XOR algorithm. For $r = 10$ the input register states are $\ket{c_{1}} = \ket{00} + \ket{10}$ or $\ket{c_{2}} = \ket{01} + \ket{11}$, and for $r = 11$ the input register states are $\ket{b_{1}} = \ket{00} + \ket{11}$ or $\ket{b_{2}} = \ket{01} + \ket{10}$, depending on the outcome of the measurement of the output register. So the three possible periods are associated with three planes in $\hil{H}^{2}\otimes \hil{H}^{2}$, which correspond to the constant and balanced planes in Deutsch's XOR algorithm, and a third plane, all three planes intersecting in the line spanned by the vector $\ket{00}$.  In the prime basis obtained by applying the Hadamard transformation, the planes are as follows:
\begin{description}
\item[$r = 01:$] plane spanned by $\ket{0'0'}, \ket{1'0'}$
\item[$r = 10:$] plane spanned by $\ket{0'0'}, \ket{0'1'}$ (corresponds to `constant'  plane)
\item[$r = 11:$] plane spanned by $\ket{0'0'}, \ket{1'1'}$ (corresponds to `balanced' plane)
\end{description}

We could simply measure the input register in the prime basis to find the period. Alternatively, we could apply a Hadamard transformation (which amounts to dropping the primes in the above representation of the $r$-planes) and measure in the computational basis. The three planes are orthogonal, except for their intersection in the line spanned by the vector $\ket{00}$. The three possible periods can therefore be distinguished by measuring the observable with eigenstates $\ket{00}, \ket{01}, \ket{10}, \ket{11}$, except when the state of the register is projected by the measurement (`collapses') onto the intersection state $\ket{00}$ (which occurs with probability 1/2). So the algorithm will generally have to be repeated until we find an outcome that is not 00.

The $n=2$ case of Simon's algorithm reduces to Deutsch's XOR algorithm. What about other cases? We can see what happens in the general case if we consider the case $n = 3$. There are now seven possible periods: 001, 010, 011, 100, 101, 110, 111. For the period $r = 001$, the state of the two registers after the unitary transformation $U_{f}$ is:
\begin{eqnarray}
\lefteqn{(\ket{000} + \ket{001})\ket{f(000)} + (\ket{010} + \ket{011})\ket{f(010)}} \nonumber \\
& & + (\ket{100} + \ket{101})\ket{f(100)} + (\ket{110} + \ket{111})\ket{f(110)}
\end{eqnarray}
If we measure the output register, the input register is left in one of four states, depending on the outcome of the measurement:
\begin{eqnarray*}
\ket{000} + \ket {001} & = & \ket{0'0'0'} + \ket{0'1'0'} + \ket{1'0'0'} + \ket{1'1'0'}\\
\ket{010} + \ket {011} & = & \ket{0'0'0'} - \ket{0'1'0'} + \ket{1'0'0'} - \ket{1'1'0'}\\
\ket{100} + \ket {101} & = & \ket{0'0'0'} + \ket{0'1'0'} - \ket{1'0'0'} - \ket{1'1'0'} \\
\ket{110} + \ket {111} & = & \ket{0'0'0'} - \ket{0'1'0'} - \ket{1'0'0'} + \ket{1'1'0'} 
\end{eqnarray*}
Applying a Hadamard transformation amounts to dropping the primes. So if the period is $r = 001$,  the state of the input register ends up in the 4-dimensional subspace of $\hil{H}^{2} \otimes \hil{H}^{2} \otimes \hil{H}^{2}$ spanned by the vectors: $\ket{000}, \ket{010}, \ket{100}, \ket{110}$. 

A similar analysis applies to the other six possible periods. The corresponding subspaces are spanned by the following vectors:
\begin{description}
\item{r = 001:} $\mbox{ } \ket{000}, \ket{010}, \ket{100}, \ket{110}$
\item{r = 010:} $\mbox{ } \ket{000}, \ket{001}, \ket{100}, \ket{101}$
\item{r = 011:} $\mbox{ } \ket{000}, \ket{011}, \ket{100}, \ket{111}$
\item{r = 100:} $\mbox{ } \ket{000}, \ket{001}, \ket{010}, \ket{011}$
\item{r = 101:} $\mbox{ } \ket{000}, \ket{010}, \ket{101}, \ket{111}$
\item{r = 110:} $\mbox{ } \ket{000}, \ket{001}, \ket{110}, \ket{111}$
\item{r = 111:} $\mbox{ } \ket{000}, \ket{011}, \ket{101}, \ket{110}$
\end{description}

These subspaces are orthogonal except for intersections in 2-dimensional planes. The period  can be found by measuring in the computational basis. Repetitions of the measurement will eventually yield sufficiently many distinct values to determine in which subspace out of the seven possibilities the final state lies. In this case ($n = 3$), it is clear by examining the above list that two values distinct from 000 suffice to determine the subspace, and these are just the values $y_{i}$ for which $y_{i} \cdot  r =  0$. Note that the subspaces correspond to quantum disjunctions. So determining the period of the function by Simon's algorithm amounts to determining which disjunction out of the seven alternative disjunctions is true, i.e., which subspace contains the state, without determining the truth values of the disjuncts.

\subsubsection{Shor's Algorithm\index{Shor's algorithm}}

Shor's factorization algorithm exploits the fact that the two prime factors $p, q$ of a positive integer $N = pq$ can be found by determining the period of a function 
$f(x) = a^{x}\mbox{ mod $N$}$, for any $a < N$  which is coprime to $N$, i.e., has no common factors with $N$ other than 1.  The period $r$ of $f(x)$ depends on $a$ and $N$. Once we know the period, we can factor $N$ if $r$ is even and $a^{r/2} \neq -1 \mbox{ mod $N$}$, which will be the case with probability greater than 1/2 if $a$ is chosen randomly. (If not, we choose another value of $a$.) The factors of $N$ are the greatest common factors of $a^{r/2} \pm 1$ and $N$, which can be found in polynomial time by the Euclidean algorithm. (For these number-theoretic results, see \cite[Appendix 4]{NielsenChuang}.) So the problem of factorizing a composite integer $N$ that is the product of two primes reduces to the problem of finding the period of a certain periodic function $f: Z_{s} \rightarrow Z_{N}$, where $Z_{n}$ is the additive group of integers mod $n$ (rather than $B^{n}$, the $n$-fold Cartesian product of a Boolean algebra $B$, as in Simon's algorithm). Note that $f(x+r) = f(x)$ if $x+r \leq s$. The function $f$ is periodic if $r$ divides $s$ exactly, otherwise it is almost periodic. 

Consider first the general form of the algorithm, as it is usually formulated. We begin by initializing the input register ($s$ qubits) to the state $\ket{0} \in \hil{H}^{s}$ and the output register ($N$ qubits) to the state $\ket{0} \in \hil{H}^{N}$. An $s$-fold Hadamard transformation is applied to the input register, followed by the unitary transformation $U_{f}$ which implements the function $f(x) = a^{x}\mbox{ mod $N$}$:
\begin{eqnarray}
\ket{0}\ket{0} & \stackrel{H}{\longrightarrow} & \frac{1}{\sqrt{s}}\sum_{x=0}^{s-1}\ket{x}\ket{0} \\
& \stackrel{U_{f}}{\longrightarrow} & \frac{1}{\sqrt{s}}\sum_{x=0}^{s-1}\ket{x}\ket{0} \\
& & = \frac{1}{\sqrt{s}}\sum_{x=0}^{s-1}\ket{x}\ket{x + a^{x} \mbox{ mod $N$}} 
\end{eqnarray}
Then we measure the output register in the computational basis\footnote{As in the discussion of Simon's algorithm, this measurement is purely hypothetical, introduced to simplify the analysis. Only the input register is actually measured.}  and obtain a state of the following form for the input register:
\begin{equation}
\frac{1}{\sqrt{s/r}}\sum_{j=0}^{s/r - 1} \ket{x_{i} + jr} \label{eqn:inputstate}
\end{equation}
This will be the case if $r$ divides $s$ exactly. The value $x_{i}$ is the offset, which depends on the outcome $i$ of the measurement of the output register. The sum is taken over the values of $j$ for which $f(x_{i} + jr) = i$. When $r$ does not divide $s$ exactly, the analysis is a little more complicated. 
For a discussion, see \cite[p. 164]{Barenco}, \cite{Jozsa97b}. Since the state label contains the random offset, a direct measurement of the label yields no information about the period. 

A discrete Fourier transform\index{discrete Fourier transform} for the integers mod $s$ is now applied to the input register, i.e., a unitary transformation:
\begin{equation}
\ket{x} \stackrel{U_{DFT_{s}}}{\longrightarrow} \frac{1}{\sqrt{s}}\sum_{y=0}^{s-1}e^{2\pi i \frac{xy}{s}}\ket{y}, \mbox{ for $x \in Z_{s}$} \label{eqn:Fourier}
\end{equation}
This yields the transition:
\begin{equation}
\frac{1}{\sqrt{\frac{s}{r}}}\sum_{j=0}^{\frac{s}{r} - 1} \ket{x_{i} + jr} \stackrel{U_{DFT_{s}}}{\longrightarrow}
\frac{1}{\sqrt{r}}\sum_{k=0}^{r-1}e^{2\pi i \frac{x_{i}k}{r}}\ket{ks/r} \label{eqn:Fourier2}
\end{equation}
and so shifts the offset into a phase factor and inverts the period as a multiple of $s/r$. A measurement of the input register in the computational basis yields $c = ks/r$. The algorithm is run a number of times until  a value of $k$ coprime to $r$ is obtained. Cancelling $c/s$ to lowest terms then yields $k$ and $r$ as $k/r$.

Since we don't know the value of $r$ in advance of applying the algorithm, we do not, of course, recognize when a measurement outcome yields a value of $k$ coprime to $r$. The idea is to run the algorithm, cancel $c/s$ to lowest terms to obtain a candidate value for $r$ and hence a candidate factor of $N$, which can then be tested  by division into $N$. Even when we do obtain a value of $k$ coprime to $r$, some values of $a$ will yield a period for which the method fails to yield a factor of $N$, in which case we randomly choose a new value of $a$ and run the algorithm with this value. The point is that all these steps are efficient, i.e., can be performed in polynomial time, and since only a polynomial number of repetitions are required to determine a factor with any given probability $p < 1$, the algorithm is a polynomial-time algorithm, achieving an exponential speed-up over any known classical algorithm. 

To see how the algorithm functions geometrically, consider the case $N= 15, a = 7$ and $s = 64$ discussed in \cite[p. 160]{Barenco}. In this case, the function $f(x) = a^{x}\mbox{ mod $15$}$ is:
\begin{eqnarray*}
7^{0} \mbox{ mod $15$} & = & 1\\
7^{1} \mbox{ mod $15$} & = & 7\\
7^{2} \mbox{ mod $15$} & = & 4\\
7^{3} \mbox{ mod $15$} & = & 13\\
7^{4} \mbox{ mod $15$} & = & 1\\
\vdots
\end{eqnarray*}
and the period is evidently $r=4$.\footnote{The factors 3 and 5 of 15 are derived as the greatest common factors of $a^{r/2} - 1 = 48$ and 15  and $a^{r/2} + 1 = 50$ and 15, respectively.} After the application of the unitary transformation $U_{f} = a^{x}\mbox{ mod $N$}$, the state of the two registers is:
\begin{eqnarray}
& \frac{1}{8}(\ket{0}\ket{1} + \ket{1}\ket{7}  +  \ket{2}\ket{4} + \ket{3}\ket{13}  \nonumber \\
& \mbox{} + \ket{4}\ket{1} + \ket{5}\ket{7}  +  \ket{6}\ket{4} + \ket{7}\ket{13} \nonumber \\
& \vdots \nonumber \\
& \mbox{} + \ket{60}\ket{1} + \ket{61}\ket{7}  +  \ket{62}\ket{4} + \ket{63}\ket{13})
\end{eqnarray}
This state can be expressed as:
\begin{eqnarray}
& \frac{1}{4}(\ket{0} + \ket{4} + \ket{8} + \ldots + \ket{60})\ket{1} \nonumber \\
& + \frac{1}{4}(\ket{1} + \ket{5} + \ket{9} + \ldots + \ket{61})\ket{7} \nonumber \\
& + \frac{1}{4}(\ket{2} + \ket{6} + \ket{10} + \ldots + \ket{62})\ket{4} \nonumber \\
& + \frac{1}{4}(\ket{3} + \ket{7} + \ket{11} + \ldots + \ket{63})\ket{13})
\end{eqnarray}

If we measure the output register, we obtain (equiprobably) one of four states for the input register, depending on the outcome of the measurement: 1, 7, 4, or 13:
\begin{eqnarray}
& \frac{1}{4}(\ket{0} + \ket{4} + \ket{8} + \ldots + \ket{60}) \\
& \frac{1}{4}(\ket{1} + \ket{5} + \ket{9} + \ldots + \ket{61}) \\
& \frac{1}{4}(\ket{2} + \ket{6} + \ket{10} + \ldots + \ket{62})  \\
& \frac{1}{4}(\ket{3} + \ket{7} + \ket{11} + \ldots + \ket{63}) 
\end{eqnarray}
These are the states (\ref{eqn:inputstate}) for values of the offset 0, 1, 2, 3. Application of the discrete Fourier transform yields:
\begin{description} 
\item{$x_{1} = 0:$}  \mbox{ }$\frac{1}{2}(\ket{0} + \ket{16} + \ket{32}+\ket{48})$
\item{$x_{7} = 1:$} \mbox{ }$\frac{1}{2}(\ket{0} + i\ket{16} - \ket{32} - i\ket{48})$
\item{$x_{4} = 2:$} \mbox{ }$\frac{1}{2}(\ket{0} - \ket{16} + \ket{32} - \ket{48})$
\item{$x_{13} = 3:$} \mbox{ }$\frac{1}{2}(\ket{0} - i\ket{16} - \ket{32} + i\ket{48})$
\end{description}
which are the states in (\ref{eqn:Fourier2}). So for the period $r = 4$, the state of the input register ends up in the 4-dimensional subspace spanned by the vectors $\ket{0}, \ket{16}, \ket{32}, \ket{48}$. 

Now consider all possible even periods $r$ for which $f(x) = a^{x} \mbox{ mod $15$}$, where $a$ is coprime to $15$. The other possible values of $a$  are 2, 4, 8, 11, 13, 14 and the corresponding periods turn out to be 4, 2, 4, 2, 4, 2. So we need only consider $r = 2$. \footnote{Every value of $a$ except $a = 14$ yields the correct factors for 15. For $a = 14$, the method fails: $r = 2$, so $a^{\frac{r}{2}} = -1 \mbox{ mod $15$}$.}

For $r = 2$, if we measure the output register, we will obtain (equiprobably) one of two states for the input register, depending on the outcome of the measurement (say, $a$ or $b$):
\begin{eqnarray}
& \ket{0} + \ket{2} + \ket{4} + \ldots + \ket{62} \\
& \ket{1} + \ket{3} + \ket{5} + \ldots + \ket{63}  
\end{eqnarray}
After the discrete Fourier transform, these states are transformed to:
\begin{description}
\item{$x_{a} = 0:$}  \mbox{ }$\ket{0} + \ket{32}$
\item{$x_{b} = 1:$} \mbox{ }$\ket{0}  - \ket{32}$
\end{description}

In this case, the 2-dimensional subspace $\hil{V}_{r=2}$ spanned by $\ket{0}, \ket{32}$ for $r = 2$ is included in the 4-dimensional subspace $\hil{V}_{r=4}$ for $r = 4$. A measurement can distinguish $r = 4$ from $r = 2$ reliably, i.e., whether the final state of the input register is in $\hil{V}_{r=4}$ or $\hil{V}_{r=2}$, only if the final state is in $\hil{V}_{r=4} - \hil{V}_{r=2}$, the part of $\hil{V}_{r=4}$ orthogonal to 
$\hil{V}_{r=2}$. What happens if the final state ends up in $\hil{V}_{r=2}$?

Shor's algorithm works as a randomized algorithm\index{randomized algorithm}. As mentioned above, it produces a candidate value for the period $r$ and hence a candidate factor of $N$, which can be tested (in polynomial time) by division into $N$. A measurement of the input register in the computational basis yields an outcome $c = ks/r$. The value of $k$ is chosen equiprobably by the measurement of the output register. The procedure is to repeat the algorithm  until the outcome yields a value of $k$ coprime to $r$, in which case canceling $c/s$ to lowest terms yields $k$ and $r$ as $k/r$.

For example, suppose we choose $a = 7$, in which case (unknown to us) $r = 4$. The values of $k$ coprime to $r$ are $k = 1$ and  $k = 3$ (this is  also unknown to us, because $k$ depends on the value of $r$). Then $c/s$ cancelled to lowest terms is $1/4$ and $3/4$, respectively, both of which yield the correct period. From the geometrical perspective, these values of $k$ correspond to finding  the state after measurement in the computational basis to be $\ket{16}$ or $\ket{48}$, both of which do distinguish $\hil{V}_{r=4}$ from $\hil{V}_{r=2}$. 

Suppose we choose a value of $a$ with period $r = 2$ and find the value $c = 32$. The only value of $k$ coprime to $r$ is $k = 1$. Then $c/s$ cancelled to lowest terms is $1/2$, which yields the correct period, and hence the correct factors of $N$. But  $c = 32$ could also be obtained for $a = 7$, $r = 4$, and $k = 2$, which does not yield the correct period, and hence does not yield the correct factors of $N$. Putting it geometrically: the value $k = 1$ for $r = 2$ corresponds to the same state, $\ket{32}$, as the value $k = 2$ for $r = 4$. Once we obtain the candidate period $r = 2$ (by cancelling $c/s = 32/64$ to lowest terms), we calculate the factors of $N$ as the greatest common factors of $a \pm 1$ and $N$ and test these by division into $N$. If $a = 7$, these calculated factors will be incorrect. If $a = 2$, say, the factors calculated in this way will be correct. 

We see that, with the added information provided by the outcome of a test division of a candidate factor into $N$, Shor's randomized algorithm again amounts to determining which disjunction among alternative disjunctions is true, i.e., which subspace contains the state, without determining the truth values of the disjuncts.

\subsection{Where Does the Speed-Up Come From?}

What, precisely, is the feature of a quantum computer responsible for the phenomenal efficiency over a classical computer? In the case of Simon's algorithm, the speed-up is exponential over any classical algorithm; in the case of Shor's algorithm, the speed-up is exponential over any known classical algorithm. 

Steane\index{Steane} \shortcite{Steane98} remarks: 
\begin{quotation}
The period finding algorithm appears at first sight like a conjuring trick: it is not quite clear how the quantum computer managed to produce the period like a rabbit out of a hat. \ldots I would say that the most important features are contained in [$\ket{\psi} = \frac{1}{s}\sum_{x=0}^{s-1}\ket{x}\ket{f(x)}$]. They are not only the \emph{quantum parallelism} already mentioned, but also \emph{quantum entanglement}, and, finally, quantum interference. Each value of $f(x)$ retains a link with the value of $x$ which produced it, through entanglement of the $x$ and $y$ registers in [$\ket{\psi}$]. The `magic' happens when a measurement of the $y$ register produces the special state [$\frac{1}{s/r}\sum_{j=0}^{s/r-1}\ket{x_{i} + jr}$] in the $x$-register, and it is quantum entanglement which permits this (see also \cite{Jozsa97a}). The final Fourier transform can be regarded as an interference between the various superposed states in the $x$-register (compare with the action of a diffraction grating).

Interference effects can be used for computational purposes with classical light fields, or water waves for that matter, so interference is not in itself the essentially quantum feature. Rather, the exponentially large number of interfering states, and the entanglement, are features which do not arise in classical systems.
\end{quotation}

Jozsa\index{Jozsa} points out \shortcite{Jozsa97a} that the state space (phase space) of a composite classical system is the Cartesian product of the state spaces of its subsystems, while the state space of a composite quantum system is the tensor product of the state spaces of its subsystems. For $n$ qubits, the quantum state space has $2^{n}$ dimensions. So the information required to represent a general state increases exponentially with $n$: even if we restrict the specification of the amplitudes to numbers of finite precision, a superposition will in general have  $\mathcal{O}(2^{n})$ components. For a classical composite system of $n$ two-level subsystems,  the number of possible states grows exponentially with $n$, but the information required to represent a general state is just $n$ times the information required to represent a single two-level system, i.e., the information grows only linearly with $n$ because the state of a composite system is just a product state. 

More formally, Jozsa\index{Jozsa} and Linden\index{Linden} \shortcite{JozsaLinden} have shown  that a quantum algorithm operating on pure states can  achieve an exponential speed-up over  classical algorithms only  if the quantum algorithm involves multi-partite entanglement that increases unboundedly with the input size. Similarly, Vidal\index{Vidal} \shortcite {Vidal} has shown that a classical computer can simulate the evolution of a pure state of $n$ qubits with computational resources that grow linearly with $n$ and exponentially in multi-partite entanglement. 

The essential feature of the quantum computations discussed above in \S \ref{sec:algorithms} is the selection of a disjunction, representing a global property of a function, among alternative possible disjunctions without computing the truth values of the disjuncts, which is  redundant information in a quantum computation but essential information classically. Note that a quantum disjunction is represented by a subspace of entangled states in the tensor product Hilbert space of the input and output registers. This is analogous to the procedure involved in the key observation underlying  the proof of the quantum bit commitment theorem discussed in \S \ref{sec:bittheoremidea}. The series of operations described by equations (\ref{eq:bc1})--(\ref{eq:V}), in which the channel particle is entangled with ancilla systems and the ancillas are subsequently measured, effectively constitute a quantum computaton.

The first stage of a quantum algorithm involves the creation of a state in which every input value to the function is correlated with a corresponding output value. This is referred to as `quantum parallelism'\index{quantum parallelism@`quantum parallelism'} and is sometimes cited as the source of the speed-up in a quantum computation. The idea is that a quantum computation is something like a massively parallel classical computation, for all possible values of a function. This appears to be Deutsch's\index{Deutsch} view, with the parallel computations taking place in parallel 
universes\index{parallel universes}. For a critique, see \cite{Steane03}\index{Steane}, who defends a view similar to that presented here. Of course, all these different values are inaccessible: a measurement in the computational basis will only yield (randomly) one correlated input-output pair. Further processing is required, including the final discrete Fourier transform for the three algorithms discussed in \S \ref{sec:algorithms}. It would be incorrect to attribute the efficiency of these quantum algorithms to the interference in the input register produced by  the Fourier transform. The role of the Fourier transform\index{discrete Fourier transform} is simply to allow a measurement in the computational basis to reveal which subspace  representing the target disjunction contains the state. 

One might wonder, then, why the discrete Fourier transform\index{discrete Fourier transform} is even necessary. We could, of course, simply perform an equivalent measurement in a different basis. But note that a computation would have to be performed to determine this basis. This raises the question of precisely how to assess the speed-up of a quantum algorithm relative to a rival classical algorithm. What are the relevant computational steps to be counted in making this assessment for a quantum computation? Since any sequence of unitary transformations is equivalent to a single unitary transformation, and a unitary transformation followed by a measurement in a certain basis is equivalent to simply performing a measurement in a different basis, any quantum computation can always be reduced to just one step: a measurement in a particular basis! 

Of course, this observation is hardly illuminating, since a computation at least as difficult as the original computation would have to be performed to determine the required basis, but it does indicate that some convention is required about what steps to count in a quantum computation. The accepted convention is to require the unitary transformations in a quantum computation to be constructed from elementary quantum gates that form a universal set (e.g., the CNOT gate\index{CNOT gate}, the Hadamard gate\index{Hadamard gate}, the phase gate\index{phase gate}, and the $\pi/8$ gate\index{pi@$\pi/8$ gate} discussed in \S \ref{sec:complexity}) and to count each such gate as one step. In addition, all measurements are required to be performed in the computational basis, and these are counted as additional steps. The final discrete Fourier transforms\index{discrete Fourier transform} in the Deutsch-Jozsa algorithm, Simon's algorithm, and Shor's algorithm are indispensable in transforming the state so that the algorithms can be completed by measurements in the computational basis, and it is an important feature of these algorithms that the Fourier transform can be implemented efficiently with elementary unitary gates. To claim that a quantum algorithm is exponentially faster than a  classical algorithm is to claim that the number of steps counted in this way for the quantum algorithm is a polynomial function of the size of the input (the number of qubits required to store the input), while the classical algorithm involves a number of steps that increases exponentially with the size of the input (the number of bits required to store the input).

\section{Quantum Foundations from the Perspective of Quantum Information}

\label{sec:foundations}
 
Does the extension of the classical theory of information to quantum states shed new light on the foundational problems of quantum mechanics underlying the Bohr-Einstein debate mentioned in \S 1, in particular the measurement problem? Researchers in the area of quantum information and quantum computation often suggest a positive answer to this question, with a promissory note for how the story is supposed to go. More fully worked-out (generally, rather different) positive responses have been proposed by various authors, notably Fuchs\index{Fuchs} \shortcite{FuchsInfo1,FuchsInfo2,FuchsInfo3,FuchsInfo4} and Brukner and Zeilinger \shortcite{BZ2001,BZ2002}. For a very thorough analysis and critique of the Brukner-Zeilinger position\index{Brukner}\index{Zeilinger}, see \cite{Timpson}\index{Timpson}. See also Hall \shortcite{Hall} and the response by Brukner and Zeilinger \shortcite{BZ2000}. Here I shall limit my discussion to the significance of a characterization theorem for quantum mechanics in terms of information-theoretic constraints by Clifton\index{Clifton}, Bub\index{Bub}, and Halvorson\index{Halvorson} (CBH)\index{CBH theorem} \shortcite{CBH}.

\subsection{The CBH Characterization Theorem\index{CBH theorem}}

CBH showed that one can derive the basic 
kinematic features of a quantum
 description of physical systems
 from three fundamental information-theoretic 
 constraints: 
 \begin{itemize}
\item the impossibility of superluminal information transfer 
 between two physical systems by performing measurements on one of 
 them, 
 \item the impossibility of perfectly broadcasting the information 
 contained in an unknown physical state (which, for pure states, amounts 
 to `no cloning'),
 \item the impossibility of communicating information 
so as to implement a bit commitment protocol
  with unconditional security (so that cheating is in principle excluded by the theory).
  \end{itemize}  
  
More precisely,  CBH formulate these information-theoretic constraints  in the general framework of $C^{*}$-algebras\index{cstar@$C^{*}$-algebra}, which  allows
 a mathematically
 abstract characterization of a physical theory that includes, as
 special cases, all classical mechanical theories of both wave and
 particle varieties, and all variations on quantum theory, including
 quantum field theories (plus any hybrids of these theories, such as 
 theories with superselection rules). Within this framework, CBH  show that the three information-theoretic constraints jointly entail three physical conditions that they take as  definitive of what it means to be a quantum theory in the most general sense. Specifically, the information-theoretic constraints entail that:
 \begin{itemize}
 \item the algebras of observables pertaining to
 distinct physical systems commute (a condition usually called
microcausality
 or, to use Summers' term \cite{Summers}, \emph{kinematic
 independence\index{kinematic
 independence}}),
 \item any individual
 system's algebra of observables is  \emph{noncommutative},
 \item the physical world is \emph{nonlocal}, in that
 spacelike separated systems  can occupy entangled
states that persist as the systems separate.
 \end{itemize}

CBH also partly demonstrated the 
  converse derivation, leaving open a question concerning nonlocality 
  and  bit commitment. This remaining
 issue was later resolved by Hans Halvorson\index{Halvorson} \shortcite{Halvorson1}, so the CBH theorem is a 
  characterization theorem for quantum theory in terms of the three 
  information-theoretic constraints.

Note that the $C^{*}$-algebraic framework is not restricted to the
 standard quantum mechanics of a system represented on a single
 Hilbert space with a unitary dynamics, but is general 
 enough to cover
 cases of systems with an infinite number of degrees of freedom
  that arise in quantum field theory and the thermodynamic limit
 of quantum statistical mechanics (in which the number of
 microsystems and the volume they occupy goes to infinity, while the
 density defined by their ratio remains constant). The $C^{*}$-algebraic framework has even been applied to quantum field theory on curved spacetime and so is applicable to the  quantum theoretical
 description of exotic phenomena
 such as Hawking radiation (black hole evaporation); see \cite{Wald}. The Stone-von Neumann theorem, 
 which guarantees the existence of a
 unique representation (up to unitary equivalence) of the canonical
 commutation relations for systems with a finite number of degrees of
 freedom, breaks down for such cases, and there will
 be many unitarily inequivalent
 representations of the canonical commutation relations.

 One could, of course, consider weaker 
mathematical structures, but it seems that the $C^{*}$-algebraic 
machinery suffices for all physical theories that have been found to 
be empirically successful to date, including phase space theories and 
Hilbert space theories \cite{Landsmann}, and theories based on a 
manifold \cite{Connes}. For further discussion of this point, see Halvorson and Bub \shortcite{HalvorsonBub}. See also Halvorson (this vol., chap. 8), Emch (this vol., ch. 8), and Landsman (this vol., ch. 5).

A $C^{*}$-algebra\index{cstar@$C^{*}$-algebra} 
is essentially an abstract generalization of the structure of the 
algebra of operators on a Hilbert space. Technically, a (unital) 
$C^{*}$-algebra is a Banach $^{*}$-algebra over the 
complex numbers containing the identity, where the involution 
operation $^{*}$ and the norm are related by $\norm{A^{*}A} = \norm{A}^{2}$. 
So the algebra $\alg{B}(\hil{H})$ of all bounded operators on a 
Hilbert space 
$\hil{H}$ is a $C^{*}$-algebra, 
with $^{*}$  
the adjoint operation and $\norm{\cdot}$ the 
standard operator norm. 

In standard quantum theory, as discussed in \S \ref{sec:entanglement}, a state on $\alg{B}(\hil{H})$ is
defined by
a density operator $\rho$ on $\hil{H}$ in terms of an expectation-valued
functional
$\tilde{\rho}(A)=\trace{\rho A}$ for all observables represented by
self-adjoint operators $A$ in $\alg{B}(\hil{H})$. This definition of $\tilde{\rho}(A)$
in terms of $\rho$
yields a positive normalized linear functional. So a
  state on a $C^{*}$-algebra $\alg{C}$ is defined, quite generally, as
any positive normalized
linear functional $\tilde{\rho}:\alg{C}\rightarrow\mathbb{C}$ on the algebra. Pure states can be defined by the condition that if
$\tilde{\rho}=\lambda\tilde{\rho}_{1}+(1-\lambda)\tilde{\rho}_{2}$ with $\lambda\in (0,1)$,
then
$\tilde{\rho}=\tilde{\rho}_{1}=\tilde{\rho}_{2}$;  other states are mixed. In the following, we drop the `\~{}'  in $\tilde{\rho}$, but note that a $C^{*}$-algebraic state $\rho$  is a positive linear functional on $\alg{C}$, while the density operator of standard quantum mechanics is an element of $\alg{C} = \alg{B}(\hil{H})$.

By Gleason's theorem\index{Gleason's theorem} \cite{Gleason}, every $C^{*}$-algebraic state in this sense on a $C^{*}$-algebra $\alg{C} = \alg{B}(\hil{H})$ is given by a density operator on $\alg{B}(\hil{H})$. However, because countable additivity is not
  presupposed by the $C^{*}$-algebraic notion of state (and,
  therefore, Gleason's theorem does not apply in general),
  there can be pure states of $\alg{B}(\hil{H})$ that  are not
 representable by vectors in $\hil{H}$.
 In fact,
 if $A$ is any self-adjoint element of a
$C^{*}$-algebra $\alg{A}$, and $a\in\mbox{sp}(A)$,
then there always exists a pure state $\rho$ of
$\alg{A}$ that assigns a \emph{dispersion-free}\index{dispersion-free} value of $a$ to $A$
\cite[Ex. 4.6.31]{KadisonRingrose}.
Since this is true even when we
consider a
point in the
continuous spectrum of a self-adjoint operator $A$ acting on a
Hilbert
space, \emph{without} any corresponding eigenvector,
it follows that there \emph{are} pure states of $\alg{B}(\hil{H})$ in
the $C^{*}$-algebraic sense
that cannot be vector states (nor, in fact, representable by any
density
operator
$\hil{H}$).

As we saw in \S \ref{sec:operations}, the general evolution of a quantum system resulting from a combination of unitary interactions and selective or nonselective measurements can be described by a quantum operation, i.e., a completely positive linear map. Accordingly, a
completely positive linear map $T: \alg{C} \rightarrow \alg{C}$,
where $0\leq T(I)\leq I$ is taken as describing the general evolution of a system
represented by a $C^{*}$-algebra of observables. The map or operation $T$ is called selective
if $T(I)<I$
and nonselective if $T(I)=I$. Recall that a yes-no measurement of some idempotent
observable represented by a projection operator $P$ is an example of a
selective operation. Here,  $T(A)=PAP$
for all $A$ in the $C^{*}$-algebra $\alg{C}$,
and $\rho^{T}$, the transformed (`collapsed') state, is the final state
obtained after measuring $P$ in the
state $\rho$
  and ignoring all elements of the ensemble that do not
yield the eigenvalue 1 of $P$
(so $\rho^{T}(A)=\rho(T(A))/\rho(T(I))$ when $\rho(T(I))\not=0$, and
$\rho^{T}=0$ otherwise). The time evolution in the Heisenberg picture
induced by a unitary operator $U\in\alg{C}$ is an example of a
nonselective operation. Here, $T(A)=UAU^{-1}$. Similarly, the
measurement of an observable $O$ with spectral measure $\{P_{i}\}$,
without selecting a particular outcome, is an example of a
nonselective operation, with $T(A) = \sum _{i=1}^{n}P_{i}AP_{i}$. As in the standard quantum theory of a system with a finite-dimensional Hilbert space (cf. Eq. (\ref{eq:extended}) of \S \ref{sec:operations}), any completely positive linear map can be regarded as the 
restriction to a local 
system of a unitary map on a larger system.

A 
representation\index{cstar@$C^{*}$-algebra!representation} of a $C^{*}$-algebra $\alg{C}$ is any mapping 
$\pi:\alg{C}\rightarrow\alg{B}(\hil{H})$ that preserves the linear, 
product, and $^{*}$ structure of \alg{C}. The representation is faithful
if $\pi$ is 
one-to-one, in which case $\pi(\alg{C})$ is an isomorphic copy of 
$\alg{C}$. The Gelfand-Naimark theorem\index{Gelfand-Naimark theorem} says that every abstract $C^{*}$-algebra 
has a concrete 
faithful representation as a norm-closed $^{*}$-subalgebra of 
$\alg{B}(\hil{H})$, for some appropriate Hilbert space $\hil{H}$. 
As indicated above, in the case of systems with an 
infinite number of degrees of freedom  (e.g., quantum field theory), 
there are inequivalent representations of the 
$C^{*}$-algebra of observables defined by the commutation relations.

Every classical phase space theory 
defines a commutative $C^{*}$-algebra\index{cstar@$C^{*}$-algebra!commutative}. For example, the observables of a classical 
system of $n$ particles---the real-valued continuous functions on the phase space 
$\mathbb{R}^{6n}$---can be represented as the self-adjoint elements of 
the $C^{*}$-algebra $\alg{B}(\mathbb{R}^{6n})$ of all continuous complex-valued 
functions $f$ on $\mathbb{R}^{6n}$. The phase space 
$\mathbb{R}^{6n}$ is locally compact and can be made compact by 
adding just one point `at infinity,' or we can simply consider a 
bounded (and thus compact) subset of $\mathbb{R}^{6n}$. 
The statistical states of 
the system are given by probability 
measures $\mu$ on 
$\mathbb{R}^{6n}$, and pure states, corresponding to maximally complete 
information about the particles, are given by
the individual points of 
$\mathbb{R}^{6n}$. The system's state $\rho$ in 
the $C^{*}$-algebraic sense is the 
expectation functional corresponding to $\mu$, defined by
$\rho(f)=\int_{\mathbb{R}^{6n}}f\mbox{d}\mu$. 
Conversely \cite[Thm. 4.4.3]{KadisonRingrose}, every  
\textit{commutative} $C^{*}$-algebra $\alg{C}$ is isomorphic to the set 
$C(X)$ of 
all continuous complex-valued functions on a locally 
compact Hausdorff space 
$X$ defined by the pure states of $\alg{C}$.  If $\alg{C}$ has a multiplicative identity, the `phase space' $X$ is compact. In this `function representation' of $C$, the isomorphism
maps
an element $C\in\alg{C}$ to the function $\hat{C}$ (the Gelfand
transformation of $C$) whose value at any
$\rho$ is just the (dispersion-free) value that
$\rho$ assigns to $C$.    
So `behind' every abstract commutative $C^{*}$-algebra there is a 
classical phase space theory defined by its function representation
on the phase space $X$. This representation theorem (and its converse) justifies characterizing a $C^{*}$-algebraic theory as classical
just in case its algebra is commutative. 

As we saw above, CBH identify quantum theories with a certain 
subclass of noncommutative $C^{*}$-algebras\index{cstar@$C^{*}$-algebra!noncommutative}, where the condition of kinematic independence is satisfied by the algebras of observables of distinct systems and the states of spacelike separated systems are characterized by the sort of nonlocality associated with entanglement. 

To clarify the rationale for this characterization and the significance of the information-theoretic 
  constraints, consider a 
composite quantum system $AB$, consisting of two subsystems, $A$ and $B$. 
For simplicity, assume the systems are indistinguishable, so their 
$C^{*}$-algebras $\alg{A}$ and $\alg{B}$ are isomorphic. 
The 
observables of the component systems $A$ and $B$ are 
represented by the self-adjoint elements of $\alg{A}$ and $\alg{B}$, 
respectively. Let  $\alg{A}\vee\alg{B}$ denote the $C^{*}$-algebra 
generated by $\alg{A}$ and $\alg{B}$. The physical states of $A$, $B$, and 
$AB$, are given by positive normalized linear functionals on their 
respective algebras that encode the expectation values of all 
observables.
To capture the idea that $A$ and $B$ are \textit{physically 
distinct}\index{physically 
distinct} systems, 
CBH make the assumption that any state of $\alg{A}$ is
compatible with any state of $\alg{B}$, i.e., for any state
$\rho _{A}$ of $\alg{A}$ and $\rho _{B}$ of $\alg{B}$,
there is a state $\rho$ of $\alg{A}\vee \alg{B}$ such that $\rho
|_{\alg{A}}=\rho _{A}$ and $\rho |_{\alg{B}}=\rho _{B}$.

The sense of the `no superluminal information transfer via measurement' 
constraint 
is 
that when Alice and Bob, say, perform local measurements, Alice's 
measurements can have no influence on the statistics for the outcomes 
of Bob's measurements, and conversely. That is, merely performing
 a local measurement cannot, in and of 
itself, convey any information to a physically distinct system, so 
that everything `looks the same' to that system after the 
measurement operation as before, in terms of the expectation values 
for the outcomes of measurements. CBH show \shortcite[Thm. 1]{CBH} that it 
follows from this constraint that 
A and B are 
\textit{kinematically independent}\index{kinematic independence} systems if they are physically 
distinct in the above sense, i.e., every element of $\alg{A}$ 
commutes pairwise with every element of $\alg{B}$. (More precisely, an operation $T$ on $\alg{A}\vee \alg{B}$ conveys no
information to
  Bob just in case $(T^{*}\rho )|_{\alg{B}}=\rho |_{\alg{B}}$ for all
  states $\rho$ of $\alg{B}$, where $T^{*}$ is the map on the states, i.e., the positive linear functionals on  $\alg{A}\vee \alg{B}$, induced by T.  Clearly, the kinematic independence of $\alg{A}$ and $\alg{B}$
entails that Alice's local measurement operations cannot convey
any
information to Bob, i.e., $T(B)= 
\sum_{i=1}^{n}E_{i}^{1/2}BE_{i}^{1/2}= B$
 for $B\in \alg{B}$ if $T$ is implemented by a POVM in \alg{A}. CBH prove that if Alice cannot convey any information
to Bob by performing local measurement operations, then $\alg{A}$
and $\alg{B}$ are kinematically independent.)

The `no broadcasting' condition now ensures that the individual algebras  
$\alg{A}$ and  $\alg{B}$ are noncommutative. Recall that for pure states, broadcasting 
reduces to cloning, and that in elementary quantum 
mechanics, neither cloning nor broadcasting is possible in general (see section \ref{sec:Neumann}). 
CBH show that broadcasting and cloning are always possible for 
classical systems, i.e., in the commutative case there \emph{is} a 
universal broadcasting map that clones 
any pair of input pure states and broadcasts any pair of input mixed 
states \cite[Thm. 2]{CBH}. Conversely, they show that if any two states can be (perfectly) 
broadcast, 
then any two 
pure states can be cloned; and if two pure states of a $C^{*}$-algebra 
can be cloned, then they must be orthogonal. So, if any 
two states can be broadcast, then all pure states are orthogonal, 
from which it follows that the algebra is commutative.

The quantum mechanical phenomenon of interference  is the 
physical manifestation of the noncommutativity of quantum observables 
or, equivalently, the superposition 
of quantum states. 
So the impossibility of perfectly broadcasting the 
information contained in an unknown physical state, or of cloning or 
copying the information in an unknown pure state, is the 
information-theoretic counterpart of interference. 

Now, if $\alg{A}$ and $\alg{B}$ are noncommutative and mutually commuting, 
it can be shown that there are nonlocal entangled states on the
$C^{*}$-algebra
$\alg{A}\vee\alg{B}$ they generate (see \cite{Landau,Summers,Bacciagaluppi}, and---more relevantly here, in 
terms of a specification of the range of entangled states that can be 
guaranteed to exist---\cite{Halvorson1}). So it seems that 
entanglement\index{entanglement}---what Schr\"{o}dinger \shortcite[p. 555]{Schr1} identified as
 `\textit{the} characteristic trait of
quantum mechanics, the one that enforces its entire departure from
classical lines of thought,' as we saw in \S \ref{sec:teleportation}--- follows automatically in 
any theory with a noncommutative algebra of observables. That is, it 
seems that once we assume `no superluminal information transfer via 
measurement,' and `no broadcasting,' the class of allowable physical 
theories is restricted to those theories in which physical systems manifest 
both interference \textit{and} nonlocal entanglement. But in terms of physical interpretation this 
conclusion is a bit too quick, since the derivation of entangled 
states depends on formal
properties of the
$C^{*}$-algebraic machinery. Moreover, we have no assurance that two systems in an entangled state will maintain their entanglement indefinitely as they separate in space, which is the case for quantum entanglement. But this is precisely what is required by the cheating strategy that thwarts secure bit commitment, since Alice will have to keep one system of such a pair and send the other system to Bob, whose degree of spatial separation from Alice is irrelevant, in principle, to the implementation of the protocol. In an information-theoretic characterization of quantum theory, the fact that entangled states of composite systems can be instantiated, and instantiated nonlocally so that the entanglement of composite system is maintained as the subsystems separate in space, should be shown to follow from some information-theoretic principle. The role of the `no bit commitment' constraint  is to guarantee the persistence of entanglement over distance, i.e., the existence of a certain class of nonlocal entangled states---hence it gives us nonlocality, not merely `holism.'

As shown in \S \ref{sec:bit}, unconditionally secure quantum bit commitment\index{bit commitment!unconditionally secure} is impossible because  a generalized version of
the EPR cheating strategy\index{EPR! cheating strategy (attack)} can always be applied  by introducing additional
ancilla particles and enlarging the
Hilbert space  in a suitable way. That is,  for a quantum
mechanical system
consisting of two (separated) subsystems represented by
 the $C^{*}$-algebra
$\alg{B}(\hil{H}_{1}) \otimes \alg{B}(\hil{H}_{2})$, any mixture of
states on $\alg{B}(\hil{H}_{2})$ can be generated
from a distance by performing an appropriate generalized measurement on the
system represented by $\alg{B}(\hil{H}_{1})$, for an appropriate
entangled state of the composite system. 
This is what Schr\"{o}dinger\index{Schr\"{o}dinger}
 called `remote steering'\index{remote steering} and found so physically 
 counterintuitive that he speculated \shortcite[p. 451]{Schr2} 
 (wrongly, as it turned out) 
 that experimental 
 evidence would eventually show that this was simply an artefact 
 of the theory, not instantiated in our world. He suggested that an entangled state of a composite system would
almost instantaneously decay to a mixture as the component systems
separated.\footnote{A similar possibility was raised and rejected by Furry
\shortcite{Furry}.} There would still be correlations between the states of the
component systems, but remote steering would no longer be possible.
\begin{quote}
    It seems worth noticing that the [EPR] paradox could be avoided
by a
    very simple assumption, namely if the situation after separating
    were described by [the entangled state $\Psi(x,y) = \sum_{k}a_{k}g_{k}(x)f_{k}(y)$], 
    but with the additional
    statement that the knowledge of the \textit{phase relations}
    between the complex constants $a_{k}$ has been entirely lost in
    consequence of the process of separation. This would mean that
    not only the parts, but the whole system, would be in the
    situation of a mixture, not of a pure state. It would not preclude
    the possibility of determining the state of the first system by
    \textit{suitable} measurements in the second one or \textit{vice
    versa}. But it would utterly eliminate the experimenters influence
    on the state of that system which he does not touch.
\end{quote}
 
Schr\"{o}dinger regarded the phenomenon of
interference associated with noncommutativity
in quantum mechanics as unproblematic, because he saw this as
reflecting the fact that particles are
wavelike.
But he did not believe that we live
in a world in which physical systems can exist nonlocally in
entangled
states, because such states would allow Alice to steer Bob's system into any
mixture of pure states compatible with Bob's reduced density operator and he did not expect that
experiments would bear this out.  Of course, it was an experimental question in 1935 whether
Schr\"{o}dinger's conjecture was correct or not.
We now know that the conjecture is false. A
wealth of experimental evidence, including the 
confirmed violations of Bell's inequality\index{Bell's inequality} \cite{Aspect1,Aspect2} and the confirmations of quantum teleportation \cite{BPM+97,BBM+98,FSB+98,NKL98},
testify to this. The
relevance of Schr\"{o}dinger's conjecture here is this: it
raises the possibility of a quantum-like world in which there is
interference but
no nonlocal entanglement. Can we exclude this possibility on
information-theoretic grounds?

Now although unconditionally secure bit commitment is no less  impossible for
classical
systems, in which the algebras of observables are 
commutative, than for quantum systems, the insecurity of a bit commitment protocol
in a noncommutative setting depends on considerations entirely
different from those in a classical commutative setting. As we saw in \S \ref{sec:bit}, the security of a classical bit commitment protocol is a matter of computational complexity and cannot be unconditional. 

 By contrast, if, as Schr\"{o}dinger speculated,
 we lived in a world in which the
algebras of observables are noncommutative but composite physical systems
cannot exist in nonlocal entangled states,
 if Alice sends Bob one of two mixtures associated with the same
 density operator to establish her commitment, then she
is, in effect, sending Bob evidence for the
truth of an exclusive disjunction that is not based on the selection
of a particular disjunct. (Bob's reduced density operator
is associated ambiguously with both mixtures, and
hence with the truth of the exclusive disjunction: `0 or 1'.) 
Noncommutativity allows the possibility of different mixtures associated with
the same density operator. What thwarts the possibility of using the
ambiguity of mixtures in this way to implement an unconditionally
secure bit commitment\index{bit commitment!unconditionally secure} protocol is the existence of nonlocal entangled
states between Alice and Bob. This allows Alice to cheat by preparing
a suitable entangled state instead of one of the mixtures, where the
reduced density operator for Bob is the same as that of the mixture.
Alice is
then able to steer Bob's systems remotely into either of the two mixtures
associated with the alternative commitments at will. 

So what \textit{would} allow unconditionally secure bit commitment\index{bit commitment!unconditionally secure} in a
noncommutative theory is the absence of physically occupied
 nonlocal entangled states, or the spontaneous destruction of entanglement as systems separate.
One can therefore take Schr\"{o}dinger's remarks as relevant to the
question of whether or not secure bit
commitment is possible in our world. In effect, Schr\"{o}dinger\index{Schr\"{o}dinger}
raised the possibility
that we live in a quantum-like world in which  unconditionally secure bit commitment\index{bit commitment!unconditionally secure}
is possible! It follows that the 
impossibility of unconditionally secure bit commitment\index{bit commitment!unconditionally secure} entails that, 
for any mixed state that Alice and Bob can prepare by following some 
(bit commitment) protocol, there is a corresponding nonlocal 
entangled state that can be physically occupied by Alice's and Bob's 
particles and persists indefinitely as the particles move apart. 

To sum up: the content of the CBH theorem is that a quantum theory---a $C^{*}$-algebraic  theory
 whose observables and states satisfy conditions of kinematic independence, noncommutativity, and nonlocality---can be characterized by
three information-theoretic constraints: no superluminal communication 
of information via measurement, no (perfect) broadcasting, and no 
(unconditionally secure) bit commitment.

\subsection{Quantum Mechanics as a Theory of Information}

Consider Einstein's\index{Einstein} view,\protect \footnote{The following discussion is adapted from \cite{BubWhy} and \cite{BubCushing}, but the argument here is developed somewhat differently.} mentioned in \S 1, that quantum mechanics is incomplete. Essentially, Einstein based his argument for this claim on the demand that a complete physical theory should satisfy certain principles of realism (essentially, a locality principle and a separability principle), which amounts to the demand that statistical correlations between spatially separated systems should have a common causal explanation in terms of causal factors obtaining at the common origin of the systems. Roughly thirty years after the publication of the Einstein-Podolsky-Rosen\index{Einstein, Podolsky, and Rosen (EPR)} paper \shortcite{EPR}, John Bell\index{Bell} \shortcite{BellEPR} showed that the statistical correlations of the entangled Einstein-Podolsky-Rosen state for spatially separated particles are inconsistent with any explanation in terms of a classical probability distribution over common causal factors originating at the source of the particles before they separate. But the fact that quantum mechanics allows the possibility of correlations that are not reducible to common causes is a \emph{virtue} of the theory. It is precisely the nonclassical correlations of entangled states that underlie the possibility of an exponential speed-up of quantum computation over classical computation, the possibility of unconditionally secure key distribution but the impossibility of unconditionally secure quantum bit commitment, and phenomena such as quantum teleportation and other nonclassical entanglement-assisted communication protocols. 

While Einstein's argument for incompleteness fails, there is another sense, also associated with entangled states, in which quantum mechanics might be said to be incomplete. In a typical (idealized) quantum mechanical measurement interaction, say an interaction in which the two possible values, $0$ and $1$, of an observable of a qubit in a certain quantum state become correlated with the two possible positions of a macroscopic pointer observable, $p_{0}$ and $p_{1}$, the final state is an entangled state, a linear superposition of the states $\ket{0}\ket{p_{0}}$ and $\ket{1}\ket{p_{1}}$  with coefficients derived from the initial quantum state of the qubit. To dramatize the problem, Schr\"{o}dinger\index{Schr\"{o}dinger's cat} \shortcite{Schr2} considered the case where $\ket{p_{0}}$ and $\ket{p_{1}}$ represent the states of a cat being alive and a cat being dead in a closed box, which is only opened by the observer some time after the measurement interaction. On the standard way of relating the quantum state of a system to what propositions about the system are determinately (definitely) true or false, and what propositions have no determinate truth value, some correlational proposition about the composite system (microsystem $+$ cat) is true in this entangled state, but the propositions asserting that the cat is alive (and the value of the qubit observable is $0$), or that the cat is dead (and the value of the qubit observable is $1$), are assigned no determinate truth value. Moreover, if we assume that the quantum propositions form an algebraic structure isomorphic to the structure of subspaces of the Hilbert space of the composite system---the representational space for quantum states and observables---then it is easy to derive a formal contradiction from the assumption that the correlational proposition corresponding to the entangled state is true, and that the cat is either definitely alive or definitely dead.\footnote{This also follows from the Bub-Clifton theorem discussed below. The sublattice of determinate quantum propositions defined by the identity and the EPR state is maximal: adding any proposition involves a contradiction.} Schr\"{o}dinger thought that it was absurd to suppose that quantum mechanics requires us to say that the cat in such a situation (a macrosystem) is neither alive nor dead (does not have a determinate macroproperty of this sort) until an observer opens the box and looks, in which case the entangled state `collapses' nonlinearly and stochastically, with probabilities given by the initial quantum state of the microsystem, onto a product of terms representing a definite state of the cat and a definite state of the microsystem. Einstein\index{Einstein} \shortcite[p. 39]{Einstein3} concurred and remarked in a letter to Schr\"{o}dinger: `If that were so then physics could only claim the interest of shopkeepers and engineers; the whole thing would be a wretched bungle.' 

This is the standard `measurement problem' of quantum mechanics. Admittedly, the formulation of the problem is highly idealized, but the fundamental problem arises from the way in which quantum mechanics represents correlations via entangled states and does not disappear entirely in less idealized formulations (even though the problem is somewhat altered by considering the macroscopic nature of the instrument, and the r\^{o}le of the environment). (See Dickson, this vol., ch. 4, and \cite{Bubbook} for further discussion.) I shall refer to this problem---the Schr\"{o}dinger incompleteness of the theory---as Schr\"{o}dinger's problem\index{Schr\"{o}dinger's problem}. It is a \textit{problem about truth} (or the instantiation of properties), as opposed to a distinct \textit{problem about probabilities}. 

Before formulating the probability problem, consider what was involved in the transition from classical to quantum mechanics. Quantum mechanics first appeared as Heisenberg's matrix mechanics\index{Heisenberg's matrix mechanics}\index{Heisenberg} in 1925, following the `old quantum theory,' a patchwork of \textit{ad hoc} modifications of classical mechanics to accommodate Planck's quantum postulate. Essentially, Heisenberg modified the kinematics of classical mechanics by replacing certain classical dynamical variables, like position and momentum, with mathematical representatives---matrices---which do not commute. Shortly afterwards, Schr\"{o}dinger developed a wave mechanical version of quantum mechanics and proved the formal equivalence of the two theories. It is common to understand the significance of the transition from classical to quantum mechanics in terms of `wave-particle duality,' the idea that a quantum system like an electron, unlike a classical system like a stone, manifests itself as a wave under certain circumstances and as a particle under other circumstances. This picture obscures far more than it illuminates. We can see more clearly what is going on conceptually if we consider the implications of Heisenberg's move for the way we think about objects and their properties in the most general sense. 

Heisenberg\index{Heisenberg} replaced the commutative algebra of dynamical variables of classical mechanics---position, momentum, angular momentum, energy, etc.---with a noncommutative algebra. Some of these dynamical variables take the values $0$ and $1$ only and correspond to properties. For example, we can represent the property of a particle being in a certain region of space by a dynamical variable that takes the value $1$ when the particle is in the region and $0$ otherwise. A dynamical variable like position corresponds to a set of such 2-valued dynamical variables or physical properties. In the case of the position of a particle, these are the properties associated with the particle being in region $R$, for 
all regions $R$. If, for 
all regions $R$, you know whether or not the particle is in that region, you know the position of the particle, and conversely. The 2-valued dynamical variables or properties of a classical system form a Boolean algebra, a subalgebra of the commutative algebra of dynamical variables.

Replacing the commutative algebra of dynamical variables with a noncommutative algebra is equivalent to replacing the Boolean algebra of 2-valued dynamical variables or properties with a non-Boolean algebra. The really essential thing about the classical mode of representation of physical systems in relation to quantum mechanics is that the properties of classical systems are represented as having the structure of a Boolean algebra or Boolean lattice. Every Boolean lattice is isomorphic to a lattice of subsets of a set.\footnote{A lattice\index{lattice} is a partially ordered set in which every pair of elements has a greatest lower bound (or infimum) and least upper bound (or supremum) with respect to the ordering, a minimum element (denoted by 0), and a maximum element (denoted by 1). A Boolean lattice\index{lattice!Boolean} is a complemented, distributive lattice, i.e., every element has a complement (the lattice analogue of set-theoretic complementation) and the distributive law holds for the infimum and supremum. The partial ordering in a Boolean lattice represented by the subsets of a set $X$ corresponds to the partial ordering defined by set inclusion, so the infimum corresponds to set intersection, the supremum corresponds to set union, $0$ corresponds to the null set, and $1$ corresponds to the set $x$. A Boolean algebra\index{Boolean algebra}, defined in terms of algebraic sum ($+$) and product ($.$) operations, is equivalent to a Boolean lattice defined as a partially ordered structure.} To say that the properties of a classical 
system form a Boolean lattice is to say that they can be represented as the subsets of a set, the phase space or state space of classical mechanics. To say that a physical system has a certain property is to associate the system with a certain set in a representation space where the elements of the space---the points of the set---represent all possible states of the system. A state picks out a collection of sets, the sets to which the point representing the state belongs, as the properties of the system in that state. The dynamics of classical mechanics is described in terms of a law of motion describing how the state moves in the state space. As the state changes with time, the set of properties selected by the state changes. (For an elaboration, see \cite{Hughes} and \cite{Bubbook}.)

So the transition from classical to quantum mechanics involves replacing the representation of properties as a Boolean lattice, i.e., as the subsets of a set, with the representation of properties as a certain sort of non-Boolean lattice. Dirac and von Neumann developed Schr\"{o}dinger's equivalence proof into a representation theory for the properties of quantum systems as subspaces in a linear vector space over the complex numbers: Hilbert space. The non-Boolean lattice in question is the lattice of subspaces of this space. Instead of representing properties as the subsets of a set, quantum mechanics  represents properties as  the subspaces of a linear space---as lines, or planes, or hyperplanes, i.e., as a projective geometry. Algebraically, this is the central structural change in the transition from classical to quantum mechanics---although there is more to it: notably the fact that the state space for quantum systems is a Hilbert space over the complex numbers,  not the reals, which is reflected in physical phenomena associated with the possibility of superposing states with different relative phases.

Instead of talking about properties, we can talk equivalently about propositions. (We say that a given property is instantiated if and only if the corresponding proposition is true.) In a Boolean propositional structure, there exist 2-valued homomorphisms on the structure that correspond to truth-value assignments to the propositions. In fact, each 
point in phase space---representing a classical state---defines a truth-value assignment to the subsets representing the propositions: each subset to which the point belongs represents a true proposition or a property that is instantiated by the system, and each subset to which the point does not belong represents a false proposition or a property that is not instantiated by the system. So a classical state corresponds to a complete assignment of truth values to the propositions, or a maximal consistent  `list' of properties of the 
system, and all possible states correspond to all possible maximal consistent lists. 

Probabilities can be introduced on such a classical property structure as measures on the subsets representing the properties. Since each phase space point defines a truth-value assignment, the probability of a property is the measure of the set of truth-value assignments that assign a $1$ (`true') to the property---in effect, we `count' (in the measure-theoretic sense) the relative number of state descriptions in which the property is instantiated (or the corresponding proposition is true), and this number represents the probability of the property. So it makes sense to interpret the probability of a property as a measure of our ignorance as to whether or not the property is instantiated. Probability distributions over classical states represented as phase space points are sometimes referred to as `mixed states,' in which case states corresponding to phase space points are distinguished as `pure states.' 

The problem for a quantum property structure, represented by the lattice of subspaces of a Hilbert space, arises because 2-valued homomorphisms do not exist on these structures (except in the special case of a 2-dimensional Hilbert space). If we take the subspace structure of Hilbert space seriously as the structural feature of quantum mechanics corresponding to the Boolean property structure or propositional structure of classical mechanics, the non-existence of 2-valued homomorphisms on the lattice of subspaces of a Hilbert space means that there is no partition of the  totality of properties of the assocated quantum system into two sets: the properties that are instantiated by the system, and the properties that are not instantiated by the system; i.e., there is no partition of the totality of propositions into true propositions and false propositions. (Of course, other ways of associating propositions with features of a Hilbert space are possible, and other ways of assigning truth values, including multi-valued truth value assignments and contextual truth value assignments. Ultimately, the issue here concerns what we take as the salient structural change involved in the transition from classical to quantum mechanics.)

It might appear that, on the standard interpretation, a pure quantum state  represented by a 1-dimensional subspace in Hilbert space---a minimal element in the subspace structure---defines a truth-value assignment on quantum propositions in an analogous sense to the truth-value assignment on classical propositions defined by a pure classical state. Specifically, on the standard interpretation, a pure quantum state selects the  propositions represented subspaces containing the state as true, and the propositions represented by subspaces  orthogonal to the state as false. (Note that orthogonality is the analogue of set-complement, or negation, in the subspace structure;  the set-theoretical complement of a subspace is not in general a subspace.)

There is, however, an important difference between the two situations. In the case of a classical state, every possible property represented by a phase space subset is selected as either instantiated by the system or not; equivalently, every proposition is either true or false. In the case of a quantum state, the properties represented by Hilbert space subspaces are not partitioned into two such mutually exclusive and collectively exhaustive sets: some propositions are assigned no truth value. Only propositions represented by subspaces that contain the state are assigned the value `true,' and only propositions represented by subspaces orthogonal to the state are assigned the value `false.' This means that propositions represented by subspaces that are at some non-zero or non-orthogonal angle to the ray representing the quantum state are not assigned any truth value in the state, and the corresponding properties must be regarded as \textit{indeterminate} or \emph{indefinite}\index{indeterminate (indefinite) property}: according to the theory, there \textit{can be no fact of the matter} about whether these properties are instantiated or not. 

It turns out that there is only one way to assign (generalized) probabilities to quantum properties, i.e., weights that satisfy the usual Kolmogorov axioms\index{Kolmogorov axioms} for a probability measure on Boolean sublattices of the non-Boolean lattice of quantum properties. This is the content of Gleason's theorem \cite{Gleason}. For a quantum state $\rho$, a property $p$ represented by a projection operator $P$ is assigned the probability $\trace{\rho P}$. If $\rho$ is a pure state $\rho = \ket{\psi}\bra{\psi}$, the probability of $p$ is $\abs{\braket{\psi_{p}}{\psi}}^{2}$, where $\ket{\psi_{p}}$ is the orthogonal projection of $\ket{\psi}$ onto the subspace $P$, i.e., the probability of $p$ is the square of the cosine of the angle between the ray $\ket{\psi}$ and the subspace $P$. This means that properties represented by subspaces containing the state are assigned probability 1, properties represented by subspaces orthogonal to the state are assigned probability 0, and all other properties, represented by subspaces at a non-zero or non-orthogonal angle to the state are assigned a probability between 0 and $1$. So quantum probabilities are not represented as measures over truth-value assignments and cannot be given an ignorance interpretation in the obvious way.

The question now is: \textit{what do these `angle probabilities' or, perhaps better, `angle weights' mean?} The orthodox answer is that the probability assigned to a property of a system by a quantum state is to be understood as the probability of finding the property in a measurement process designed to ascertain whether or not that property obtains. A little thought will reveal that this proposal is very problematic. When the system is represented by a quantum state that assigns a certain property the probability $1/2$, say, this property is indeterminate. Physicists would say that ascribing the property to the system in that state is `meaningless.' But somehow it makes sense to design an experiment to ascertain whether or not the property is instantiated by the system. And in such a measurement, the probability is asserted to be $1/2$ that the experiment will yield the answer `yes,' and $1/2$ that the experiment will yield the answer `no.' Clearly, a measurement process in quantum mechanics is not simply a procedure for ascertaining whether or not a property is instantiated in any straightforward sense. Somehow, a measurement process enables an indeterminate property, that is neither instantiated nor not instantiated by a system in a given quantum state, to either instantiate itself or not with a certain probability; or equivalently, a proposition that is neither true nor false can become true or false with a certain probability in a suitable measurement process. 

The probability problem (as opposed to the truth problem, Schr\"{o}dinger's problem\index{Schr\"{o}dinger's problem}) is the problem of interpreting the `angle weights' as probabilities in some sense (relative frequencies? propensities? subjective Bayesian betting probabilities?) that does not reduce to a purely instrumentalist interpretation of quantum mechanics, according to which the theory is simply regarded as a remarkably accurate instrument for prediction. (Recall Einstein's remark about quantum mechanics being of interest only to shopkeepers and engineers on the Copenhagen interpretation.) The problem arises because of the unique way in which probabilities can be introduced in quantum mechanics, and because the notion of measurement or observation is utterly mysterious on the Copenhagen interpretation. 

In classical theories, we measure to find out what \textit{we don't know}, but in principle a measurement does not change \textit{what is} (and even if it does change what is, this is simply a change or disturbance from one state of being to another that can be derived on the basis of the classical theory itself). In quantum mechanics, measurements apparently bring into being something that was \textit{indeterminate}, not merely unknown, before, i.e., a proposition that was neither true nor false becomes true in a measurement process, and the way in which this happens according to the theory is puzzling, given our deepest assumptions about objectivity, change, and intervention.

Now, \textit{we know how to solve Schr\"{o}dinger's problem}, i.e., we know all the possible ways of modifying quantum mechanics to solve this problem. The problem arises because of the linear dynamics of the theory, which yields a certain entangled state as the outcome of a measurement interaction, and the interpretation of this entangled state as representing a state of affairs that makes certain propositions true, certain propositions false, and other propositions indeterminate. Either we change the linear dynamics in some way, or we keep the linear dynamics and say something non-orthodox about the relation between truth and indeterminateness and the quantum state. Both options have been explored in various ways and in great detail: \textit{we understand the solution space for Schr\"{o}dinger's problem, and the consequences of adopting a particular solution}. 

`Collapse' theories\index{collapse@`collapse' theories}, like the theory developed by Ghirardi\index{Ghirardi}, Rimini\index{Rimini}, and Weber\index{Weber} (GRW\index{GRW}), and extended by Pearle\index{Pearle} \cite{Ghirardi}, solve the problem by modifying the linear dynamics of quantum mechanics. (See Dickson, this vol., ch. 4, for an account.) In the modified theory, there is a certain very small probability that the wavefunction of a particle (the function defined by the components of the quantum state with respect to the position basis in Hilbert space) will spontaneously `collapse'  after being multiplied by a peaked Gaussian of a specified width. For a macroscopic system consisting of many particles, this probability can be close to $1$ for very short time intervals. In effect, this collapse solution modifies the linear dynamics of standard quantum mechanics by adding uncontrollable noise. When the stochastic terms of the modified dynamics become important at the mesoscopic and macroscopic levels, they tend to localize the wave function in space. So measurement interactions involving macroscopic pieces of equipment (or cats) can be distinguished from elementary quantum processes, insofar as they lead to the almost instantaneous collapse of the wave function and the correlation of the measured observable with the position of a localized macroscopic pointer observable.
 
`No collapse' solutions\index{no collapse@`no collapse' theories} are constrained by certain `no go' theorems\index{no go@`no go' theorems} that restrict the assignment of properties, or values to observables, under very general assumptions about the algebra of observables \cite{KochenSpecker}, or restrict the assignment of values to observables under certain assumptions about how distributions of values are related to quantum probabilities \cite{BellEPR}. A theorem by Bub and Clifton \shortcite{BubClifton}\index{Bub}\index{Clifton}\index{Bub-Clifton theorem} shows that if you assume that the set of definite-valued observables has a certain structure (essentially allowing quantum probabilities to be recovered as classical measures over distributions defined by different possible sets of values or properties), and the pointer observable in a measurement process belongs to the set of definite-valued observables, then the class of such theories---so-called `modal interpretations'\index{modal interpretations}---is uniquely specified. More precisely, the sublattice associated with any single observable $R$ is a Boolean lattice, $\hil{B}$, and a quantum state $\ket{\psi}$ defines a classical probability measure on $\hil{B}$, in the sense that all the single and joint probabilities assigned by $\ket{\psi}$ to elements in $\hil{B}$ can be recovered as measures on a Kolmogorov probability space defined on the `phase space' $X$ of 2-valued homomorphisms on $\hil{B}$. The Bub-Clifton theorem characterizes the maximal lattice extension, $\hil{L}$, of any such Boolean sublattice associated with an observable $R$ and a given quantum state $\ket{\psi}$, under the assumption that $\hil{L}$ is an ortholattice,\footnote{I.e., an orthogonal complement exists for every element of $\hil{L}$.} invariant under lattice automorphisms that preserve $R$ and $\ket{\psi}$, for which the probabilities assigned by $\ket{\psi}$ to elements in $\hil{L}$ can be similarly recovered as measures on a Kolmogorov probability space defined on the `phase space' $Y$ of 2-valued homomorphisms on $\hil{L}$. In this sense, the theorem characterizes the limits of classicality in a quantum propositional structure. It turns out that different modal interpretations can be associated with different `determinate sublattices' $\hil{L}$, i.e., with different choices of a `preferred observable' $R$. For standard quantum mechanics, $R$ is the identity, and the determinate sublattice $\hil{L}$ consists of all quantum propositions represented by subspaces containing the state $\ket{\psi}$ (propositions assigned probability 1 by $\ket{\psi}$) and subspaces orthogonal to $\ket{\psi}$ (propositions assigned probability 0 by $\ket{\psi}$. Bohm's hidden variable theory\index{Bohm's theory} can be regarded as a modal interpretation in which the preferred observable is position in configuration space. (See Dickson, this vol., ch. 4, and \cite{Goldstein} for an account of Bohm's theory\index{Bohm's theory}.)
 
An alternative type of `no-collapse' solution to the Schr\"{o}dinger problem is provided by the
Everett interpretation\index{Everett interpretation} \cite{Everett}. (See Dickson, this vol., ch. 4, for an account.) There are a variety of Everettian interpretations in the literature, the common theme being that all possible outcomes of a measurement are regarded as actual in some indexical sense, relative to different terms in the global entangled state (with respect to a certain preferred basis in Hilbert space), which are understood to be associated with different worlds or different minds, depending on the version. The most sophisticated formulation of Everett's interpretation is probably the Saunders-Wallace version\index{Everett interpretation!Saunders-Wallace version} \cite{Saunders,Wallace}\index{Saunders}\index{Wallace}. Here the preferred basis is selected by decoherence (see below), and probabilities are introduced as rational betting probabilities in the Bayesian sense via a decision-theoretic argument originally due to Deutsch \shortcite{Deutsch1999}. 

To sum up: any solution to Schr\"{o}dinger's measurement problem involves either modifying the linear dynamics of the theory (`collapse' theories), or taking some observable in addition to the identity as having a determinate value in every quantum state, and modifying what the standard theory says about what propositions are true, false, and indeterminate in a quantum state (modal interpretations, `no collapse' hidden variable theories), so that at the end of a measurement interaction that correlates macroscopic pointer positions with possible values of a measured observable, the pointer propositions and propositions referring to measured values end up having determinate truth values. Alternatively (Everettian interpretations), we can interpret quantum mechanics so that every measurement outcome becomes determinate in some indexical sense (with respect to different worlds, or different minds, or different branches of the entangled state, etc.). 
	
 We know in considerable detail what these solutions look like, in terms of how quantum mechanics is modified. It was a useful project to explore these solutions, because we learnt something about quantum mechanics in the process, and perhaps there is more to learn by exploring the solution space further. But the  point to note here is that all these solutions to the `truth problem' of measurement distort quantum mechanics in various ways by introducing additional structural features that obscure rather than illuminate our understanding of the phenomena involved in information-theoretic applications of entanglement, such as quantum teleportation, the possibility and impossibility of certain quantum cryptographic protocols relative to classical protocols, the exponential speed-up of quantum computation algorithms relative to classical algorithms,  and so on. 
 
 Consider again the Bohr-Einstein dispute about the interpretation of quantum mechanics. One might say that what separated Einstein (and Schr\"{o}dinger) and Bohr was their very different answers to what van Fraassen\index{van Fraassen} \shortcite[p. 4]{Fraassen} has called `the foundational question \emph{par excellence}: \emph{how could the world possibly be the way quantum theory says it is?} This would be misleading. Einstein\index{Einstein} answered this question by arguing that the world couldn't be the way quantum theory says it is, unless the theory is not the whole story (so a `completion' of the theory---perhaps Einstein's sought-after unified field theory---would presumably answer the question). But Bohr's complementarity interpretation\index{Bohr}\index{complementarity} is not intended to be an answer to this question. Rather, complementarity should be understood as suggesting an answer to a different question: \emph{why must the world be the way quantum theory says it is?}
 
To bring out the difference between these two questions, consider Einstein's distinction between what he called `principle' versus `constructive' theories\index{principle vs constructive theories}. Einstein introduced this distinction in an article
 on the significance of the special and 
general theories of relativity that he wrote for the London 
\textit{Times}, which appeared in the issue of November 28, 1919 
\shortcite{EinsteinTimes}:

\begin{quote}
    We can distinguish various kinds of theories in physics. Most of 
    them are constructive. They attempt to build up a picture of the 
    more complex phenomena out of the material of a relatively simple 
    formal scheme from which they start out. Thus the kinetic theory 
    of gases seeks to reduce mechanical, thermal, and diffusional 
    processes to movements of molecules---i.e., to build them up out 
    of the hypothesis of molecular motion. When we say that we have 
    succeeded in understanding a group of natural processes, we 
    invariably mean that a constructive theory has been found which 
    covers the processes in question. 
    
    Along with this most important class of theories there exists a 
    second, which I will call `principle theories.' These employ the 
    analytic, not the synthetic, method. The elements which form their 
    basis and starting-point are not hypothetically constructed but 
    empirically discovered ones, general characteristics of natural 
    processes, principles that give rise to mathematically formulated 
    criteria which the separate processes or the theoretical 
    representations of them have to satisfy. Thus the science of 
    thermodynamics seeks by analytical means to deduce necessary 
    conditions, which separate events have to satisfy, from the 
    universally experienced fact that perpetual motion is impossible.
\end{quote} 

Einstein's point was that 
relativity theory is to be understood 
as a principle theory\index{principle vs constructive theories}. He returns to this theme in his 
`Autobiographical Notes' \shortcite[pp. 51--52]{EinsteinBiog},
 where he remarks that he first tried to find a 
constructive theory\index{principle vs constructive theories} that would account for the known properties of 
mater and radiation, but eventually became convinced that the 
solution to the problem was to be found in a principle theory that
 reconciled the constancy of the velocity 
of light in vacuo for all inertial frames of reference, and the 
equivalence of inertial frames for all physical laws (mechanical 
as well as electromagnetic):

\begin{quote}
    Reflections of this type made it clear to me as long ago as 
shortly 
    after 1900, i.e., shortly after Planck's trailblazing work, that 
    neither mechanics nor electrodynamics could (except in limiting 
    cases) claim exact validity. By and by I despaired of the 
    possibility of discovering the true laws by means of constructive 
    efforts based on known facts. The longer and the more 
despairingly 
    I tried, the more I came to the conviction that only the 
    discovery of a universal formal principle could lead us to 
    assured results. The example I saw before me was thermodynamics. 
    The general principle was there given in the theorem: the laws of 
    nature are such that it is impossible to construct a 
    \textit{perpetuum mobile} (of the first and second kind). How, 
    then, could such a universal principle be found?
\end{quote}

A little later \shortcite[p. 57]{EinsteinBiog}, he adds:

\begin{quote}
    The universal principle of the special theory of relativity is 
    contained in the postulate: The laws of physics are invariant 
    with respect to the Lorentz-transformations (for the transition 
    from one inertial system to any other arbitrarily chosen system 
    of inertia). This is a restricting principle for natural laws, 
    comparable to the restricting principle for the non-existence of 
    the \textit{perpetuum mobile} which underlies thermodyamics.
\end{quote} 

According to Einstein, two very different sorts of 
theories should be distinguished in physics. One sort involves the 
reduction of a domain of 
relatively complex phenomena to the properties of simpler 
elements, as in the kinetic theory, which reduces
the mechanical and thermal behavior of gases to the motion of 
molecules, the elementary building blocks of the constructive theory\index{principle vs constructive theories}. 
The other sort of theory is formulated in terms of `no go' principles 
that 
impose constraints on physical processes or events, as in 
thermodynamics (`no perpetual motion machines'). 
For an illuminating account of the role 
played by this distinction in Einstein's work, see the discussion by 
Martin Klein in \shortcite{Klein}.

The special theory of relativity\index{relativity theory} is a principle theory\index{principle vs constructive theories}, formulated in 
terms of two 
principles: the equivalence of inertial frames for all physical laws 
(the laws of electromagnetic phenomena as well as the laws of 
mechanics), and the constancy of the velocity of light in vacuo for 
all inertial frames. These principles are irreconcilable in the 
geometry of Newtonian space-time, where inertial frames are 
related by Galilean transformations. The required revision yields
Minkowski geometry, where inertial 
frames are related by Lorentz transformations. Einstein characterizes
the special principle of relativity, that the laws of physics are 
invariant with respect to Lorentz transformations from one inertial 
system to another, as `a restricting principle for 
natural laws, comparable to the restricting principle for the 
non-existence of the \textit{perpetuum mobile} which underlies 
thermodynamics.' (In the case of the general theory of relativity, 
the group of allowable transformations includes all differentiable 
transformations of the space-time manifold onto itself.) 
By contrast, the Lorentz theory\index{Lorentz theory} \cite{Lorentz}, 
which 
derives the Lorentz transformation from the electromagnetic 
properties of the aether, 
and assumptions about the transmission of molecular forces through 
the aether, 
is a constructive theory. 

The question:
\begin{quote}
How could the world possibly be the way the quantum theory says it is?
\end{quote}
is motivated by a difficulty in interpreting quantum mechanics as a constructive theory, and the appropriate response is some constructive repair to the theory that resolves the difficulty, or the demonstration that the puzzling features of quantum mechanics at the phenomenal level (the phenomena of interference and entanglement) can be derived from a physically unproblematic constructive theory. 

The question:
\begin{quote}
Why must the world be the way the quantum theory says it is?
\end{quote}
does not ask for a `bottom-up' explanation of quantum phenomena in terms of a physical ontology and dynamical laws. Rather, the question concerns a `top-down' derivation of quantum mechanics as a principle theory, in terms of  operational constraints on the possibilities of manipulating phenomena. In the case of quantum mechanics, the relevant phenomena concern information.

This shift in perspective between the two questions is highlighted in a remark by Andrew Steane\index{Steane} in his review article on `Quantum Computing' \shortcite[p.119]{Steane98}:
\begin{quote}
Historically, much of fundamental physics has been concerned with
discovering the fundamental particles of nature and the equations
which describe their motions and interactions. It now appears that a
different programme may be equally important: to discover the ways
that nature allows, and prevents, \emph{information} to be expressed and
manipulated, rather than particles to move.
\end{quote}
Steane concludes his review with the following proposal \shortcite[p. 171]{Steane98}:
\begin{quote}
To conclude with, I would like to propose a more wide-ranging theoretical task: to arrive at a set of principles like energy and momentum conservation, but which apply to information, and from which much of quantum mechanics could be derived. Two tests of such ideas would be whether the EPR-Bell correlations thus became transparent, and whether they rendered obvious the proper use of terms such as `measurement' and `knowledge.'
\end{quote} 

A similar shift in perspective is implicit in Wheeler's\index{Wheeler} question `Why the quantum?,' one of Wheeler's `Really Big Questions' \shortcite{Wheeler}. Steane's suggestion is to answer the question by showing how quantum mechanics can be derived from information-theoretic principles. A more specific proposal along these lines originates with Gilles Brassard\index{Brassard} and Chris Fuchs\index{Fuchs}. As remarked in \S \ref{sec:bithistory}, Brassard and
Fuchs \cite{Brassard,Fuchs1,Fuchs2,FuchsJacobs} speculated that
  quantum mechanics could be derived from information-theoretic
constraints formulated in terms of certain primitive cryptographic protocols:
specifically, the possibility of unconditionally secure key 
distribution, and the
impossibility of unconditionally secure bit commitment.

The CBH theorem\index{CBH theorem} (motivated by the Brassard-Fuchs conjecture) shows that quantum 
mechanics  can be regarded as a principle theory\index{principle vs constructive theories} in Einstein's sense, where the principles are 
information-theoretic constraints. So we have an answer to the question: why must the world be the way quantum mechanics says it is? The phenomena of interference and nonlocal entanglement are bound to occur in a world in which there are certain constraints on the acquisition, communication, and processing  of information.  

Consider, for comparison, relativity theory\index{relativity theory}, the other pillar of modern physics. A relativistic theory is a theory with certain 
symmetry or invariance properties, defined in terms of a group of 
space-time transformations. Following Einstein's formulation of 
special relativity as a principle theory, we understand this 
invariance to be a consequence of the fact that we live in a world in 
which natural processes are subject to certain constraints: roughly (as Hermann Bondi\index{Bondi} \shortcite{Bondi} puts it), `no overtaking of light by light,' and `velocity doesn't matter' (for electromagnetic as well as mechanical phenomena). Recall 
Einstein's characterization of the special principle of relativity as 
`a restricting principle for natural laws, comparable to the 
restricting principle of the non-existence of the \textit{perpetuum 
mobile} which underlies thermodynamics.') Without Einstein's analysis, the transformations of Minkowski space-time would simply be a rather puzzling algorithm for relativistic kinematics 
and the Lorentz transformation, which is incompatible with the 
kinematics of Newtonian space-time. What Einstein's analysis provides is a 
rationale for taking the structure of space and time as Minkowskian: we 
see that this is required for the consistency of the two principles of 
special relativity. 

A quantum theory is a 
theory in which the observables and states have a certain 
characteristic algebraic structure. Unlike relativity theory, quantum 
mechanics was born as a recipe or algorithm for caclulating the 
expectation values of observables measured by macroscopic measuring 
instruments. A theory with a commutative $C^{*}$-algebra has a phase space 
representation---not necessarily the phase space of 
classical mechanics, but a theory in which the observables of the 
$C^{*}$-algebra are replaced by `beables' (Bell's term, see \shortcite{BellBeables}), 
and the $C^{*}$-algebraic states are replaced by beable-states representing complete 
lists of properties (idempotent quantities). In this case, it is 
possible to extend the theory to include the measuring instruments 
that are the source of the $C^{*}$-algebraic statistics, so that they are no 
longer `black boxes' but constructed out of systems that are 
characterized by properties and states of the phase space theory. 
That is, the $C^{*}$-algebraic theory can be replaced by a `detached observer' 
theory of the 
physical processes underlying the phenomena, to use Pauli's term 
\cite[p. 218]{Born}, 
including the processes involved in the functioning of measuring 
instruments. Note that this depends on a representation theorem. In the 
noncommutative case, we are guaranteed only the existence of a 
Hilbert space representation of the $C^{*}$-algebra, and it is an open 
question whether a `detached observer' description of the phenomena is 
possible. 

Solving Schr\"{o}dinger's problem\index{Schr\"{o}dinger's problem}---the truth problem---amounts to a proposal to treat quantum mechanics as a failed or incomplete constructive theory  in need of constructive repair. In effect, the problem is \emph{how to account for quantum information}---the puzzling features of interference and nonlocal entanglement---in a theoretical framework in which only classical information is meaningful in a fundamental sense. If we treat quantum mechanics as a principle theory of information, the core foundational problem  is the probability problem. From this perspective, the problem is \emph{how to account for the appearance of classical information} in a quantum world characterized by information-theoretic constraints. 

One might complain that treating quantum mechanics as a principle theory amounts to simply postulating what is ultimately \emph{explained} by a constructive theory like the GRW theory or Bohm's theory. This would amount to rejecting the idea that a principle theory\index{principle vs constructive theories} can be explanatory. From the perspective adopted here, Bohm's constructive theory\index{Bohm's theory} in relation to quantum mechanics is like Lorentz's constructive theory\index{Lorentz theory} of the electron in relation to special relativity. Cushing \shortcite[p. 204]{Cushing} quotes Lorentz\index{Lorentz} 
(from the conclusion of the 1916 edition of 
\textit{The Theory of Electrons}) as complaining similarly that `Einstein simply 
postulates what we have deduced.'
\begin{quotation}
I cannot speak here of the many highly interesting applications which 
Einstein has made of this principle [of relativity]. 
His results concerning electromagnetic and optical phenomena ... agree in 
the main with those which we have obtained in the preceding pages, 
the chief difference being that Einstein simply postulates what we have deduced, 
with some difficulty and not altogether satisfactorily, 
from the fundamental equations of the electromagnetic field. By doing 
so, he may certainly take credit for making us see in the negative 
result of experiments like those of Michelson, Rayleigh and Brace, 
not a fortuitous compensation of opposing effects, but the 
manifestation of a general and fundamental principle. 

Yet, I think, something may also be claimed in favour of the form in which 
I have presented the theory. I cannot but regard the aether, which can be 
the seat of an electromagnetic field with its energy and its vibrations, 
as endowed with a certain degree of substantiality, however different 
it may be from all ordinary matter. In this line of thought, it seems natural 
not to assume at starting that it can never make any difference whether a 
body moves through the aether or not, and to measure distances and lengths 
of time by means of rods and clocks having a fixed position relative to 
the aether.
\end{quotation}

Note that  Lorentz's theory is constrained by the  principles of special relativity, which 
means that the aether as a rest frame for electromagnetic phenomena 
must, in principle, be undetectable. So such a theory can have no excess 
empirical content over special relativity. Cushing \shortcite[p. 
193]{Cushing} 
also quotes Maxwell as asking whether `it is not more philosophical to 
admit the existence of a medium which we cannot at present perceive, 
than to assert that a body can act at a place where it is not.' Yes, 
but not if we also have to admit that, in principle, as a matter of 
physical law, if we live in a world in 
which events are constrained by the two relativistic principles, the medium 
must remain undetectable.

You can, if you like, 
tell a constructive story about quantum phenomena, but such an account, if constrained by the 
information-theoretic principles, will have no excess 
empirical content over quantum mechanics. Putting this differently, a solution to Schr\"{o}dinger's truth problem that has excess empirical content over quantum mechanics must violate one or more of the CBH information-theoretic constraints.
So, e.g., a Bohmian theory of quantum phenomena is like an aether theory 
for electromagnetic fields. Just as 
the aether theory attempts to make sense of the behaviour of fields by 
proposing an aether that is a sort of \textit{sui generis} mechanical system 
different from all other mechanical systems, so Bohm's theory\index{Bohm's theory}
attempts to make sense of quantum phenomena by introducing a field 
(the quantum potential or guiding field) that is a sort of \textit{sui 
generis} field different from other physical fields. 

The crucial distinction here is between a constructive theory formulated in terms of a physical ontology and dynamical laws (`bottom-up') and a principle theory formulated in terms of operational constraints at the phenomenal level (`top-down')\index{principle vs constructive theories}. A constructive theory introduces an algebra of beables and beable-states. A principle theory introduces an algebra of  observables and observable-states, which are essentially probability measures. 

It seems clear that the algebra of observables will be non-trivially distinct from the algebra of beables if cloning\index{cloning!and measurement} is impossible. For if a constructive theory for a certain domain of phenomena allows dynamical interactions in which a beable of one system, designated as the measuring instrument, can become correlated with a beable of another system, designated as the measured system, without disturbing the values of other beables of the measured system, we can take such an interaction as identifying the value of the beable in question (in the sense that the value of a beable of one system is recorded in the value of a beable of a second system). If this is possible, then it will be possible to  simultaneously measure any number of beables of a system by concatenating measurement interactions, and so it will be possible in principle to  identify any arbitrary beable state. If we assume that we can prepare any state, then the possibility of identifying an arbitrary state means that we can construct a device that could copy any arbitrary state. So if we \emph{cannot} construct such a device, then measurement in this sense must also be impossible. It follows that a `measurement' in the constructive theory will be something other than the mere identification of a beable value of a system, without disturbance,  and the question of what the observables are in such a theory will require a non-trivial analysis. 

Such an analysis is indeed given by Bohm in Part II of his two-part 1952 paper on hidden variables \shortcite{Bohm2}, and a more careful and sophisticated analysis is given by \cite{DGZ} for their `Bohmian mechanics' version of Bohm's theory. As one would expect (given the equilibrium distribution assumption  which ensures that Bohm's theory is empirically indistinguishable from quantum mechanics), while the \emph{beables} are functions of position in configuration space (and  form a commutative algebra), the \emph{observables} of the theory are just the observables of quantum theory and form a noncommutative algebra.

The CBH theorem assumes that, for the theories we are concerned with, the \emph{observables} form a $C^{*}$-algebra. The content of the CBH theorem is that, given certain information-theoretic constraints, the $C^{*}$-algebra of \emph{observables} and \emph{observable states} takes a certain form characteristic of quantum theories. The theorem says nothing about beables and beable-states, and  does not address the measurement problem (Schr\"{o}dinger's truth problem), let alone solve it. But from the perspective adopted here, the measurement problem is simply the observation that cloning is impossible, and a `solution to the measurement problem' is the proposal of a physical ontology and dynamics and an analysis of measurement that yields the observables and observable-states of standard quantum mechanics. Such theories provide possible explanations for the impossibility of cloning. But since there are now a variety of such explanations available, and---assuming the CBH information-theoretic principles---there are no empirical constraints, in principle, that could distinguish these explanations, there seems little point in pursuing the question further. A constructive theory whose \emph{sole} motivation is to `solve the measurement problem' seems unlikely to survive fundamental advances in physics driven by other theoretical or experimental problems

The probability problem---the core foundational problem for the interpretation of quantum mechanics as a principle theory of information---can be put this way: From the information-theoretic constraints, we get a noncommutative (or non-Boolean) theory of correlations for which there is no phase space representation. One can define, in a unique way (according to Gleason's theorem) generalized `transition probabilities' or `transition weights' associated with certain structural features of the noncommutative structure: the angles between geometrical elements representing quantum `propositions.' The problem is  how to understand these weights as representing probabilities, \emph{without reducing the problem to a solution of the truth problem.}

It seems clear that we need to take account of the phenomenon of decoherence\index{decoherence} (see Landsmann, this vol., ch. 5; Dickson, this vol., ch. 4; \cite{Zurek1,Zurek2}): an extremely fast process that occurs in the spontaneous interaction between a macrosystem and its environment that leads to the virtually instantaneous suppression of quantum interference. What happens, roughly, is that a macrosystem like Schr\"{o}dinger's cat typically becomes correlated with the environment---an enormous number of stray dust particles, air molecules, photons, background radiation, etc.---in an entangled state that takes a certain form with respect to a preferred set of basis states, which remain stable as the interaction develops and includes more and more particles. It is as if the environment is `monitoring' the system via a measurement  of properties associated with the preferred states, in such a way that information about these properties is stored redundantly in the environment. This stability, or robustness, of the preferred basis, and the redundancy of the information in the environment, allows one to identify certain emergent structures in the overall pattern of correlations---such as macroscopic pointers and cats and information-gatherers in general---as classical-like: the correlational information required to reveal quantum interference for these structures is effectively lost in the environment. So it appears that the information-theoretic constraints are consistent with both (i)  the conditions for the existence of measuring instruments as sources of classical information, and (ii) the existence of information-gatherers with the ability to use measuring instruments to apply and test quantum mechanics, given a characterization of part of the overall system as the environment. That is, decoherence provides an explanation for the emergence of classical information in a quantum correlational structure. 

If something like the above account of decoherence is acceptable, then the probability problem reduces to showing that the probabilities assigned to measurement outcomes by these information-gatherers, in the subjective Bayesian sense, are just the Gleason generalized transition probabilities. That is, we need to show that, while quantum theory, at the fundamental level, is a noncommutative theory of correlations for which there is no phase space representation,  it is \emph{also} a theory of the probabilistic behavior of information-gatherers, certain emergent structures in the pattern of correlations when correlational information in their environment is ignored. For an argument along these lines, see  \cite{Pitowsky}. 

On the view proposed here, no measurement outcomes are certified as determinate by the theory. Rather, measuring instruments are sources of classical information, where the individual occurrence of a particular distinguishable event (`symbol') produced stochastically by the information source lies outside the theory. In this sense, a measuring instrument, insofar as it functions as a classical information source, is still ultimately a `black box' in the theory. So a quantum description will
have to introduce a `cut' between what we take to be the ultimate measuring 
instrument in a given measurement process
and the quantum phenomenon revealed by the instrument. But this `cut' is no longer \emph{ad hoc}, or mysterious, or in some other way problematic, as it is in the Copenhagen interpretation\index{Copenhagen interpretation} (see Landsmann, this vol., ch. 5). For here the `cut' just reflects the fundamental interpretative claim: that quantum mechanics is \textit{a theory about the 
  representation and manipulation of information} constrained by 
  the possibilities and 
impossibilities of information-transfer in our world, rather than a theory 
  about the ways in which nonclassical waves and particles move.

\bibliography{Handbook}

\printindex

\end{document}